\documentclass[
aps,prd,
reprint,
nofootinbib,
superscriptaddress,
amsfonts,amsmath,amssymb,
floatfix
]{revtex4-2}
\usepackage{graphicx}
\usepackage{hyperref}
\usepackage{physics}
\usepackage[capitalize]{cleveref}
\usepackage{multirow}
\usepackage[dvipsnames]{xcolor}
\usepackage{orcidlink}
\usepackage{aas_macros}

\hypersetup{colorlinks=true,allcolors=blue}

\newcommand{\lcdm}{$\Lambda$CDM}
\newcommand{\planck}{\textit{Planck}}
\newcommand{\euclid}{\textit{Euclid}}
\newcommand{\agora}{\textsc{Agora}}
\newcommand{\camb}{\textsc{CAMB}}
\newcommand{\hmcode}{\textsc{HMCode}}
\newcommand{\healpix}{\textsc{HEALPix}}
\newcommand{\namaster}{\textsc{NaMASTER}}
\newcommand{\tensiometer}{\textsc{tensiometer}}

\newcommand{\kcmb}{\kappa_\text{CMB}}
\newcommand{\nside}{N_{\text{side}}}
\newcommand{\lmax}{\ell_\text{max}}
\newcommand{\kmax}{k_\text{max}}
\newcommand{\sqdeg}{deg$^2$}
\newcommand{\ukam}{$\mu$K-arcmin}
\newcommand{\logTagn}{\log T_\text{AGN}}
\newcommand{\Om}{\Omega_\text{m}}
\newcommand{\Ob}{\Omega_\text{b}}
\newcommand{\Npeff}{N_{p,\text{eff}}}

\newcommand{\desfull}{\texttt{Full}}
\newcommand{\desblue}{\texttt{Blue}}
\newcommand{\gmv}{\texttt{GMV}}
\newcommand{\pol}{\texttt{Pol}}
\newcommand{\prof}{\texttt{GMVprof}}
\newcommand{\mh}{\texttt{GMVtszdpj}}
\newcommand{\xilc}{\texttt{GMVxilc}}

\newcommand{\Oinprep}{Y. Omori et al., in preparation}

\begin{document}

\title{Cross-correlation of SPT-3G D1 CMB lensing and DES Y3 galaxy lensing}

\author{A.~Ouellette\,\orcidlink{0000-0003-0170-5638}}
\email{aaronjo2@illinois.edu}
\affiliation{Department of Physics, University of Illinois Urbana-Champaign, 1110 West Green Street, Urbana, IL, 61801, USA}
\author{Y.~Omori\,\orcidlink{0000-0002-0963-7310}}
\affiliation{Department of Astronomy and Astrophysics, University of Chicago, 5640 South Ellis Avenue, Chicago, IL, 60637, USA}
\affiliation{Kavli Institute for Cosmological Physics, University of Chicago, 5640 South Ellis Avenue, Chicago, IL, 60637, USA}
\affiliation{NSF-Simons AI Institute for the Sky (SkAI), 172 E. Chestnut St., Chicago, IL 60611, USA}
\author{E.~Anderes\,\orcidlink{0009-0003-3245-3979}}
\affiliation{Department of Statistics, University of California, One Shields Avenue, Davis, CA 95616, USA}
\author{A.~J.~Anderson\,\orcidlink{0000-0002-4435-4623}}
\affiliation{Fermi National Accelerator Laboratory, MS209, P.O. Box 500, Batavia, IL, 60510, USA}
\affiliation{Kavli Institute for Cosmological Physics, University of Chicago, 5640 South Ellis Avenue, Chicago, IL, 60637, USA}
\affiliation{Department of Astronomy and Astrophysics, University of Chicago, 5640 South Ellis Avenue, Chicago, IL, 60637, USA}
\author{B.~Ansarinejad}
\affiliation{School of Physics, University of Melbourne, Parkville, VIC 3010, Australia}
\author{M.~Archipley\,\orcidlink{0000-0002-0517-9842}}
\affiliation{Department of Astronomy and Astrophysics, University of Chicago, 5640 South Ellis Avenue, Chicago, IL, 60637, USA}
\affiliation{Kavli Institute for Cosmological Physics, University of Chicago, 5640 South Ellis Avenue, Chicago, IL, 60637, USA}
\author{L.~Balkenhol\,\orcidlink{0000-0001-6899-1873}}
\affiliation{Sorbonne Universit\'e, CNRS, UMR 7095, Institut d'Astrophysique de Paris, 98 bis bd Arago, 75014 Paris, France}
\author{D.~R.~Barron\,\orcidlink{0000-0002-1623-5651}}
\affiliation{Department of Physics and Astronomy, University of New Mexico, Albuquerque, NM, 87131, USA}
\author{K.~Benabed}
\affiliation{Sorbonne Universit\'e, CNRS, UMR 7095, Institut d'Astrophysique de Paris, 98 bis bd Arago, 75014 Paris, France}
\author{A.~N.~Bender\,\orcidlink{0000-0001-5868-0748}}
\affiliation{High-Energy Physics Division, Argonne National Laboratory, 9700 South Cass Avenue, Lemont, IL, 60439, USA}
\affiliation{Kavli Institute for Cosmological Physics, University of Chicago, 5640 South Ellis Avenue, Chicago, IL, 60637, USA}
\affiliation{Department of Astronomy and Astrophysics, University of Chicago, 5640 South Ellis Avenue, Chicago, IL, 60637, USA}
\author{B.~A.~Benson\,\orcidlink{0000-0002-5108-6823}}
\affiliation{Fermi National Accelerator Laboratory, MS209, P.O. Box 500, Batavia, IL, 60510, USA}
\affiliation{Kavli Institute for Cosmological Physics, University of Chicago, 5640 South Ellis Avenue, Chicago, IL, 60637, USA}
\affiliation{Department of Astronomy and Astrophysics, University of Chicago, 5640 South Ellis Avenue, Chicago, IL, 60637, USA}
\author{F.~Bianchini\,\orcidlink{0000-0003-4847-3483}}
\affiliation{Kavli Institute for Particle Astrophysics and Cosmology, Stanford University, 452 Lomita Mall, Stanford, CA, 94305, USA}
\affiliation{Department of Physics, Stanford University, 382 Via Pueblo Mall, Stanford, CA, 94305, USA}
\affiliation{SLAC National Accelerator Laboratory, 2575 Sand Hill Road, Menlo Park, CA, 94025, USA}
\author{L.~E.~Bleem\,\orcidlink{0000-0001-7665-5079}}
\affiliation{High-Energy Physics Division, Argonne National Laboratory, 9700 South Cass Avenue, Lemont, IL, 60439, USA}
\affiliation{Kavli Institute for Cosmological Physics, University of Chicago, 5640 South Ellis Avenue, Chicago, IL, 60637, USA}
\affiliation{Department of Astronomy and Astrophysics, University of Chicago, 5640 South Ellis Avenue, Chicago, IL, 60637, USA}
\author{S.~Bocquet\,\orcidlink{0000-0002-4900-805X}}
\affiliation{University Observatory, Faculty of Physics, LMU Munich, Scheinerstr.~1, 81679 Munich, Germany}
\author{F.~R.~Bouchet\,\orcidlink{0000-0002-8051-2924}}
\affiliation{Sorbonne Universit\'e, CNRS, UMR 7095, Institut d'Astrophysique de Paris, 98 bis bd Arago, 75014 Paris, France}
\author{E.~Camphuis\,\orcidlink{0000-0003-3483-8461}}
\affiliation{Sorbonne Universit\'e, CNRS, UMR 7095, Institut d'Astrophysique de Paris, 98 bis bd Arago, 75014 Paris, France}
\author{M.~G.~Campitiello}
\affiliation{High-Energy Physics Division, Argonne National Laboratory, 9700 South Cass Avenue, Lemont, IL, 60439, USA}
\author{J.~E.~Carlstrom\,\orcidlink{0000-0002-2044-7665}}
\affiliation{Kavli Institute for Cosmological Physics, University of Chicago, 5640 South Ellis Avenue, Chicago, IL, 60637, USA}
\affiliation{Enrico Fermi Institute, University of Chicago, 5640 South Ellis Avenue, Chicago, IL, 60637, USA}
\affiliation{Department of Physics, University of Chicago, 5640 South Ellis Avenue, Chicago, IL, 60637, USA}
\affiliation{High-Energy Physics Division, Argonne National Laboratory, 9700 South Cass Avenue, Lemont, IL, 60439, USA}
\affiliation{Department of Astronomy and Astrophysics, University of Chicago, 5640 South Ellis Avenue, Chicago, IL, 60637, USA}
\author{J.~Carron\,\orcidlink{0000-0002-5751-1392}}
\affiliation{Istituto ricerche solari Aldo e Cele Dacc\`o (IRSOL), Faculty of Informatics, Universit\`a della Svizzera italiana, 6605 Locarno, Switzerland}
\affiliation{Universit\'e de Gen\`eve, D\'epartement de Physique Th\'eorique, 24 Quai Ansermet, CH-1211 Gen\`eve 4, Switzerland}
\author{C.~Chang\,\orcidlink{0000-0002-7887-0896}}
\affiliation{Department of Astronomy and Astrophysics, University of Chicago, 5640 South Ellis Avenue, Chicago, IL, 60637, USA}
\affiliation{Kavli Institute for Cosmological Physics, University of Chicago, 5640 South Ellis Avenue, Chicago, IL, 60637, USA}
\affiliation{NSF-Simons AI Institute for the Sky (SkAI), 172 E. Chestnut St., Chicago, IL 60611, USA}
\author{C.~L.~Chang}
\affiliation{High-Energy Physics Division, Argonne National Laboratory, 9700 South Cass Avenue, Lemont, IL, 60439, USA}
\affiliation{Kavli Institute for Cosmological Physics, University of Chicago, 5640 South Ellis Avenue, Chicago, IL, 60637, USA}
\affiliation{Department of Astronomy and Astrophysics, University of Chicago, 5640 South Ellis Avenue, Chicago, IL, 60637, USA}
\author{P.~M.~Chichura\,\orcidlink{0000-0002-5397-9035}}
\affiliation{Department of Physics, University of Chicago, 5640 South Ellis Avenue, Chicago, IL, 60637, USA}
\affiliation{Kavli Institute for Cosmological Physics, University of Chicago, 5640 South Ellis Avenue, Chicago, IL, 60637, USA}
\author{A.~Chokshi}
\affiliation{Department of Astronomy and Astrophysics, University of Chicago, 5640 South Ellis Avenue, Chicago, IL, 60637, USA}
\author{T.-L.~Chou\,\orcidlink{0000-0002-3091-8790}}
\affiliation{Department of Astronomy and Astrophysics, University of Chicago, 5640 South Ellis Avenue, Chicago, IL, 60637, USA}
\affiliation{Kavli Institute for Cosmological Physics, University of Chicago, 5640 South Ellis Avenue, Chicago, IL, 60637, USA}
\affiliation{National Taiwan University, No. 1, Sec. 4, Roosevelt Road, Taipei 106319, Taiwan}
\author{A.~Coerver\,\orcidlink{0000-0002-2707-1672}}
\affiliation{Department of Physics, University of California, Berkeley, CA, 94720, USA}
\author{T.~M.~Crawford\,\orcidlink{0000-0001-9000-5013}}
\affiliation{Department of Astronomy and Astrophysics, University of Chicago, 5640 South Ellis Avenue, Chicago, IL, 60637, USA}
\affiliation{Kavli Institute for Cosmological Physics, University of Chicago, 5640 South Ellis Avenue, Chicago, IL, 60637, USA}
\author{C.~Daley\,\orcidlink{0000-0002-3760-2086}}
\affiliation{Universit\'e Paris-Saclay, Universit\'e Paris Cit\'e, CEA, CNRS, AIM, 91191, Gif-sur-Yvette, France}
\affiliation{Department of Astronomy, University of Illinois Urbana-Champaign, 1002 West Green Street, Urbana, IL, 61801, USA}
\author{T.~de~Haan\,\orcidlink{0000-0001-5105-9473}}
\affiliation{High Energy Accelerator Research Organization (KEK), Tsukuba, Ibaraki 305-0801, Japan}
\author{K.~R.~Dibert}
\affiliation{Department of Astronomy and Astrophysics, University of Chicago, 5640 South Ellis Avenue, Chicago, IL, 60637, USA}
\affiliation{Kavli Institute for Cosmological Physics, University of Chicago, 5640 South Ellis Avenue, Chicago, IL, 60637, USA}
\author{M.~A.~Dobbs}
\affiliation{Department of Physics and McGill Space Institute, McGill University, 3600 Rue University, Montreal, Quebec H3A 2T8, Canada}
\affiliation{Canadian Institute for Advanced Research, CIFAR Program in Gravity and the Extreme Universe, Toronto, ON, M5G 1Z8, Canada}
\author{M.~Doohan}
\affiliation{School of Physics, University of Melbourne, Parkville, VIC 3010, Australia}
\author{D.~Dutcher\,\orcidlink{0000-0002-9962-2058}}
\affiliation{Joseph Henry Laboratories of Physics, Jadwin Hall, Princeton University, Princeton, NJ 08544, USA}
\author{C.~Feng}
\affiliation{Department of Astronomy, University of Science and Technology of China, Hefei 230026, China}
\affiliation{School of Astronomy and Space Science, University of Science and Technology of China, Hefei 230026}
\affiliation{Department of Physics, University of Illinois Urbana-Champaign, 1110 West Green Street, Urbana, IL, 61801, USA}
\author{K.~R.~Ferguson\,\orcidlink{0000-0002-4928-8813}}
\affiliation{Department of Physics and Astronomy, University of California, Los Angeles, CA, 90095, USA}
\affiliation{Department of Physics and Astronomy, Michigan State University, East Lansing, MI 48824, USA}
\author{N.~C.~Ferree\,\orcidlink{0000-0002-7130-7099}}
\affiliation{California Institute of Technology, 1200 East California Boulevard., Pasadena, CA, 91125, USA}
\affiliation{Kavli Institute for Particle Astrophysics and Cosmology, Stanford University, 452 Lomita Mall, Stanford, CA, 94305, USA}
\affiliation{Department of Physics, Stanford University, 382 Via Pueblo Mall, Stanford, CA, 94305, USA}
\author{K.~Fichman}
\affiliation{Department of Physics, University of Chicago, 5640 South Ellis Avenue, Chicago, IL, 60637, USA}
\affiliation{Kavli Institute for Cosmological Physics, University of Chicago, 5640 South Ellis Avenue, Chicago, IL, 60637, USA}
\author{A.~Foster\,\orcidlink{0000-0002-7145-1824}}
\affiliation{Joseph Henry Laboratories of Physics, Jadwin Hall, Princeton University, Princeton, NJ 08544, USA}
\author{S.~Galli}
\affiliation{Sorbonne Universit\'e, CNRS, UMR 7095, Institut d'Astrophysique de Paris, 98 bis bd Arago, 75014 Paris, France}
\author{A.~E.~Gambrel}
\affiliation{Kavli Institute for Cosmological Physics, University of Chicago, 5640 South Ellis Avenue, Chicago, IL, 60637, USA}
\author{A.~K.~Gao}
\affiliation{Department of Physics, University of Illinois Urbana-Champaign, 1110 West Green Street, Urbana, IL, 61801, USA}
\author{F.~Ge}
\affiliation{California Institute of Technology, 1200 East California Boulevard., Pasadena, CA, 91125, USA}
\affiliation{Kavli Institute for Particle Astrophysics and Cosmology, Stanford University, 452 Lomita Mall, Stanford, CA, 94305, USA}
\affiliation{Department of Physics, Stanford University, 382 Via Pueblo Mall, Stanford, CA, 94305, USA}
\affiliation{Department of Physics \& Astronomy, University of California, One Shields Avenue, Davis, CA 95616, USA}
\author{F.~Guidi\,\orcidlink{0000-0001-7593-3962}}
\affiliation{Department of Physics \& Astronomy, University of California, One Shields Avenue, Davis, CA 95616, USA}
\affiliation{Sorbonne Universit\'e, CNRS, UMR 7095, Institut d'Astrophysique de Paris, 98 bis bd Arago, 75014 Paris, France}
\author{S.~Guns}
\affiliation{Department of Physics, University of California, Berkeley, CA, 94720, USA}
\author{N.~W.~Halverson}
\affiliation{CASA, Department of Astrophysical and Planetary Sciences, University of Colorado, Boulder, CO, 80309, USA }
\affiliation{Department of Physics, University of Colorado, Boulder, CO, 80309, USA}
\author{E.~Hivon\,\orcidlink{0000-0003-1880-2733}}
\affiliation{Sorbonne Universit\'e, CNRS, UMR 7095, Institut d'Astrophysique de Paris, 98 bis bd Arago, 75014 Paris, France}
\author{G.~P.~Holder\,\orcidlink{0000-0002-0463-6394}}
\affiliation{Department of Physics, University of Illinois Urbana-Champaign, 1110 West Green Street, Urbana, IL, 61801, USA}
\author{W.~L.~Holzapfel}
\affiliation{Department of Physics, University of California, Berkeley, CA, 94720, USA}
\author{J.~C.~Hood}
\affiliation{Kavli Institute for Cosmological Physics, University of Chicago, 5640 South Ellis Avenue, Chicago, IL, 60637, USA}
\author{A.~Hryciuk}
\affiliation{Department of Physics, University of Chicago, 5640 South Ellis Avenue, Chicago, IL, 60637, USA}
\affiliation{Kavli Institute for Cosmological Physics, University of Chicago, 5640 South Ellis Avenue, Chicago, IL, 60637, USA}
\author{N.~Huang\,\orcidlink{0000-0003-3595-0359}}
\affiliation{Department of Physics, University of California, Berkeley, CA, 94720, USA}
\author{T.~Jhaveri}
\affiliation{Department of Astronomy and Astrophysics, University of Chicago, 5640 South Ellis Avenue, Chicago, IL, 60637, USA}
\affiliation{Kavli Institute for Cosmological Physics, University of Chicago, 5640 South Ellis Avenue, Chicago, IL, 60637, USA}
\author{F.~K\'eruzor\'e}
\affiliation{High-Energy Physics Division, Argonne National Laboratory, 9700 South Cass Avenue, Lemont, IL, 60439, USA}
\author{A.~R.~Khalife\,\orcidlink{0000-0002-8388-4950}}
\affiliation{Sorbonne Universit\'e, CNRS, UMR 7095, Institut d'Astrophysique de Paris, 98 bis bd Arago, 75014 Paris, France}
\author{L.~Knox}
\affiliation{Department of Physics \& Astronomy, University of California, One Shields Avenue, Davis, CA 95616, USA}
\author{K.~Kornoelje}
\affiliation{Department of Astronomy and Astrophysics, University of Chicago, 5640 South Ellis Avenue, Chicago, IL, 60637, USA}
\affiliation{Kavli Institute for Cosmological Physics, University of Chicago, 5640 South Ellis Avenue, Chicago, IL, 60637, USA}
\affiliation{High-Energy Physics Division, Argonne National Laboratory, 9700 South Cass Avenue, Lemont, IL, 60439, USA}
\author{C.-L.~Kuo}
\affiliation{Kavli Institute for Particle Astrophysics and Cosmology, Stanford University, 452 Lomita Mall, Stanford, CA, 94305, USA}
\affiliation{Department of Physics, Stanford University, 382 Via Pueblo Mall, Stanford, CA, 94305, USA}
\affiliation{SLAC National Accelerator Laboratory, 2575 Sand Hill Road, Menlo Park, CA, 94025, USA}
\author{K.~Levy}
\affiliation{School of Physics, University of Melbourne, Parkville, VIC 3010, Australia}
\author{Y.~Li\,\orcidlink{0000-0002-4820-1122}}
\affiliation{Kavli Institute for Cosmological Physics, University of Chicago, 5640 South Ellis Avenue, Chicago, IL, 60637, USA}
\author{A.~E.~Lowitz\,\orcidlink{0000-0002-4747-4276}}
\affiliation{Kavli Institute for Cosmological Physics, University of Chicago, 5640 South Ellis Avenue, Chicago, IL, 60637, USA}
\author{C.~Lu}
\affiliation{Department of Physics, University of Illinois Urbana-Champaign, 1110 West Green Street, Urbana, IL, 61801, USA}
\author{G.~P.~Lynch\,\orcidlink{0009-0004-3143-1708}}
\affiliation{Department of Physics \& Astronomy, University of California, One Shields Avenue, Davis, CA 95616, USA}
\author{T.~J.~Maccarone\,\orcidlink{0000-0003-0976-4755}}
\affiliation{Department of Physics \& Astronomy, Box 41051, Texas Tech University, Lubbock TX 79409-1051, USA}
\author{A.~S.~Maniyar\,\orcidlink{0000-0002-4617-9320}}
\affiliation{Kavli Institute for Particle Astrophysics and Cosmology, Stanford University, 452 Lomita Mall, Stanford, CA, 94305, USA}
\affiliation{Department of Physics, Stanford University, 382 Via Pueblo Mall, Stanford, CA, 94305, USA}
\affiliation{SLAC National Accelerator Laboratory, 2575 Sand Hill Road, Menlo Park, CA, 94025, USA}
\author{E.~S.~Martsen}
\affiliation{Department of Astronomy and Astrophysics, University of Chicago, 5640 South Ellis Avenue, Chicago, IL, 60637, USA}
\affiliation{Kavli Institute for Cosmological Physics, University of Chicago, 5640 South Ellis Avenue, Chicago, IL, 60637, USA}
\author{F.~Menanteau}
\affiliation{Department of Astronomy, University of Illinois Urbana-Champaign, 1002 West Green Street, Urbana, IL, 61801, USA}
\affiliation{Center for AstroPhysical Surveys, National Center for Supercomputing Applications, Urbana, IL, 61801, USA}
\author{M.~Millea\,\orcidlink{0000-0001-7317-0551}}
\affiliation{Department of Physics, University of California, Berkeley, CA, 94720, USA}
\author{J.~Montgomery}
\affiliation{Department of Physics and McGill Space Institute, McGill University, 3600 Rue University, Montreal, Quebec H3A 2T8, Canada}
\author{Y.~Nakato}
\affiliation{Department of Physics, Stanford University, 382 Via Pueblo Mall, Stanford, CA, 94305, USA}
\author{T.~Natoli}
\affiliation{Kavli Institute for Cosmological Physics, University of Chicago, 5640 South Ellis Avenue, Chicago, IL, 60637, USA}
\author{Z.~Pan\,\orcidlink{0000-0002-6164-9861}}
\affiliation{High-Energy Physics Division, Argonne National Laboratory, 9700 South Cass Avenue, Lemont, IL, 60439, USA}
\affiliation{Kavli Institute for Cosmological Physics, University of Chicago, 5640 South Ellis Avenue, Chicago, IL, 60637, USA}
\affiliation{Department of Physics, University of Chicago, 5640 South Ellis Avenue, Chicago, IL, 60637, USA}
\author{P.~Paschos}
\affiliation{Enrico Fermi Institute, University of Chicago, 5640 South Ellis Avenue, Chicago, IL, 60637, USA}
\author{K.~A.~Phadke\,\orcidlink{0000-0001-7946-557X}}
\affiliation{Department of Astronomy, University of Illinois Urbana-Champaign, 1002 West Green Street, Urbana, IL, 61801, USA}
\affiliation{Center for AstroPhysical Surveys, National Center for Supercomputing Applications, Urbana, IL, 61801, USA}
\affiliation{NSF-Simons AI Institute for the Sky (SkAI), 172 E. Chestnut St., Chicago, IL 60611, USA}
\author{A.~W.~Pollak}
\affiliation{Department of Astronomy and Astrophysics, University of Chicago, 5640 South Ellis Avenue, Chicago, IL, 60637, USA}
\author{K.~Prabhu}
\affiliation{Department of Physics \& Astronomy, University of California, One Shields Avenue, Davis, CA 95616, USA}
\author{W.~Quan\,\orcidlink{0009-0002-2589-5501}}
\affiliation{High-Energy Physics Division, Argonne National Laboratory, 9700 South Cass Avenue, Lemont, IL, 60439, USA}
\affiliation{Department of Physics, University of Chicago, 5640 South Ellis Avenue, Chicago, IL, 60637, USA}
\affiliation{Kavli Institute for Cosmological Physics, University of Chicago, 5640 South Ellis Avenue, Chicago, IL, 60637, USA}
\author{M.~Rahimi}
\affiliation{School of Physics, University of Melbourne, Parkville, VIC 3010, Australia}
\author{A.~Rahlin\,\orcidlink{0000-0003-3953-1776}}
\affiliation{Department of Astronomy and Astrophysics, University of Chicago, 5640 South Ellis Avenue, Chicago, IL, 60637, USA}
\affiliation{Kavli Institute for Cosmological Physics, University of Chicago, 5640 South Ellis Avenue, Chicago, IL, 60637, USA}
\author{C.~L.~Reichardt\,\orcidlink{0000-0003-2226-9169}}
\affiliation{School of Physics, University of Melbourne, Parkville, VIC 3010, Australia}
\author{M.~Rouble}
\affiliation{Department of Physics and McGill Space Institute, McGill University, 3600 Rue University, Montreal, Quebec H3A 2T8, Canada}
\author{J.~E.~Ruhl}
\affiliation{Department of Physics, Case Western Reserve University, Cleveland, OH, 44106, USA}
\author{A.~C.~Silva~Oliveira\,\orcidlink{0000-0001-5755-5865}}
\affiliation{California Institute of Technology, 1200 East California Boulevard., Pasadena, CA, 91125, USA}
\affiliation{Kavli Institute for Particle Astrophysics and Cosmology, Stanford University, 452 Lomita Mall, Stanford, CA, 94305, USA}
\affiliation{Department of Physics, Stanford University, 382 Via Pueblo Mall, Stanford, CA, 94305, USA}
\author{A.~Simpson}
\affiliation{Department of Astronomy and Astrophysics, University of Chicago, 5640 South Ellis Avenue, Chicago, IL, 60637, USA}
\affiliation{Kavli Institute for Cosmological Physics, University of Chicago, 5640 South Ellis Avenue, Chicago, IL, 60637, USA}
\author{J.~A.~Sobrin\,\orcidlink{0000-0001-6155-5315}}
\affiliation{Department of Physics, Villanova University, 800 E Lancaster Ave, Villanova, PA 19085, USA}
\affiliation{Fermi National Accelerator Laboratory, MS209, P.O. Box 500, Batavia, IL, 60510, USA}
\author{A.~A.~Stark}
\affiliation{Center for Astrophysics \textbar{} Harvard \& Smithsonian, 60 Garden Street, Cambridge, MA, 02138, USA}
\author{J.~Stephen}
\affiliation{Enrico Fermi Institute, University of Chicago, 5640 South Ellis Avenue, Chicago, IL, 60637, USA}
\author{C.~Tandoi\,\orcidlink{0000-0002-2077-6004}}
\affiliation{Department of Astronomy, University of Illinois Urbana-Champaign, 1002 West Green Street, Urbana, IL, 61801, USA}
\author{C.~Trendafilova}
\affiliation{Center for AstroPhysical Surveys, National Center for Supercomputing Applications, Urbana, IL, 61801, USA}
\author{J.~D.~Vieira\,\orcidlink{0000-0001-7192-3871}}
\affiliation{Department of Astronomy, University of Illinois Urbana-Champaign, 1002 West Green Street, Urbana, IL, 61801, USA}
\affiliation{Department of Physics, University of Illinois Urbana-Champaign, 1110 West Green Street, Urbana, IL, 61801, USA}
\affiliation{Center for AstroPhysical Surveys, National Center for Supercomputing Applications, Urbana, IL, 61801, USA}
\author{A.~G.~Vieregg\,\orcidlink{0000-0002-4528-9886}}
\affiliation{Kavli Institute for Cosmological Physics, University of Chicago, 5640 South Ellis Avenue, Chicago, IL, 60637, USA}
\affiliation{Department of Astronomy and Astrophysics, University of Chicago, 5640 South Ellis Avenue, Chicago, IL, 60637, USA}
\affiliation{Enrico Fermi Institute, University of Chicago, 5640 South Ellis Avenue, Chicago, IL, 60637, USA}
\affiliation{Department of Physics, University of Chicago, 5640 South Ellis Avenue, Chicago, IL, 60637, USA}
\author{A.~Vitrier\,\orcidlink{0009-0009-3168-092X}}
\affiliation{Sorbonne Universit\'e, CNRS, UMR 7095, Institut d'Astrophysique de Paris, 98 bis bd Arago, 75014 Paris, France}
\author{Y.~Wan}
\affiliation{Department of Astronomy, University of Illinois Urbana-Champaign, 1002 West Green Street, Urbana, IL, 61801, USA}
\affiliation{Center for AstroPhysical Surveys, National Center for Supercomputing Applications, Urbana, IL, 61801, USA}
\author{N.~Whitehorn\,\orcidlink{0000-0002-3157-0407}}
\affiliation{Department of Physics and Astronomy, Michigan State University, East Lansing, MI 48824, USA}
\author{W.~L.~K.~Wu\,\orcidlink{0000-0001-5411-6920}}
\affiliation{California Institute of Technology, 1200 East California Boulevard., Pasadena, CA, 91125, USA}
\affiliation{Kavli Institute for Particle Astrophysics and Cosmology, Stanford University, 452 Lomita Mall, Stanford, CA, 94305, USA}
\affiliation{SLAC National Accelerator Laboratory, 2575 Sand Hill Road, Menlo Park, CA, 94025, USA}
\author{M.~R.~Young}
\affiliation{Fermi National Accelerator Laboratory, MS209, P.O. Box 500, Batavia, IL, 60510, USA}
\affiliation{Kavli Institute for Cosmological Physics, University of Chicago, 5640 South Ellis Avenue, Chicago, IL, 60637, USA}
\author{J.~A.~Zebrowski}
\affiliation{Kavli Institute for Cosmological Physics, University of Chicago, 5640 South Ellis Avenue, Chicago, IL, 60637, USA}
\affiliation{Department of Astronomy and Astrophysics, University of Chicago, 5640 South Ellis Avenue, Chicago, IL, 60637, USA}
\affiliation{Fermi National Accelerator Laboratory, MS209, P.O. Box 500, Batavia, IL, 60510, USA}
\collaboration{SPT-3G Collaboration}
\noaffiliation

\date{\today}

\begin{abstract}
    Measurements of the weak lensing of galaxies and of the cosmic microwave background (CMB) provide direct probes of the cosmic matter density field, but the two observables are sensitive to different spatial scales, redshift ranges, and survey systematics. Their cross-correlation thus enables consistency checks of the theoretical model and of potential systematics in either dataset. We present measurements of the cross-correlation between CMB lensing and cosmic shear over $\sim$1,300 \sqdeg\ of the sky using the SPT-3G D1 CMB lensing maps and the Dark Energy Survey Year 3 (DES Y3) shear catalogs. For the first time, we measure this cross-correlation at high significance ($\sim 14\sigma$) when using a polarization-only CMB lensing reconstruction that is expected to be robust against biases induced by extragalactic foregrounds. We test a variety of other CMB lensing estimators that include temperature information and exhibit different tradeoffs between foreground biases and noise, as well as a shear sample that consists of blue, star-forming galaxies and has been shown to be less impacted by galaxy intrinsic alignments. Assuming \lcdm\ and marginalizing over uncertainties in intrinsic alignments, baryonic feedback, and various nuisance parameters, we obtain a constraint on the amplitude of matter clustering $S_8 \equiv \sigma_8 \sqrt{\Om / 0.3} = 0.833^{+0.047}_{-0.061}$, consistent with both the primary CMB results from \planck\ and shear-only results from DES Y3. By combining our measurement with \planck, we find mild constraints on the astrophysical processes that impact the cross-correlation. We obtain a constraint on the intrinsic alignment amplitude of the DES sample that is competitive with that from shear-only analyses, and we find a lower limit on the strength of baryonic feedback.
\end{abstract}

\maketitle

\section{Introduction}
Gravitational lensing of distant light sources by the intervening large-scale structure (LSS) provides a wealth of information about the contents and evolution of the Universe since the lensing signal is sensitive to both the cosmic expansion rate and the growth of structure \cite{Bartelmann2001,Kilbinger2015}. One of the main strengths of weak lensing as a cosmological probe is that it provides a direct tracer of the matter density field, in contrast to galaxy clustering which has a complicated connection to the underlying matter field.
The weak lensing signal has now been measured at high significance using both galaxies and the cosmic microwave background (CMB) as sources. On the galaxy lensing (also referred to as cosmic shear) side, Stage-III surveys, such as the Dark Energy Survey (DES), the Kilo-Degree Survey (KiDS), the Hyper Suprime-Cam (HSC) survey, and the Dark Energy Camera All Data Everywhere (DECADE) survey, have been placing tight constraints on the amplitude of matter clustering at low redshift \cite{Abbott2022, DESandKIDS2023, Wright2025, Dalal2023, Anbajagane2025}. On the CMB lensing side, the \planck, Atacama Cosmology Telescope (ACT), and South Pole Telescope (SPT) collaborations have all achieved comparable constraints that are sensitive to significantly higher redshifts and larger physical scales \cite{Carron2022,Madhavacheril2024,Omori2017,Wu2019,Ge2025,Qu2025}.

With increasingly precise measurements of the amplitude of matter clustering across redshifts, a potential mild tension has emerged in the inferred value of the parameter $S_8 \equiv \sigma_8 \sqrt{\Om / 0.3}$, where $\Om$ is the fractional matter density and $\sigma_8$ parametrizes the amplitude of present-day density fluctuations. This so-called $S_8$-tension is typically presented as a consistent trend where cosmic shear surveys measure the value of $S_8$ to be $\sim2$ to $3\sigma$ low relative to the $\Lambda$CDM prediction based on measurements of the primary CMB \cite[see, for example,][]{DiValentino2025,Pantos2026}, while measurements of CMB lensing are generally fully consistent with the primary CMB \cite{Qu2025}. One possible way of reframing this tension is as a mismatch between the amplitude of clustering on large and on small scales at the low redshifts that cosmic shear is sensitive to \cite{Amon2022,Preston2023}, but there are a variety of factors that could contribute towards alleviating this tension, including: more flexible modeling of baryonic feedback \cite[e.g.,][]{Arico2023}, more consistent modeling of small scales \cite{DeRose2025,Chen2024}, improved survey systematics, or beyond-standard-model physics \cite[e.g.,][]{Terasawa2025, Sabogal2024, Tanimura2023, Rogers2023, He2023}.
These unresolved questions about the impact of nonlinear physics on lensing observables motivate the measurement of $S_8$ using a variety of probes that are sensitive to different systematics, spatial scales, and redshift ranges.

The combination of CMB lensing and cosmic shear provides a unique opportunity to address these questions.
While cosmic shear is highly sensitive to nonlinear scales, which are challenging to model robustly, CMB lensing is mainly sourced by linear and quasi-linear scales (see \cite{Doux2026} for a visualization of the $z$- and $k$-sensitivity of various weak lensing experiments). Their cross-correlation is then expected to be mostly sensitive to scales straddling the nonlinear transition, allowing for simpler modeling compared to cosmic shear alone while retaining some sensitivity to small-scale physics compared to CMB lensing alone.
The cross-correlation of these two completely independent datasets with different measurement systematics also provides a useful consistency test of the two probes. Nonetheless, there are two main physical effects that are expected to contaminate the cross-correlation: galaxy intrinsic alignments \cite[IA, see][for reviews]{Troxel2015,Lamman2024,Chisari2025} and extragalactic foregrounds in CMB data that affect lensing reconstructions \cite{Osborne2014,vanEngelen2014}.
In this work, we attempt to mitigate both of these effects using tailored data selections.

Until recently, measurements of the cross-correlation between CMB lensing ($\kcmb$) and cosmic shear ($\gamma$) have had relatively low signal-to-noise ratios (SNR) \citep[e.g.,][]{Namikawa2019, Marques2020}. The advent of high-resolution ground-based CMB experiments and deeper cosmic shear surveys has now established the potential of the $\kcmb\gamma$ cross-correlation as a useful probe of cosmology. Recent measurements include: ACT-DR4+\planck\ $\times$ KiDS-1000 \cite[$7.7\sigma$,][]{Robertson2021}, ACT-DR4+\planck\ $\times$ DES Y3 \cite[$7\sigma$,][]{Shaikh2024}, SPT-SZ+\planck\ $\times$ DES Y3 \cite[$18\sigma$,][]{Chang2023}, and \planck\ PR4 $\times$ DES Y3 \cite[$20\sigma$,][]{Xu2025}.
CMB lensing has also been combined with cosmic shear and galaxy clustering to produce the so-called 6$\times$2-point analysis which combines the six unique cross-correlations between these three probes in a joint analysis \citep{Omori2023, Chang2023, Abbott2023, Xu2025}. This has been shown to be a powerful cosmological probe that breaks various parameter degeneracies and probes the growth of structure across a wide range of cosmic time.

In this work, we measure the cross-correlation between CMB lensing and cosmic shear over roughly 1,300 \sqdeg\ of the sky using new CMB lensing reconstructions based on data from SPT-3G (\Oinprep, hereafter O26) and the DES Y3 shear catalog \cite{Gatti2021}. We obtain a $\sim 14 \sigma$ measurement of this cross-correlation using a CMB polarization-only lensing reconstruction. This is a significant improvement over the last polarization-only cross-correlation measurement which used data from the POLARBEAR CMB experiment \cite{Kermish2012} and HSC to obtain a $3.5\sigma$\footnote{Note that HSC is significantly deeper than DES, but covers a smaller sky area.} detection over 11 \sqdeg\ of the sky \cite{Namikawa2019}. Our high-significance polarization-only measurement effectively allows us to sidestep the issue of extragalactic foregrounds, and we use this measurement as a baseline to then assess a variety of foreground mitigation techniques when including CMB temperature data in the lensing reconstruction. We also include rigorous validation tests using the multicomponent \agora\ simulations \citep{Omori2024}, which fully model the correlations between CMB lensing, galaxy lensing, and extragalactic foreground sources. We additionally use a high-purity sample of blue star-forming galaxies selected from the DES Y3 catalog \cite{McCullough2024} that has been shown to have an IA amplitude consistent with zero as a consistency check on our modeling of IA.

This paper is structured as follows. We provide an overview of the data used in \cref{sec:data}. We describe the simulations used to validate our analysis in \cref{sec:sims}. In \cref{sec:measurement} we describe our measurement of the cross-correlation between CMB lensing and cosmic shear. \cref{sec:analysis} provides details on the modeling and parameter inference. We present the results in \cref{sec:results} and conclude in \cref{sec:conclusion}.

Throughout, when assuming a fiducial cosmology, we use values consistent with the \planck\ 2018 flat \lcdm\ results \cite{Planck2020} as listed in \cref{tab:priors}.

\section{Data}\label{sec:data}

\begin{figure}
    \centering
    \includegraphics[width=\linewidth]{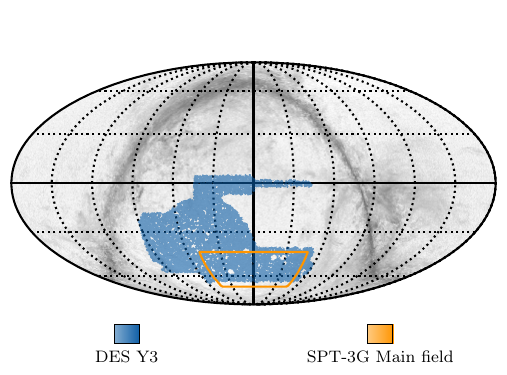}
    \caption{Survey footprints for the SPT-3G Main field and DES Y3 in equatorial coordinates. Galactic dust as measured by \planck\ is shown in the background. The overlapping region covers approximately 1,300 \sqdeg\ and contains roughly 31 million source galaxies.}
    \label{fig:footprints}
\end{figure}

\subsection{SPT-3G CMB lensing maps}\label{sec:data_cmb}
In this work we use new CMB lensing convergence maps derived from SPT-3G D1 data. These maps, made from different combinations of temperature and polarization data, along with a full analysis of their auto-spectra, are presented in O26. Here we provide a quick overview of the data used to construct the lensing maps and the different reconstruction techniques used to mitigate foreground biases.

SPT-3G \cite{Sobrin2022} is the third-generation camera installed on the South Pole Telescope \cite{Carlstrom2011} and has been in operation since 2017. SPT-3G observes in three frequency bands (centered at 95, 150, and 220 GHz) with roughly $1'$ angular resolution. The lensing maps used here are based on the D1 dataset which consists of temperature and polarization observations of the SPT-3G Main field, a $\sim$1,500 \sqdeg\ patch in the southern sky shown as the orange contour in \cref{fig:footprints}, during the 2019 and 2020 winter observing seasons.  These observations achieve coadded white noise levels of 3.3 (5.1) \ukam\ in temperature (polarization) \cite{Camphuis2025}. The full map-making procedure for this dataset is described in \cite{Quan2026}.

CMB lensing maps are often reconstructed from observations of the primary CMB using some form of a quadratic estimator (QE) that combines two maps of the primary CMB \cite{Hu2002}.\footnote{Various beyond-QE methods also exist \cite[e.g.,][]{Hirata2003,Hirata2003b,Carron2017,Millea2022}. See \cite{Ge2025} for a recent application to SPT-3G data.}
The SPT-3G QE $\kcmb$ maps are based on primary CMB temperature (polarization) multipoles in the range $500 \le \ell^T \le 3500$ ($500 \le \ell^P < 3000$)\footnote{Beyond $\ell^T \sim 3500$, extragalactic foregrounds become dominant and significantly bias the lensing reconstruction. The lower value of $\lmax^P$ reflects the fact that high polarization multipoles do not contribute significantly to the lensing reconstruction SNR.} and include reconstructed lensing multipoles in the range $30 \le L \le 3500$.\footnote{Hereafter, we only use $\ell$ to refer to multipoles, whether of CMB lensing or of cosmic shear.}

A major concern in CMB lensing measurements is contamination from extragalactic foregrounds on small scales in the observed CMB temperature field that can mimic the lensing signal \cite{Osborne2014}. The main foregrounds include the Sunyaev-Zeldovich effects (both thermal, tSZ, and kinetic, kSZ), the cosmic infrared background (CIB), and radio sources.
All of these foregrounds are correlated with other tracers of LSS, including cosmic shear. This has the effect of changing the nature of foreground-induced biases in cross-correlations; specifically, the biases become determined by the bispectra between the foreground fields and low-redshift LSS \cite{vanEngelen2014,Baleato-Lizancos2025}. These considerations motivate our tests of foreground-induced biases separate from those for the $\kcmb$ auto-spectrum presented in O26.
On the other hand, the SZ effects and the CIB are all essentially unpolarized \cite{Smith2009, Deutsch2018, Feng2020}, meaning that the issue of extragalactic foregrounds can be avoided through the use of polarization-only lensing reconstructions. This is especially attractive for low-noise CMB experiments such as SPT-3G, where polarization becomes the dominant source of reconstruction SNR on large scales.

In this work, we consider five different QE variants that have different tradeoffs between foreground mitigation and noise.
\begin{enumerate}
\item Global minimum variance (GMV) map. In contrast to the standard quadratic estimator (SQE) \cite{Hu2002,Okamoto2003}, which constructs individual estimators from pairs of the CMB $T$, $E$, and $B$ fields before combining them into a minimum-variance estimator, the GMV QE \cite{Hirata2003,Maniyar2021} combines the fields directly into a single estimator, accounting for the full covariance between $T$, $E$, and $B$. The GMV estimator has been shown to result in improved reconstruction noise compared to the SQE for high-sensitivity experiments such as SPT-3G and Simons Observatory \cite{Maniyar2021}.
\item Polarization-only SQE map. This reconstruction only uses polarization information and is thus expected to be the most robust against biases due to extragalactic foregrounds.
\item Profile-hardened GMV map. Profile-hardening (or bias-hardening) \cite{Namikawa2013, Osborne2014, Sailer2020} modifies the QE to make it insensitive to bias contributions from some assumed source profiles such as those of galaxy clusters, radio sources, and CIB galaxies.
\item tSZ-deprojected GMV map. tSZ-deprojection (also referred to as gradient-cleaning) \cite{Madhavacheril2018} replaces one of the QE input $T$ fields with a tSZ-nulled map, which is made through a linear combination of individual frequency maps by exploiting the different spectral energy distributions (SEDs) of the CMB and tSZ effect. By reducing the overlap in foregrounds between the two input maps, this effectively removes tSZ-sourced biases from the lensing reconstruction with only a small penalty in reconstruction noise.
\item Cross-ILC GMV map. The cross-ILC estimator \cite{Raghunathan2023} uses the same approach as tSZ-deprojection, exploiting the difference in SEDs between the CMB and foregrounds to reduce correlated foregrounds in the input maps. One $T$ map is replaced with a tSZ-nulled map as before, but the second input map is chosen to be a CIB-minimized map. This suppresses both the tSZ- and CIB-sourced biases at the cost of a slightly larger increase in reconstruction noise compared to the tSZ-deprojected estimator.
\end{enumerate}
Both the gradient-cleaned and cross-ILC estimators used in the SPT-3G D1 reconstructions are implemented on top of the GMV (rather than SQE) estimator using the formalism of \cite{Nakato2025}, see O26 for full details on the various estimator implementations.

We refer to these five lensing reconstruction variants as \gmv, \pol, \prof, \mh, and \xilc, respectively. Of these variants, the \gmv\ reconstruction is expected to have the lowest noise, especially on small scales, but is the most impacted by foregrounds. The \pol\ reconstruction is expected to be the most robust to foregrounds while having significantly higher noise on small scales. The other variants all include temperature information and employ some foreground mitigation technique in order to minimize reconstruction noise while reducing foreground biases. We note that the \gmv\ reconstruction is the fiducial choice in the SPT-3G D1 CMB lensing auto-correlation analysis, where an emulator-based approach is used to marginalize over foreground biases (O26). Since the foreground biases are significantly different in a low-redshift cross-correlation, we do not implement this mitigation strategy here and instead rely on the other QE variants that implement different levels of foreground cleaning.

\subsection{DES Y3 shear catalog}

\begin{figure}
    \centering
    \includegraphics[width=3.4in]{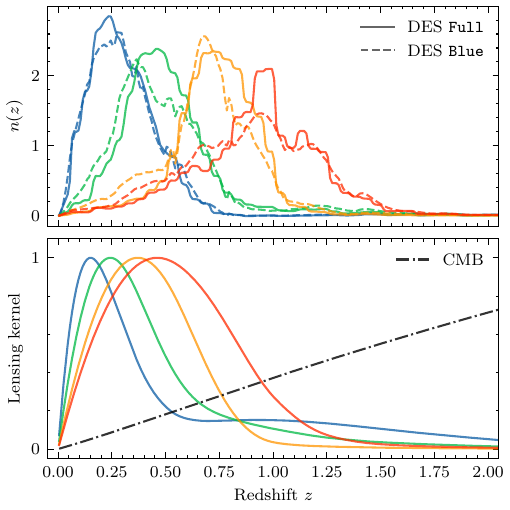}
    \caption{\textit{Top panel:} Normalized redshift distributions of the four DES Y3 tomographic bins. Solid lines represent the \desfull\ sample while dashed lines represent the \desblue\ subsample. \textit{Bottom panel:} The corresponding lensing kernels (normalized to unit height) for the \desfull\ sample compared to the lensing kernel of the CMB (black dot-dashed line). The kernels for the \desblue\ subsample are qualitatively similar to those of the \desfull\ sample.}
    \label{fig:dndz}
\end{figure}

The Dark Energy Survey has observed over $\sim$5,000 \sqdeg\ of the southern sky in five photometric bands ($grizY$) \cite{DES2005,Abbott2018,Abbott2021}. The observations were taken using the DECam \cite{Flaugher2015} camera on the Blanco telescope at the Cerro Tololo Inter-American Observatory in Chile. In this work we use the publicly released catalogs\footnote{\url{https://des.ncsa.illinois.edu/releases/y3a2/Y3key-catalogs}. Note that we use the updated \textsc{SOMPZ} v0.50 photometric catalog which corrects the tomographic bin placement for a small fraction of galaxies. See also Footnote 5 in \cite{McCullough2024}.\label{foot:sompz}} based on data from the first three years (2013 - 2016) of observations (Y3), covering an effective area of 4,143 \sqdeg\ (survey footprint shown as the blue region in \cref{fig:footprints}). This is a subset of the Gold catalog \cite{Sevilla-Noarbe2021} which consists of 100 million galaxies with shape measurements \cite{Gatti2021} and is divided into four tomographic bins based on photometric redshifts. The redshift distributions of the tomographic bins have been characterized using the \textsc{SOMPZ} algorithm and calibrated using clustering redshifts and shear ratios \cite{Myles2021}. The redshift distributions of the DES Y3 tomographic bins are shown in the top panel of \cref{fig:dndz}.

Additionally, we use the blue subsample of the DES Y3 data derived and analyzed by \cite{McCullough2024}. This subsample was developed with the goal of obtaining a pure sample of blue star-forming galaxies that are expected to be less impacted by intrinsic alignments than a mixed sample. We use the publicly released sample selection\footnote{\url{https://jamiemccullough.github.io/data/blueshear/}} to generate the blue catalogs and use the updated redshift distributions and calibrations that are presented in \cite{McCullough2024} when analyzing this sample. Throughout this work, we refer to the full DES Y3 shear sample and the blue subsample as \desfull\ and \desblue\ respectively.

The DES catalogs contain galaxy shape measurements $\vb*{e} = (e_1, e_2)$, which must be first calibrated in order to obtain estimates of the shear $\vb*{\gamma} = (\gamma_1, \gamma_2)$. We calibrate each of the shear catalogs using the \textsc{Metacalibration} algorithm \cite{Sheldon2017,Huff2017}. For each galaxy (represented by an index $i$) in a given tomographic bin, we subtract the mean ellipticity and correct by the multiplicative shear bias:
\begin{equation}
    \widehat{\vb*{\gamma}}_i = \frac{1}{\langle R \rangle} (\vb*{e}_i - \langle\vb*{e}\rangle),
\end{equation}
where $\langle . \rangle$ indicates an average over all the galaxies in the tomographic bin and $R = R_\gamma + R_s$ is the diagonalized combination of the shear response and selection response. This calibration factor is calculated from the catalogs as detailed in \cite{Gatti2021}.

In \cref{tab:shear_samples} we provide an overview of the properties and calibration values of the four tomographic bins for both the \desfull\ and \desblue\ samples. \Cref{fig:footprints} shows the survey footprints of both DES Y3 and the SPT-3G Main field. The overlapping region contains $\sim$31 million source galaxies out of $\sim$100 million in the full DES Y3 catalog.

\begin{table*}
    \centering
    \begin{tabular}{ccccccccc}
        \hline
        Sample & Bin & $N_\text{gal}$ & $R_\gamma$ & $R_s$ & $\langle z\rangle$ & $\sigma_{\Delta z}$ & $\langle m\rangle$ & $\sigma_m$ \\
        \hline \hline
        \multirow{5}{0.25in}{\desfull}
        & 1 & 24,882,718 & 0.763 & 0.005 & 0.33 & 0.018 & $-0.0063$ & 0.0091 \\
        & 2 & 25,224,643 & 0.719 & 0.008 & 0.52 & 0.015 & $-0.0198$ & 0.0078 \\
        & 3 & 24,831,367 & 0.693 & 0.012 & 0.74 & 0.011 & $-0.0241$ & 0.0076 \\
        & 4 & 25,265,298 & 0.609 & 0.015 & 0.93 & 0.017 & $-0.0369$ & 0.0076 \\
        & Total & 100,204,026 \\
        \hline
        \multirow{5}{0.25in}{\desblue}
        & 1 & 18,031,829 & 0.760 & 0.007 & 0.36 & 0.018 & $-0.0129$ & 0.0091 \\
        & 2 & 16,670,470 & 0.713 & 0.012 & 0.52 & 0.015 & $-0.0180$ & 0.0078 \\
        & 3 & 12,233,530 & 0.696 & 0.018 & 0.70 & 0.011 & $-0.0203$ & 0.0076 \\
        & 4 & 18,130,765 & 0.618 & 0.015 & 0.90 & 0.017 & $-0.0356$ & 0.0076 \\
        & Total & 65,066,594 \\
        \hline
    \end{tabular}
    \caption{Properties and calibration parameters for the DES Y3 shear samples (\desfull\ and \desblue). For each tomographic bin we list the number of galaxies ($N_\text{gal}$), the average shear and selection responses ($R_\gamma$, $R_s$), the average redshift ($\langle z\rangle$), the uncertainty on the shift in the mean redshift ($\sigma_{\Delta z}$), the residual shear calibration ($\langle m\rangle$), and the uncertainty on the shear calibration ($\sigma_m$). The calibration numbers for the \desfull\ sample are from \cite{Amon2022b} while those for the \desblue\ sample are from \cite{McCullough2024}. Note that the number of galaxies in each bin and the response factors for the \desfull\ sample are slightly different from those reported by \cite{Amon2022b} since they have been recomputed based on the updated galaxy bin assignments (see \cref{foot:sompz}).}
    \label{tab:shear_samples}
\end{table*}

\section{Simulations}\label{sec:sims}
To validate our measurement and analysis pipelines, we use the \agora\footnote{\url{https://yomori.github.io/agora/index.html}} simulations \citep{Omori2024} which are based on the \textsc{Multidark-Planck2} (MDPL2) \citep{Klypin2016} $N$-body simulation. \agora\ uses the halos and particles from MDPL2 to generate lightcones, full-sky fields, and mock observations for a variety of observables, including CMB lensing and cosmic shear. \agora\ consists of a single full-sky realization which we split into 10 independent patches that each cover the same sky area as the SPT-3G Main field.

\subsection{Mock SPT-3G lensing maps}
\agora\ includes a $\kcmb$ realization by raytracing the entire MDPL2 lightcone (a small Gaussian component is included to account for the high redshifts and large scales that are not modeled by the simulation lightcone). This lensing field is then used to produce a realization of the lensed CMB temperature and polarization fields. \agora\ also includes correlated non-Gaussian realizations of the tSZ, kSZ, radio, and CIB fields on the full MDPL2 lightcone in order to properly model the impact of foreground biases on cross-correlations with low-redshift tracers. These fields are generated using a variety of empirical relations and are calibrated against a mixture of hydrodynamic simulations and observational data. The combination of lensed CMB fields and foreground fields are then mock-observed to capture the effects of noise and filtering that are present in the SPT-3G data. 
Realistic mock CMB lensing reconstructions are then generated from the mock-observed CMB and foreground fields using each of the estimators listed in \cref{sec:data_cmb}. Full details on the construction of these mock lensing maps can be found in \cite[][O26]{Omori2024}.

\subsection{Mock DES Y3 shear catalogs}
The \agora\ full-sky galaxy weak lensing signal maps \cite{Omori2024} were generated to match the actual DES Y3 redshift distributions and include a level of intrinsic alignments that is consistent with the results of \cite{Abbott2018b}. We generate simulated shear catalogs from the full-sky maps by sampling the signal at the locations of galaxies in the DES Y3 catalog. These noiseless catalogs are used to test for potential bias in the measured cross-correlation (see \cref{sec:val-bias}). We additionally incorporate realistic shape noise by randomly rotating each of the phases of the shear measurements in the DES Y3 catalog (thus breaking the correlation in the data, leaving only shape noise) and adding the resulting noise to the noiseless catalogs. These realistic catalogs allow us to test the estimated covariance matrix of the measured cross-correlations (see \cref{sec:val-cov}).

\section{Measurement}\label{sec:measurement}

\subsection{Shear map making}
We generate shear maps and associated masks from the calibrated DES Y3 catalogs following \cite{Nicola2021,Doux2022}. Throughout, we use the \healpix\ pixelization scheme as implemented in \textsc{Healpy} \citep{Gorski2005,Zonca2019} with a resolution parameter of $\nside = 2048$.

The average shear $\vb*{\gamma}$ at pixel $p$ is given by
\begin{equation}
    \left. \vb*{\gamma}_p = \sum_{i \in p} v_i \vb*{\gamma}_i \middle/ \sum_{i \in p} v_i \right.,
\end{equation}
where the sum is over all galaxies $i$ inside the pixel, $\vb*{\gamma}_i$ is the estimated shear (calibrated from the catalog ellipticity), and $v_i$ is the measurement weight.

The mask associated with the shear maps is simply given by the sum-of-weights map
\begin{equation}\label{eq:wp}
    w_p = \sum_{i \in p} v_i.
\end{equation}

\subsection{Power spectra}\label{sec:namaster}
Cosmic shear $\vb*{\gamma}$ is a spin-2 field and can be decomposed into $E$- and $B$-modes \cite{Zaldarriaga1997}. The harmonic-space cross-correlation of cosmic shear with the scalar CMB lensing field $\kcmb$ then has two components: $C_\ell^{\kcmb\gamma_E}$ and $C_\ell^{\kcmb\gamma_B}$. At linear order, weak lensing by LSS does not generate $B$-modes, so our measurement mainly concerns the $E$-mode cross-spectrum, while the $B$-mode cross-spectrum is expected to be consistent with zero. For simplicity, we will generally write $C_\ell^{\kappa\gamma}$ to refer to the $E$-mode cross-spectrum between CMB lensing and cosmic shear, unless otherwise noted.

We use \namaster\footnote{\url{https://namaster.readthedocs.io/en/latest/}} \citep{Alonso2019}, which implements the pseudo-$C_\ell$ or \textsc{Master} algorithm \cite{Hivon2002} for arbitrary spin fields, to measure the mask-deconvolved cross-spectra $C_\ell^{\kappa\gamma}$ from the masked shear $\gamma$ and CMB lensing $\kappa$ maps.
The full details of the pseudo-$C_\ell$ method for general fields are given in \cite{Alonso2019} and its specific application to cosmic shear is described in \cite{Nicola2021}. Here we provide a brief description. For masked observed fields $\tilde{\vb*{\gamma}}$ and $\tilde{\kappa}$, the $E$-mode pseudo-cross-spectrum is defined as
\begin{equation}
    \tilde{C}_\ell^{\kappa\gamma} = \frac{1}{2\ell + 1} \sum_m \tilde{\gamma}^E_{\ell m} \tilde{\kappa}^\ast_{\ell m}.
\end{equation}
The pseudo-cross-spectrum is related to the underlying true cross-spectrum through a mode-coupling matrix $M_{\ell \ell'}$:
\begin{equation}
    \langle \tilde{C}_\ell^{\kappa\gamma} \rangle = \sum_{\ell'} M_{\ell \ell'} C_\ell^{\kappa\gamma}.
\end{equation}
The coupling matrix depends only on the masks applied to the two fields and the field spin values (see \cite{Alonso2019} for the full expressions).

Using \namaster, we apply separate masks to the $\kcmb$ and $\gamma$ fields (the cross-spectrum is effectively measured from the overlap of these two masks). The mask applied to the $\kcmb$ maps is a combination of an apodized border mask representing the SPT-3G Main field and a source mask that depends on whether temperature is included in the reconstruction. The source mask applied to all GMV variants masks all point sources detected above 6 mJy at 150 GHz and all tSZ-detected clusters above $10\sigma$ and is apodized using a $3'$ Gaussian kernel, resulting in a fractional area lost of 4\% compared to the full SPT-3G Main field (O26). A less restrictive source mask which only includes polarization-detected sources (137 in total) is applied to the \pol\ reconstruction. The mask applied to the shear field is the map of inverse-variance weights as given by \cref{eq:wp}. It is important to note that since the temperature source mask contains galaxy clusters, it must be correlated with the DES shear maps, breaking the usual pseudo-$C_\ell$ assumption that masks are uncorrelated with the fields. We find a correlation of at most $\sim 10$\% between the temperature source mask and the fourth DES shear bin. This correlation is expected to produce biases that depend on bispectra and trispectra between the fields and masks \cite{Surrao2023}. We neglect this correction in this analysis since we find negligible changes in the final results whether we include the source mask or not, but this will likely need to be more carefully considered in future analyses with higher SNR.

We additionally correct the pseudo-cross-spectra by one power of the \healpix\ pixel window function to account for the pixelization of the shear maps (at most a 13\% correction at the highest measured multipoles). The $\kcmb$ maps are computed directly from the $a_{\ell m}$ coefficients, so no pixelization correction is necessary. We have also confirmed that our measurement is robust whether we use higher-resolution \healpix\ maps ($\nside = 4096$) or use the catalog-based pseudo-$C_\ell$ algorithm implemented in \namaster\ \cite{Wolz2025}. Since our measurement is not sensitive to these analysis choices, we choose to use the $\nside = 2048$ resolution maps for computational efficiency.

In order to invert the mode-coupling matrix, \namaster\ bins the cross-spectra into bandpowers. We choose to use the following binning scheme: 6 linear bins over the range $30 \le \ell \le 246$ and 18 logarithmic bins over the range $246 \le \ell \le 3500$. The minimum and maximum values of $\ell$ are chosen based on the reconstructed $\kcmb$ maps.
The mix of linear and logarithmic binning allows us to have somewhat finer bins on large scales where the signal is the largest and wider bins on small scales where the CMB lensing reconstruction noise increases substantially.

We calculate the Gaussian (or ``disconnected'') part of the bandpower covariance matrices using the improved narrow kernel approximation (iNKA) as described in \cite{Nicola2021,Garcia-Garcia2019} and implemented in \namaster. The covariances are calculated directly from the mode-coupled pseudo-$C_\ell$s and do not rely on a theory prediction for the cross-spectra in order to avoid any mis-modeling of signal and noise on small scales. This calculation assumes that all involved fields are fully Gaussian, but includes corrections for the non-trivial mode-coupling introduced by masking. Additional terms that contribute to the full covariance matrix are addressed in \cref{sec:cov}.

\begin{figure*}
    \centering
    \includegraphics[width=7in]{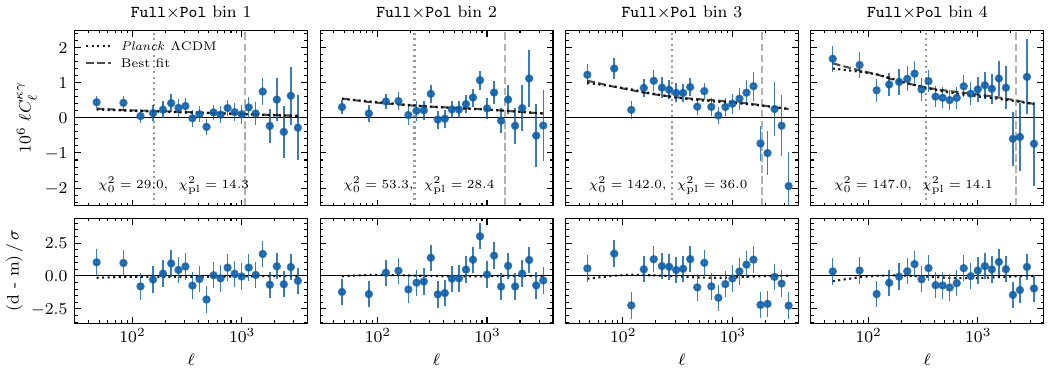}
    \caption{Measured $C_\ell^{\kappa\gamma}$ cross-correlation bandpowers using the SPT-3G polarization-only lensing reconstruction and the full DES Y3 shear sample (\desfull$\times$\pol). The black dashed lines show the best fit model using all scales (see \cref{sec:results}) and the bottom panels show the residuals relative to this model. The black dotted lines show the \planck\ \lcdm\ prediction for this cross-correlation. We also list the $\chi^2$ values relative to zero ($\chi^2_0$) and relative to the \planck\ prediction ($\chi^2_\text{pl}$) for each tomographic bin. The number of data points in each bin is 24. The total $\chi^2$ relative to the \planck\ model is 96.6 with 96 degrees of freedom, while the total $\chi^2$ relative to the best fit is 94.2 with 93.0 degrees of freedom (accounting for the effective number of fit parameters). The gray vertical lines indicate the $\ell$ values corresponding to scale cuts of $k_\text{max} = 5h$ Mpc$^{-1}$ (dashed) and $1h$ Mpc$^{-1}$ (dotted) (see \cref{sec:scale_cuts}).}
    \label{fig:data_bpws}
\end{figure*}

\begin{figure*}
    \centering
    \includegraphics[width=7in]{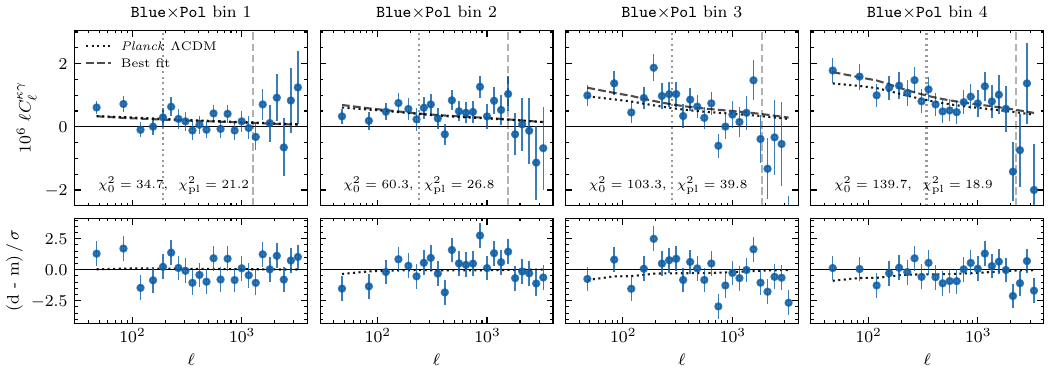}
    \caption{Same as \cref{fig:data_bpws}, but using the DES blue subsample (\desblue$\times$\pol). The total $\chi^2$ relative to the \planck\ model is 107.3 with 96 degrees of freedom, while the total $\chi^2$ relative to the best fit is 105.5 with 93.7 degrees of freedom.}
    \label{fig:data_bpws_blue}
\end{figure*}

We show the resulting cross-correlation bandpowers for the \desfull$\times$\pol\ cross-correlation in \cref{fig:data_bpws} and for the \desblue$\times$\pol\ cross-correlation in \cref{fig:data_bpws_blue}. We find no evidence for non-zero cross-spectra with the shear $B$-modes. The full shear $B$-mode cross-spectra are shown in \cref{app:bmodes}.

\section{Analysis}\label{sec:analysis}

\subsection{Modeling}

In the Limber approximation \citep{Limber1953,Loverde2008} (valid for multipoles $\ell \gtrsim 20$) lensing cross-correlations are given by
\begin{equation}
    C^{ab}_\ell = \int \frac{d\chi}{\chi^2} W_a(\chi)W_b(\chi) P_m\left[k_\ell(\chi), z(\chi)\right],
\end{equation}
where $k_\ell(\chi) \equiv (\ell + 1/2) / \chi$, $\chi$ is the comoving radial distance, and $P_m(k, z)$ is the 3D matter power spectrum. The indices $a, b \in [\kcmb, \gamma_i]$ represent either CMB lensing or a tomographic bin of cosmic shear and $W(\chi)$ is the corresponding lensing kernel.\footnote{Throughout we use natural units such that $c=1$.}

For a sample of galaxies with a redshift distribution $n_i(z)$, the lensing kernel is
\begin{equation}\label{eq:lensing_kernel}
    W_{\gamma_i}(\chi) = \frac{3}{2}\Om H_0^2 \frac{\chi}{a(\chi)} \int_\chi^{\infty} d\chi' n_i(\chi') \frac{\chi' - \chi}{\chi'}.
\end{equation}
We note that this equation ignores the relation between underlying observable of interest, lensing convergence or shear, and the direct observable, reduced shear.
But in the weak lensing regime ($\kappa \ll 1$) and at the current levels of sensitivity, the difference between reduced shear and shear can be neglected \citep{Doux2022}. Additionally, the difference between the $E$-mode shear power spectrum and the convergence power spectrum is only an $\ell$-dependent prefactor that is $\sim1$ for multipoles $\ell \gtrsim 20$ and can also be neglected in this analysis. Finally, as in most analyses, we do not model the shear $B$-mode power spectrum since it is expected to be consistent with zero.

The CMB is well approximated by a single source at $\chi = \chi_\star$, where $\chi_\star$ is the distance to the surface of last scattering, so the CMB lensing kernel becomes
\begin{equation}
    W_{\kcmb}(\chi) = \frac{3}{2}\Om H_0^2 \frac{\chi}{a(\chi)} \frac{\chi_\star - \chi}{\chi_\star}.
\end{equation}

We use \camb\ \cite{Lewis2000,Lewis2011} and \textsc{HMCode} \cite{Mead2021} to calculate the nonlinear matter power spectrum.

\subsubsection{Intrinsic alignments}
Due to tidal forces, galaxies are not randomly oriented, but tend to align with the large-scale tidal field, leading to an additional contribution to the shear power spectrum. Intrinsic alignments are one of the main sources of systematic uncertainty in the modeling of shear data and many methods have been developed to model their contribution (see, for example, \cite{Troxel2015,Lamman2024,Chisari2025} for reviews).

The DES Y3 analysis used the tidal alignment and tidal torquing (TATT) model \cite{Blazek2019}, which includes second-order effects, as their fiducial analysis choice, but showed that the simpler nonlinear linear alignment (NLA) model \citep{Hirata2004,Bridle2007} was sufficient to model the DES Y3 data \cite{Secco2022}. Since the cross-correlations with CMB lensing are less constraining than the shear auto-spectra, we use the simpler NLA model in this analysis.

The NLA model assumes that the IA power spectrum is linearly proportional to the nonlinear matter power spectrum. In this model, we simply add an extra term to the galaxy lensing kernel:
\begin{equation}
    W_{\gamma_i}(\chi) \rightarrow W_{\gamma_i}(\chi) + W_{\text{IA}}^i(\chi).
\end{equation}
The IA projection kernel $W_{\text{IA}}^i$ is given by:
\begin{equation}\label{eq:ia_kernel}
    W_{\text{IA}}^i(\chi) = - A_1 C_1 \rho_{\text{crit}} \frac{\Om}{D(z)}n_i(\chi) \left(\frac{1+z}{1+z_0}\right)^{\alpha_1},
\end{equation}
where $D(z)$ is the linear growth factor, $C_1 = 5\times10^{-14} h^{-2}$ $M_\odot^{-1}$ Mpc$^3$ is a conventional normalization constant \cite{Bridle2007}, and $z_0 = 0.62$ is the pivot redshift chosen by previous DES analyses. $A_1$ and $\alpha_1$ are free parameters that determine the IA amplitude and redshift evolution respectively.

\subsubsection{Baryonic feedback}
Baryonic feedback is a major source of theoretical uncertainty in the nonlinear matter power spectrum on small scales ($k \agt 1h$ Mpc$^{-1}$) \cite{Chisari2019b}. We account for this uncertainty by marginalizing over the phenomenological $\logTagn$ parameter in \hmcode. This parameterization was calibrated to reproduce the matter power spectrum observed in the \textsc{Bahamas} hydrodynamical simulations \cite{McCarthy2017} for $z < 1$ and $k < 20h$ Mpc$^{-1}$. While more flexible models of feedback have been developed \cite[e.g.,][]{Arico2021}, the level of small-scale noise in our measurement does not warrant the use of more complicated models.

\subsubsection{Shear measurement uncertainties}\label{sec:cal}
As in the official DES Y3 analysis \citep{Doux2022}, we marginalize over uncertainties in the shear calibration and the redshift distributions of the source galaxies.
Uncertainties in the redshift distributions are parameterized through a shift $\Delta z_i$ of their means:
\begin{equation}
    n_i(z) \rightarrow n_i(z - \Delta z_i).
\end{equation}
Residual uncertainties in the shear calibration are parameterized using a constant amplitude $m_i$ for each bin:
\begin{eqnarray}
    C_\ell^{\kappa\gamma_i} \rightarrow& (1 + m_i) C_\ell^{\kappa\gamma_i}.
\end{eqnarray}
All of the redshift shift parameters $\Delta z_i$ and shear calibration parameters $m_i$ have tight Gaussian priors from detailed simulations of the DES data. Extensive testing in the official DES Y3 analyses has shown that these parameterizations of the measurement uncertainties are sufficiently accurate \cite{Amon2022b}.

Since the shear calibration parameters $m_i$ enter only as linear amplitudes on the cross-spectra, we choose to analytically marginalize over them in a similar manner to what is done in \cite{Doux2021,Joachimi2021}. This is accomplished by fixing $m_i$ to the means of their priors and adding an additional term to the data covariance:
\begin{equation}
    \mathrm{Cov}_m(C_\ell^{\kappa\gamma_i}, C_{\ell'}^{\kappa\gamma_j}) = \sigma_i^2 \delta_{ij} C_\ell^{\kappa\gamma_i} C_{\ell'}^{\kappa\gamma_j},
\end{equation}
where $\sigma_i$ is the Gaussian prior on $m_i$ and the cross-spectra are computed at the fiducial model. This expression can be straightforwardly derived from the expressions in \cite{Doux2021,Joachimi2021} with slight modifications to account for the DES Y3 assumption of uncorrelated calibration priors per tomographic bin (in contrast to KiDS which assumes correlated priors) and the fact that $m = 0$ for the $\kcmb$ field. We have verified that we get essentially identical results whether we use this analytic marginalization method or directly sample the $m_i$ parameters.

\subsubsection{Non-Gaussian covariance terms}\label{sec:cov}
We model the total covariance of the cross-spectra as the sum of the Gaussian covariance computed using \namaster\ ($\mathrm{C}_\text{G}$, \cref{sec:namaster}), the shear calibration term ($\mathrm{C}_m$, \cref{sec:cal}), and two additional terms that account for mode coupling from the non-Gaussianity of the lensing field:
\begin{equation}
    \mathrm{C} = \mathrm{C}_{\text{G}} + \mathrm{C}_m + \mathrm{C}_{\text{SSC}} + \mathrm{C}_{\text{cNG}}.
\end{equation}
$\mathrm{C}_{\text{SSC}}$ is the super-sample covariance term \citep{Takada2013} that captures the connected covariance arising from modes larger than the survey mask. $\mathrm{C}_{\text{cNG}}$ is the connected non-Gaussian term \citep{Takada2004} that arises from the non-Gaussian matter trispectrum.

\begin{figure}
    \centering
    \includegraphics[width=3.4in]{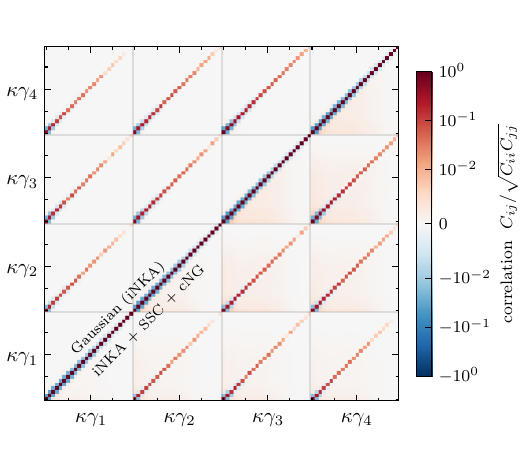}
    \caption{Normalized data covariance matrix for the \desfull$\times$\pol\ cross-correlation. The upper half of the matrix shows only the Gaussian part computed using \namaster, while the bottom half shows the full covariance, including the non-Gaussian components. The sub-blocks correspond to the cross-correlations between the CMB lensing ($\kcmb$) and the individual DES tomographic bins ($\gamma_i$) and show strong correlations due to the overlapping lensing kernels of the DES tomographic bins.}
    \label{fig:cov}
\end{figure}

The SSC and cNG terms are calculated through a halo model approximation implemented using the Core Cosmology Library (\textsc{CCL}) \citep{Chisari2019}. To construct the halo model, we assume a NFW profile \citep{Navarro1996} for dark matter halos, use the halo mass function and halo bias parameterizations from \citep{Tinker2010}, and use the halo concentration relation from \citep{Duffy2008}. We note that this halo model does not incorporate baryonic effects and is thus not expected to be fully accurate on small scales, but this is not expected to cause issues since the small scales are fully noise-dominated. Additionally, the SSC term is computed by approximating the survey mask as a circular aperture. This is expected to be sufficient, given that the SSC term is significantly sub-dominant to the Gaussian term.

There are additional contributions to the covariance due to various uncertainties in the CMB lensing reconstructions that we neglect here since they are expected to be negligibly small compared to the Gaussian covariance of the $\kcmb\gamma$ cross-spectra. These contributions include uncertainties in the SPT-3G calibration and beam, and uncertainties in the primary CMB spectra which are used to compute the lensing response function. These are all small uncertainties in the analysis of the lensing auto-spectrum (O26) and so are not expected to impact the cross-spectra at the current level of significance.

The full normalized covariance matrix for the \desfull$\times$\pol\ data vector is shown in \cref{fig:cov}. We find significant off-diagonal blocks in the covariance due to the overlapping lensing kernels of the DES tomographic bins. Small negative off-diagonal elements result from \namaster\ reversing the mode-coupling due to the survey masks. Both the SSC and cNG terms are significantly sub-dominant to the Gaussian term, but the SSC term adds small off-diagonal correlations on large scales. The cNG term is negligible compared to the others, but is still included for completeness.

\subsection{Parameter inference}\label{sec:inference}

\begin{table}
    \centering
    \begin{tabular}{cll}
        \hline
        Parameter & Fiducial value & Prior \\ \hline \hline
        \multicolumn{3}{c}{Cosmology} \\ \hline
        $\Om$ & 0.311 & $\mathcal{U}(0.1, 0.9)$ \\
        $\Ob$ & 0.049 & $\mathcal{U}(0.03, 0.07)$ \\
        $h$ & 0.677 & $\mathcal{U}(0.55, 0.91)$ \\
        $10^9 A_s$ & 2.105 & $\mathcal{U}(0.5, 5)$ \\
        $n_s$ & 0.967 & $\mathcal{U}(0.87, 1.07)$ \\
        $m_\nu$ & 0.06 eV &  \\ \hline

        \multicolumn{3}{c}{Intrinsic alignments} \\ \hline
        $A_1$ & 0.44 / 0.0 & $\mathcal{U}(-5, 5)$ / $\mathcal{N}(0, 0.5)$ \\
        $\alpha_1$ & 0.0 & $\mathcal{N}(0, 2)$ \\ \hline

        \multicolumn{3}{c}{Baryonic feedback} \\ \hline
        $\logTagn$ & 7.8 & $\mathcal{U}(7.3, 8.3)$ \\ \hline

        \multicolumn{3}{c}{Shear calibration} \\ \hline
        $m_1$ & \multicolumn{2}{c}{\multirow{4}{1.5in}{marginalized over analytically using mean values and uncertainties given in \cref{tab:shear_samples}}} \\
        $m_2$ \\
        $m_3$ \\
        $m_4$ \\ \hline

        \multicolumn{3}{c}{Redshift uncertainties} \\ \hline
        $\Delta z_1$ & 0.0  & $\mathcal{N}(0, 0.018)$ \\
        $\Delta z_2$ & 0.0 & $\mathcal{N}(0, 0.015)$ \\
        $\Delta z_3$ & 0.0 & $\mathcal{N}(0, 0.011)$ \\
        $\Delta z_4$ & 0.0 & $\mathcal{N}(0, 0.017)$ \\ \hline
    \end{tabular}
    \caption{Summary of parameters and priors used in the analysis. $\mathcal{U}(a, b)$ indicates a uniform prior, while $\mathcal{N}(\mu, \sigma)$ indicates a Gaussian prior. For the IA amplitude $A_1$, we use an uninformative prior for the DES \desfull\ sample and an informative prior for the \desblue\ sample. When doing a joint analysis with the \planck\ likelihood, we additionally vary the optical depth $\tau$ using a prior $\mathcal{U}(0.01, 0.8)$.}
    \label{tab:priors}
\end{table}

\subsubsection{Priors}\label{sec:prior}
We list all of the fiducial parameter values and priors used in our parameter inference in \cref{tab:priors}.
We sample all parameters over the same ranges used in the DES Y3 analysis with a few small changes. We use broad uniform priors for the cosmological parameters $\Om$, $\Ob$, $h$, $A_s$, and $n_s$. We assume a single massive neutrino species with a fixed mass of $m_\nu = 0.06$ eV.

For intrinsic alignments, we use two different priors on the amplitude $A_1$ based on the shear sample used. For the \desfull\ sample, we use an uninformative prior $A_1 \sim \mathcal{U}(-5, 5)$, while for the \desblue\ sample, we use a Gaussian prior centered at zero with a standard deviation of 0.5. This informative prior is motivated by recent studies that put fairly tight limits on the IA amplitude of blue galaxies \cite{Johnston2019, Samuroff2023, Siegel2025, Navarro-Girones2026}. For both samples, we sample the power law exponent $\alpha_1$ from a Gaussian prior centered at zero with standard deviation of 2.0. We choose to use a Gaussian prior for $\alpha_1$ since this parameter is unconstrained by the data and very steep power laws are not expected to be physically plausible \cite[see, for example,][]{Han2025,Shi2021}. For the shear nuisance parameters $m_i$ and $\Delta z_i$, we use the same Gaussian priors used in the official DES Y3 analyses.

We sample the baryonic feedback parameter $\logTagn$ uniformly over the range (7.3, 8.3). While \textsc{HMCode} was calibrated such that varying $\logTagn$ between 7.6 and 8.3 roughly reproduces the strength of feedback seen in hydrodynamical simulations, we extend the range, following \cite{DESandKIDS2023}, in order to include low-feedback models that are roughly consistent with a gravity-only model.

\subsubsection{Likelihood}\label{sec:like}

We assume a Gaussian likelihood $\mathcal{L}$ for the joint distribution of the cross-spectrum bandpowers:
\begin{multline}
    -2\ln \mathcal{L} + K \equiv \\
    \chi^2 = (\vb*{d} - \vb*{m}(\vb*{\theta}))^T \mathrm{C}^{-1} (\vb*{d} - \vb*{m}(\vb*{\theta})),
\end{multline}
where $K$ is an arbitrary constant, $\vb*{d}$ is the data vector consisting of the measured cross-spectra, $\vb*{m}$ is the model computed at given values of the parameters $\vb*{\theta}$, and $\mathrm{C}$ is the covariance matrix of the data. The posterior is then proportional to the product of the likelihood and the prior:
\begin{equation}
    P(\vb*{\theta}|\vb*{d}) \propto \mathcal{L}(\vb*{d}|\vb*{\theta}) P(\vb*{\theta}).
\end{equation}

The likelihood is implemented using the \textsc{CosmoSIS}\footnote{\url{https://cosmosis.readthedocs.io/en/latest/}} framework \cite{Zuntz2015} and we sample the posterior using the nested sampler \textsc{Nautilus}\footnote{\url{https://nautilus-sampler.readthedocs.io/en/latest/}} \cite{Lange2023}.
We estimate the maximum-a-posteriori (MAP) as the highest posterior sample in each \textsc{Nautilus} chain.

For each fit, we estimate the effective number of parameters $\Npeff$ that are constrained relative to the prior using the Gaussian approximation implemented in \tensiometer\footnote{\url{https://tensiometer.readthedocs.io/en/latest/index.html}} \cite{Raveri2019}:
\begin{equation}
    \Npeff = N_p - \mathrm{Tr}(\mathrm{C}_\Pi^{-1} \mathrm{C}_p),
\end{equation}
where $N_p$ is the total number of free parameters in the fit and $\mathrm{C}_\Pi$ and $\mathrm{C}_p$ are the (approximate) prior and posterior covariances respectively.
This is then used to estimate the goodness-of-fit by comparing the $\chi^2$ value of the MAP point to the estimated degrees of freedom $N_\text{dof} = N_d - \Npeff$, where $N_d$ is the size of the data vector.

\subsubsection{Scale cuts}\label{sec:scale_cuts}

\begin{table}
    \centering
    \begin{tabular}{lllllll}
        \hline
        \multicolumn{2}{c}{\multirow{2}{0.6in}{$\kmax$ cut\newline [$h$ Mpc$^{-1}$]}} & \multicolumn{4}{c}{$\lmax$ cut} & \multirow{2}{0.25in}{$N_d$} \\
        & & bin 1 & bin 2 & bin 3 & bin 4 \\
        \hline \hline
        0.5 & \desfull & 82 & 115 & 150 & 182 & 10 \\
            & \desblue & 99 & 126 & 151 & 183 & 12 \\ \hline
        1 & \desfull & 160 & 218 & 281 & 340 & 24 \\
          & \desblue & 193 & 238 & 282 & 341 & 26 \\ \hline
        2 & \desfull & 342 & 462 & 593 & 709 & 43 \\
          & \desblue & 409 & 501 & 593 & 710 & 45 \\ \hline
        5 & \desfull & 1071 & 1451 & 1871 & 2216 & 75 \\
          & \desblue & 1266 & 1563 & 1867 & 2213 & 77 \\ \hline
    \end{tabular}
    \caption{Conversion of $\kmax$ cuts to $\lmax$ cuts for the lensing-shear cross-correlation, assuming the fiducial model listed in \cref{tab:priors}. Due to the slightly different redshift distributions of the \desfull\ and \desblue\ shear samples, a given $k$ cut translates into slightly different $\ell$ cuts, especially for the first two tomographic bins. The last column lists the number of remaining data points $N_d$ (out of the total of 96) after applying the scale cuts.}
    \label{tab:scale_cuts}
\end{table}

For our fiducial analysis, we attempt to fit all scales present in the measured data vectors ($30 \le \ell \le 3500$). To test the robustness of our results, we also test the effect of scale cuts designed to remove sensitivity to all spatial modes with $k > \kmax$. To convert a $\kmax$ cut to a corresponding $\lmax$ cut, we use a method similar to that developed by \cite{Doux2021}. Specifically, for a given cross-spectrum we calculate the $C_\ell$ using the full nonlinear $P(k)$ and using a $P(k)$ that has been exponentially suppressed for $k > \kmax$. We then define $\lmax$ as the maximum value of $\ell$ for which the two $C_\ell$s agree within a given threshold (here we choose 5\%). The resulting scale cut ensures that no more than 5\% of the signal in the data vectors depends on $k$ modes beyond $\kmax$. \cref{tab:scale_cuts} lists the specific $\lmax$ cuts that correspond to $\kmax \in \{0.5, 1, 2, 5\}$ $h$ Mpc$^{-1}$ (chosen to represent a range of quasi-linear to nonlinear scales) for each of the tomographic bins of the $\kappa\gamma$ cross-correlation for both the \desfull\ and \desblue\ samples. Note that, due to the wide lensing kernels mapping a range of $k$ modes into a given multipole $\ell$, this method of defining scale cuts will remove some sensitivity to $k < \kmax$ in addition to removing the sensitivity to $k > \kmax$. In future analyses, it will likely be helpful to use methods similar to that recently presented by \cite{DeRose2025}, which marginalizes over small-scale contributions to the lensing signal.

There are two main components of the theory model that are expected to be mis-modeled at some level in this analysis: the matter power spectrum and galaxy IA. Since \hmcode\ has been calibrated against simulations to roughly the per-cent level at $k < 20h$ Mpc$^{-1}$ and $z < 1$ \cite{Mead2021}, systematic uncertainty in the matter power spectrum is not expected to be significant compared to the statistical uncertainty of our measurement. On the other hand, it is reasonable to expect more significant problems on the IA modeling side, since IA models are generally not tested on nonlinear scales. While the model that we use (NLA) is particularly simplistic, there is evidence from simulations that it is roughly valid out to $k \sim 1h$ Mpc$^{-1}$ \cite{Shi2021}.
Using the scale cuts described above, we estimate which $k$ modes contribute the most towards the total SNR of our cross-correlation measurement. We find that our measurement is essentially insensitive to scales $k \gtrsim 5h$ Mpc$^{-1}$ and that roughly half of the SNR in the measurement comes from fairly large scales ($k < 0.5h$ Mpc$^{-1}$). Based on this, we do not expect our fiducial choice of including all multipoles to significantly bias our results.

\begin{figure*}
    \centering
    \includegraphics[width=\linewidth]{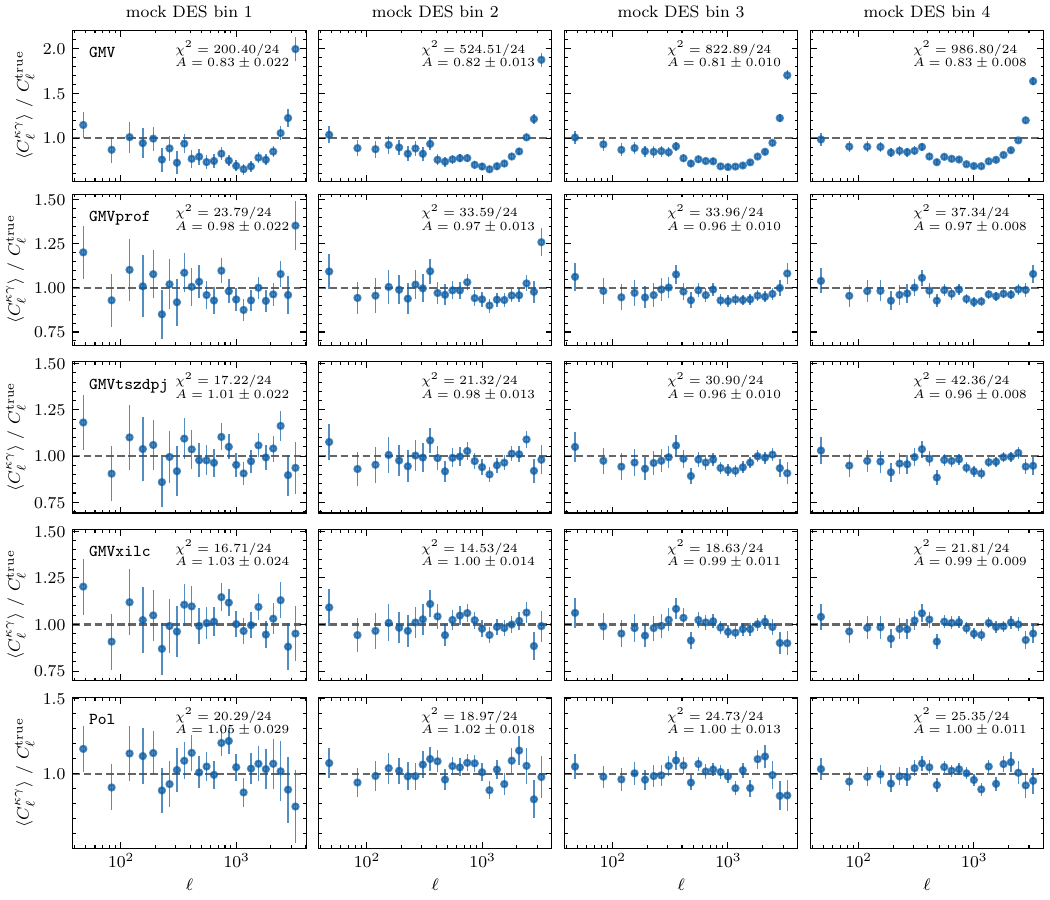}
    \caption{Measurement validation on the \agora\ simulations. Each panel compares the mean measured $\kappa\gamma$ cross-correlation to the truth (measured over the full sky with no noise). Each column corresponds to a different tomographic bin, while each row corresponds to a different CMB lensing reconstruction. The simulated shear catalogs in this test do not include shape noise to increase statistical power. In each panel we list the $\chi^2$ of the ratio relative to one and the best fit amplitude. The error bars represent the standard deviation of the mean of the 10 \agora\ patches.}
    \label{fig:agora_cls_noiseless}
\end{figure*}

\section{Validation}\label{sec:validation}

Using the \agora-simulated CMB lensing reconstructions and shear catalogs, we conduct three tests to validate our measurement and analysis. First, using the simulated shear catalogs without shape noise, we test for biases in the cross-correlation bandpowers. This test aims to validate the mitigation of extragalactic foreground biases and the pseudo-$C_\ell$ measurement pipeline. We then include a realistic level of shape noise to test our analytic estimate of the covariance matrix and to test the \textsc{CosmoSIS} modeling and inference pipeline for correct parameter recovery.

\subsection{Cross-correlation bandpower recovery}\label{sec:val-bias}
In \cref{fig:agora_cls_noiseless}, for each of the \agora-simulated CMB lensing reconstructions, we plot the ratio of the average recovered $C_\ell^{\kappa\gamma}$ cross-spectra over the ``true'' cross-spectra. Here the recovered cross-spectra are measured from all 10 of the \agora\ realizations of simulated CMB lensing reconstructions and simulated DES shear catalogs (without shape noise, to increase sensitivity to potential biases). The ``true'' cross-spectra are calculated from the noiseless full-sky \agora\ lensing maps. We also note that the ``true'' cross-spectra provide a reasonable fit to the data, meaning that, up to differences in cosmology and feedback as well as uncertainties in the modeling of foregrounds, these tests on \agora\ should provide a reasonable picture of what is happening in the data. For the lensing estimators that include temperature information, this test provides an estimate of the level of foreground-induced biases that should be expected in the data. While for the \pol\ estimator, which is expected to be unaffected by foregrounds, this test validates the measurement pipeline, including QE normalization and shear pixelization.

We find that using the \pol\ lensing maps (bottom row of \cref{fig:agora_cls_noiseless}) we are able to recover the cross-correlation bandpowers without bias, confirming that they are robust against foreground contamination and that our measurement pipeline correctly recovers the input power spectra. We also do not see any evidence for QE mis-normalization due to coupling between the SPT and DES masks. On the other hand, we find significant evidence for foreground biases in the \gmv\ maps (top row of \cref{fig:agora_cls_noiseless}), as expected since in this analysis we do not marginalize over foregrounds as done in O26. On large to intermediate angular scales, this manifests as a negative $\sim 20$ - 40\% bias, while on small angular scales ($\ell > 2000$) it becomes a large positive $\sim50$\% bias. The shape of this bias is qualitatively consistent with what is expected from tSZ contamination \cite{vanEngelen2014,Omori2024,Raghunathan2023}.

For the \prof\ and \mh\ lensing maps, which mainly attempt to remove tSZ contamination, we find hints of small (no more than a few percent) biases, potentially indicating either imperfect removal or potential CIB contamination. We note that these foreground-mitigated estimators were derived in the context of measuring the $\kcmb$ auto-spectrum and might not be optimal for cross-correlations with low-$z$ tracers. Specifically, it is possible that the profiles chosen for the \prof\ QE to minimize the bias in the $\kcmb$ auto-spectrum might not match the profiles of a low-$z$ sample. In the case of the \mh\ QE, nulling tSZ generally enhances the contribution from the CIB \cite{Raghunathan2023}, potentially leading to biases in the high-$z$ tomographic bins which are more correlated with the CIB. The \xilc\ maps, which attempt to remove both tSZ and CIB contamination, are consistent with no bias, but we again note that the weights used were chosen to minimize CIB in the CMB $TT$ spectrum and thus might not be optimal for mitigating bias in a cross-correlation with low-$z$ tracers.

Other than extragalactic foregrounds and IA, plausible sources of systematics that are correlated across the CMB lensing and cosmic shear datasets could include the correlation between galactic foreground contamination to CMB lensing and the effects of galactic dust on the optical weak lensing signal. We do not expect galactic foregrounds to introduce any significant biases into our results since our analysis is limited to a fairly small region far from the galactic plane, and the CMB lensing reconstructions exclude the largest scales that are most affected by galactic foregrounds. Individually, both the shear catalogs and the lensing reconstruction have passed rigorous validation tests and have shown no signs of possible contamination from galactic foregrounds \cite[][O26]{Gatti2021}. Additionally, \cite{Abril-Cabezas2025} showed that non-Gaussian galactic foregrounds are not expected to be a significant source of bias in CMB lensing reconstructions, even for the Simons Observatory which will cover a significantly larger sky area than the SPT-3G Main field.

\subsection{Chi-squared test}\label{sec:val-cov}
We validate our estimate of the covariance matrix $\mathrm{C}$ by testing whether it correctly describes the covariance of simulated bandpowers $\vb*{d}_i$ around the true full sky cross-correlation $\vb*{d}^\text{true}$. For each \agora\ realization $i$, we compute
\begin{equation}
\chi^2_i = (\vb*{d}_i - \vb*{d}^\text{true})^T \mathrm{C}^{-1} (\vb*{d}_i - \vb*{d}^\text{true}).
\end{equation}
Under the assumption that the bandpowers are unbiased and drawn from a multivariate Gaussian distribution with covariance $\mathrm{C}$, the $\chi^2_i$ values are expected to follow a $\chi^2$-distribution with number of degrees of freedom equal to the length of the data vector. If our analytic estimate of the covariance matrix is incorrect (or the bandpowers are significantly biased), then we would expect deviations from the expected distribution.

\begin{figure}
    \centering
    \includegraphics[width=\linewidth]{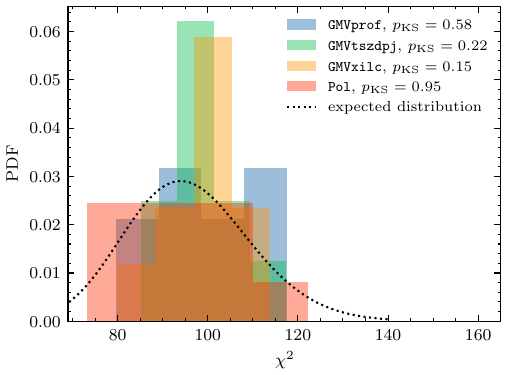}
    \caption{Distributions of $\chi^2$ values for each of the simulated SPT-3G $\times$ DES Y3 data vectors. For each variant (excluding \gmv) we list the $p$-value corresponding to the null hypothesis that the values follow the expected distribution computed using a two-sided Kolmogorov-Smirnov test. The raw \gmv\ data vectors (not shown in this plot), which do not mitigate foreground biases, result in $\chi^2$ values that are biased $\sim20$\% high relative to the other variants.}
    \label{fig:chi2_distr}
\end{figure}

Since the previous test already shows that the simulated \gmv\ bandpowers are significantly affected by foregrounds, we exclude them from this test. For all the other estimators any residual bias is significantly below the statistical noise of our measurement, so we show the distributions of $\chi_i^2$ values in \cref{fig:chi2_distr} where we also list the $p$-values corresponding to the null hypothesis that the values follow the expected distribution computed using a two-sided Kolmogorov-Smirnov (KS) test. While this test is not particularly powerful due to the limited number of \agora\ realizations, we do not find any evidence that the observed distributions deviate from the expected one for any of the non-\gmv\ cross-correlations.

\subsection{Parameter recovery}

\begin{figure}
    \centering
    \includegraphics[width=\linewidth]{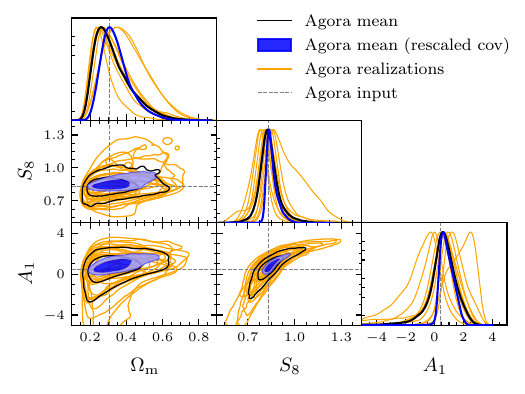}
    \caption{\agora\ parameter recovery test for the \pol\ $\kappa\gamma$ data vectors. We plot the marginal posterior distributions corresponding to the mean \agora\ data vector with the data covariance in black (representing a single realization) and with the data covariance scaled by a factor of 1/10 in blue (representing the mean of 10 realizations). The posteriors for each of the individual realizations are shown in orange. The \agora\ input parameters are shown as the dashed lines.}
    \label{fig:agora_param_pol}
\end{figure}

\begin{figure}
    \centering
    \includegraphics[width=3in]{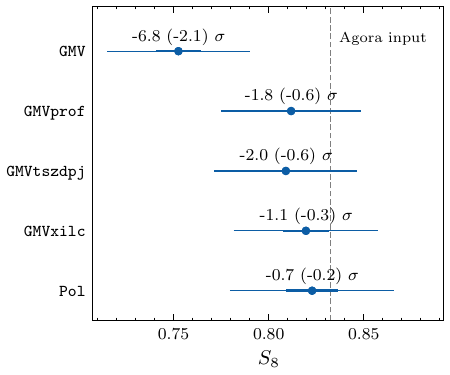}
    \caption{\agora\ $S_8$ recovery test for each of the five lensing map variants. The thick error bars represent the marginalized error bar on $S_8$ from the combination of all 10 realizations, while the thin error bars are scaled by a factor of $\sqrt{10}$ to represent the mean error of a single realization. We also list the resulting biases relative to the input value for the combined and (single realization) cases.}
    \label{fig:agora_s8}
\end{figure}

When inferring cosmological parameters from the simulated bandpowers, we use a slightly modified \textsc{CosmoSIS} pipeline compared to what is used on the data. (i) To be consistent with the MDPL2 cosmology, neutrinos are assumed to be massless. (ii) The nonlinear matter power spectrum is computed using \textsc{HMCode} without baryonic feedback, since \agora\ is based on a gravity-only $N$-body simulation.

We run the inference pipeline on all 10 \agora\ realizations for each of the five lensing map variants. We also run the pipeline on the mean data vector for each lensing map variant with a covariance corresponding to a single realization and with a covariance that has been scaled by a factor of 1/10, representing the variance of the mean.

We show the results of this test on the \pol\ $\kappa\gamma$ data vectors in \cref{fig:agora_param_pol}, where we plot the marginalized posteriors in the $\Om$-$S_8$-$A_1$ parameter space. We find that the contours for individual realizations scatter around the true input values and the contours corresponding to the mean data vectors are correctly centered on the input values.

In \cref{fig:agora_s8} we summarize the marginalized constraints on $S_8$ obtained from each of the lensing map variants. Specifically, the thick error bars represent the combined constraint obtained from all 10 realizations, while the thin error bars are scaled by a factor of $\sqrt{10}$ to represent the mean error bar of a single realization. When using the \gmv\ reconstruction with no foreground mitigation or marginalization, we find a $-6.8\sigma$ bias on $S_8$ when combining all 10 \agora\ realizations, translating to roughly an average of a $-2.1\sigma$ bias for a single realization. All of the other estimators result in constraints that are consistent with the input. Only the \prof\ and \mh\ estimators can be considered to result in a slight tension with the input at roughly 1.8/2.0$\sigma$ respectively when combining all 10 \agora\ realizations.

\section{Results}\label{sec:results}

\subsection{Measurement significance}
In \cref{tab:snr} we list the estimated SNR for each set of cross-correlation bandpowers which we define as
\begin{equation}
    \mathrm{SNR} = \sqrt{\vb*{d}^T \mathrm{C}^{-1} \vb*{d} - N_d},
\end{equation}
where $N_d$ is the length of the data vector. We also list the amplitude of the measured cross-correlations relative to the prediction assuming a \planck\ 2018 cosmology. This amplitude is calculated as
\begin{equation}
    A_\text{Planck} = \frac{\vb*{t}^T \mathrm{C}^{-1} \vb*{d}}{\vb*{t}^T \mathrm{C}^{-1} \vb*{t}},
\end{equation}
where $\vb*{t}$ is the theory vector calculated using the fiducial model listed in \cref{tab:priors}.

\begin{table}
    \centering
    \begin{tabular}{lcc}
        \hline
        Data vector & SNR & $A_\text{Planck}$ \\ \hline
        \hline
        \desfull$\times$\gmv & 13.9 & $0.86\pm0.060$ \\
        \desfull$\times$\prof & 15.5 & $0.98\pm0.062$ \\
        \desfull$\times$\mh & 15.2 & $0.98\pm0.063$ \\
        \desfull$\times$\xilc & 14.9 & $0.99\pm0.065$ \\
        \desfull$\times$\pol & 13.9 & $1.02\pm0.073$ \\
        \desblue$\times$\pol & 13.4 & $1.10\pm0.084$ \\
        \hline
    \end{tabular}
    \caption{Signal-to-noise ratios and amplitudes relative to the \planck\ \lcdm\ prediction ($A_\text{Planck}$) for each of the cross-correlation variants.}
    \label{tab:snr}
\end{table}

For all data combinations, we get high-significance ($> 13\sigma$) measurements of the $\kappa\gamma$ cross-correlation. The highest-significance measurement (15.5$\sigma$) is obtained when using the full shear sample and the profile-hardened lensing estimator. It is notable that when using the polarization-only estimator, the significance drops by less than $2\sigma$, indicative of the low polarization noise levels in the SPT-3G maps. We also see evidence of foreground biases in the \desfull$\times$\gmv\ cross-correlation, as expected due to the lack of foreground mitigation or marginalization. While the \gmv\ reconstruction results in the lowest lensing noise, the SNR of the \desfull$\times$\gmv\ data vector is roughly the same as that of the \desfull$\times$\pol\ one, due to negative foreground biases canceling the signal. We also find that the \desfull$\times$\gmv\ data vector exhibits a mild $2.3\sigma$ tension relative to the fiducial \planck\ prediction. For all other $\kcmb$ map variants we obtain values of $A_\text{Planck}$ that are fully consistent with 1.

\begin{figure*}
    \centering
    \includegraphics[width=5in]{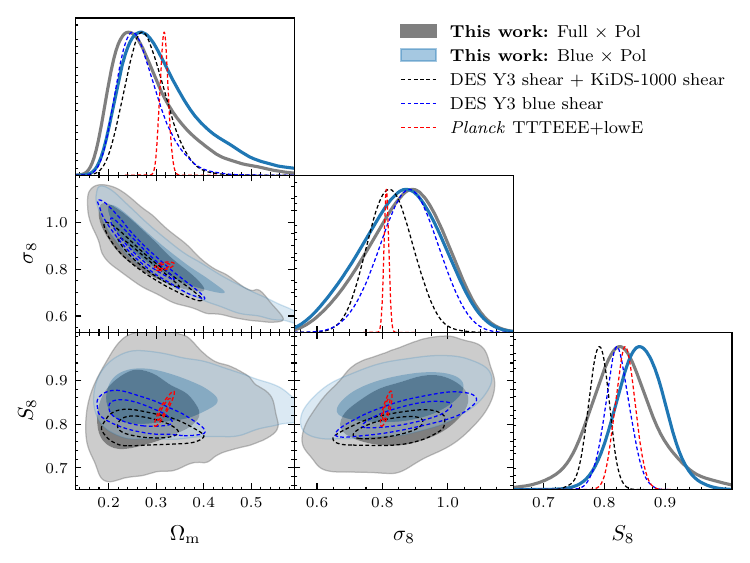}
    \caption{Marginalized posteriors in the $\Om$-$\sigma_8$-$S_8$ parameter space for our \desfull$\times$\pol\ and \desblue$\times$\pol\ data vectors. Note that the \desblue$\times$\pol\ analysis includes an informative prior on the IA amplitude that breaks the degeneracy with $S_8$. We also compare our results with published cosmic shear constraints (DES Y3 + KiDS-1000 \cite{DESandKIDS2023} and DES Y3 blue shear \cite{McCullough2024}) and the primary CMB (\planck\ TTTEEE+lowE \cite{Planck2020}).}
    \label{fig:results}
\end{figure*}

\subsection{Cosmological constraints}
The cosmological constraints from our measurement are shown in \cref{fig:results} where we show the marginalized posteriors in the  $\Om$-$\sigma_8$-$S_8$ parameter space obtained from the \desfull$\times$\pol\ and \desblue$\times$\pol\ data vectors. We also compare our results with those from three external data sets: DES Y3 + KiDS-1000 cosmic shear \cite{DESandKIDS2023}, DES Y3 blue shear \cite{McCullough2024},\footnote{Note that these shear-only results do not include shear ratios, which can improve photo-$z$ and IA constraints, in contrast to the fiducial DES Y3 analysis \citep{Amon2022b,Secco2022}.} and measurements of the primary CMB from \planck\ (TTTEEE+lowE) \cite{Planck2020}. While they are significantly less constraining, we find that our constraints using both the \desfull$\times$\pol\ and the \desblue$\times$\pol\ data vectors are fully consistent with both shear-only and primary CMB results.

Using the \desfull$\times$\pol\ data vector and assuming the NLA model for galaxy IA, we obtain:
\begin{eqnarray*}
    \Om &= 0.290^{+0.034}_{-0.097}, \\
    S_8 &= 0.833^{+0.047}_{-0.061}.
\end{eqnarray*}
The effective number of constrained parameters for this fit is $\Npeff = 3.0$ (roughly corresponding to $\Om$, $S_8$, and $A_1$).

Using the \desblue$\times$\pol\ data vector and assuming NLA with an informative prior on the IA amplitude, we obtain:
\begin{eqnarray*}
    \Om &= 0.323^{+0.041}_{-0.11}, \\
    S_8 &= 0.860\pm 0.040.
\end{eqnarray*}
In this case, the effective number of constrained parameters is $\Npeff = 2.3$, reflecting the fact that a significant portion of the constraining power on $A_1$ for the blue sample is coming from the informative prior.
While the blue sample has significantly fewer galaxies than the full sample, limiting the analysis to blue galaxies allows us to use an informative IA prior, which improves the constraint on $S_8$ by roughly 25\% by breaking the degeneracy with the IA amplitude.

\begin{figure}
    \centering
    \includegraphics[width=\linewidth]{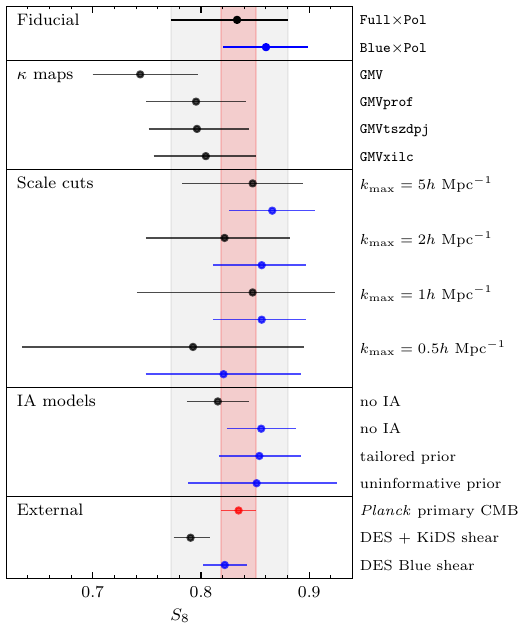}
    \caption{Effect of different analysis choices on the resulting $S_8$ constraint. Black (blue) points correspond to the \desfull\ (\desblue) DES samples. The vertical red band corresponds to the constraint from the \planck\ TTTEEE+lowE likelihood, while the gray band corresponds to the fiducial \desfull$\times$\pol\ result. The external constraints shown are the same ones as in \cref{fig:results}. See \cref{sec:results-foregrounds,sec:results-scale-cuts,sec:results-ia} for details on the different sub-panels.}
    \label{fig:results_var}
\end{figure}

We summarize the the main cosmological constraints from all considered data combinations in \cref{tab:results}. For each data vector we list the inferred $S_8$ marginalized 68\% confidence interval and estimated MAP value. We also list the $\chi^2$ value of the estimated MAP point along with the estimated probability-to-exceed (PTE).
We summarize the results from several analysis variations in the following subsections and in \cref{fig:results_var}.

\begin{table}
    \centering
    \begin{tabular}{lcc}
        \hline
        Data vector & $S_8$ & $\chi^2_\text{MAP}$ / PTE \\
        \hline \hline
        \desfull$\times$\gmv & $0.744^{+0.053}_{-0.044}$ (0.79) & 85.0 / 0.71 \\
        \desfull$\times$\prof & $0.795\pm 0.052$ (0.81) & 87.2 / 0.65 \\
        \desfull$\times$\mh  & $0.796\pm0.052$ (0.82) & 83.6 / 0.74 \\
        \desfull$\times$\xilc & $0.804\pm 0.054$ (0.80) & 84.7 / 0.71 \\
        \desfull$\times$\pol & $0.833^{+0.047}_{-0.061}$ (0.86) & 94.5 / 0.44 \\
        \desblue$\times$\pol & $0.860\pm 0.040$ (0.87) & 105.8 / 0.19 \\ \hline
    \end{tabular}
    \caption{Summary of $S_8$ constraints for all data combinations considered. For each combination we list the marginalized 68\% confidence limits as well as the MAP values in parentheses. We also list the $\chi^2$ value at the estimated MAP point and the estimated probability-to-exceed (PTE) value calculated using the number of degrees of freedom as described in \cref{sec:like}. The effective number of constrained parameters for each data combination is 3.2, 3.2, 3.3, 3.2, 3.0, and 2.3 respectively.}
    \label{tab:results}
\end{table}

\subsubsection{CMB lensing foregrounds}\label{sec:results-foregrounds}
In the second panel of \cref{fig:results_var} we compare the constraints obtained when cross-correlating the four GMV variants of the CMB lensing maps with the full DES Y3 sample. Generally, we see good agreement between all of the data combinations and the fiducial polarization-only result, except for the one that uses the raw \gmv\ map which is clearly biased towards lower $S_8$ values, consistent with the lower amplitude found relative to \planck\ (\cref{tab:snr}) compared to the other data combinations.

We find that the signs of foregrounds in the data are fairly consistent with the \agora\ foreground models. Using \tensiometer\ to calculate the Gaussian tension between marginalized $S_8$ posteriors, we find differences of $1.7\sigma$, $0.7\sigma$, $0.7\sigma$, $0.5\sigma$, and $0.02\sigma$ between \planck\ TTTEEE+lowE and the \gmv, \prof, \mh, \xilc, and \pol\ $\kappa\gamma$ cross-correlations respectively. These differences are similar to the average \agora\ single-realization biases (\cref{fig:agora_s8}).

To quantify the amount of foreground contamination in a model-independent way, we calculate the $\chi^2$ values relative to zero of the difference between each data vector and the polarization-only one.\footnote{We include the covariance between lensing reconstructions in this calculation.} Since the cross-correlations with the first two DES bins are noise-dominated, we only include the last two bins in this test. The resulting PTEs of the $\chi^2$ values are 0.05, 0.23, 0.33, and 0.21 for the \gmv, \prof, \mh, and \xilc\ difference vectors respectively. This corresponds to roughly $2\sigma$ model-independent evidence for foregrounds in the \gmv\ cross-correlation, while the other variants all differ from the polarization-only one by less than $1.5\sigma$.

We additionally quantify the level of foreground biases by fitting a free amplitude times the best fit model resulting from the \desfull$\times$\pol\ cross-correlation to each of the other cross-correlation data vectors. For the \gmv\ cross-correlation, we find an amplitude of $0.83\pm0.058$ (roughly $2.9\sigma$ different from one), while for the other variants the amplitudes are consistent with one within $1\sigma$.

\subsubsection{Scale cuts}\label{sec:results-scale-cuts}
A potential concern about this analysis is that we attempt to use all scales to extract cosmological constraints while using relatively simplistic models for IA and baryonic feedback. Since the measured $\kappa\gamma$ cross-correlation on its own is unable to constrain the parameters related to these effects, we conclude that there is not enough sensitivity in the data on small scales to warrant more complicated models. For completeness, in the third panel of \cref{fig:results_var} we show the marginalized constraints on $S_8$ obtained from the polarization-only data vectors when applying scale cuts of $\kmax = 5$, 2, 1, and $0.5h$ Mpc$^{-1}$ as described in \cref{sec:scale_cuts}. As expected, applying more conservative scale cuts significantly loosens the $S_8$ constraint, but we do not see any significant shifts in the mean values that might indicate biases due to mis-modeling of small scales. We note that the constraints based on the \desblue\ sample do not loosen as much as those for the \desfull\ sample as the scale cuts become more restrictive. This is due to the inclusion of the informative IA prior, which breaks the $S_8$-$A_1$ degeneracy.

\subsubsection{Intrinsic alignments}\label{sec:results-ia}
Under the assumption of NLA (which is expected to be a reasonable approximation on the quasi-linear scales our measurement is mostly sensitive to), we find that our results are robust against IA modeling choices. In the bottom panel of \cref{fig:results_var} we show how varying the IA priors for the \desfull$\times$\pol\ and \desblue$\times$\pol\ analyses affects the resulting $S_8$ constraints.

Consistent with the DES Y3 findings \cite{Secco2022}, we do not find any preference in the data for non-zero IA. In our fiducial \desfull$\times$\pol\ results, we obtain a weak constraint on the IA amplitude of $A_1 = 0.21_{-0.75}^{+0.98}$.\footnote{In the NLA model, the IA amplitude is formally degenerate with the amplitude of the matter power spectrum, but this degeneracy is mildly broken due to the different redshift dependence of the IA and lensing kernels, see \cref{eq:lensing_kernel,eq:ia_kernel}.} When assuming no IA, the $S_8$ constraint tightens by a factor of $2\times$ while the fit PTE improves by a marginal amount of 0.01 due to the simpler model. The central inferred value of $S_8$ shifts slightly lower when assuming no IA, but the consistency with \planck\ remains within $1\sigma$.

For the \desblue$\times$\pol\ cross-correlation, we find almost no shift in the central value of $S_8$ when varying the IA priors, consistent with our expectation that the blue shear sample is less sensitive to IA than the full sample.
We compare our fiducial informative IA prior to the ``tailored IA'' prior from \cite{Bigwood2025}. Rather than assuming a power-law evolution of the IA amplitude, \cite{Bigwood2025} assumes independent amplitudes for each tomographic bin and sets priors centered at zero with widths calibrated to external direct IA measurements from DESI, eBOSS, and PAUS \cite{Siegel2025,Samuroff2023,Navarro-Girones2026}. We find that this ``tailored IA'' prior results in a slightly tighter $S_8$ constraint ($0.854\pm0.038$) compared to our fiducial result ($0.860\pm0.040$).
Similar to the full sample, assuming no IA tightens the $S_8$ constraint (in this case roughly by 20\%) and marginally improves the fit PTE.
When using the same uninformative IA prior as used for the full shear sample, the $S_8$ constraint widens significantly (roughly $1.7\times$ compared to the fiducial result), but the central value shows no significant shift that might indicate potential IA mis-modeling.

\subsection{Combining with Planck}
We run a joint analysis of our measured $\kappa\gamma$ cross-correlation with the \planck\ 2018 TTTEEE+lowE likelihood \cite{Planck2020_like}. This has the effect of putting a tight prior on the cosmological parameters, allowing us to put constraints on astrophysical parameters (IA and baryonic feedback) or shear calibration parameters. While the recently presented CMB-SPA T\&E results \cite{Camphuis2025}, based on the combination of primary CMB measurements from SPT-3G D1, \planck, and ACT-DR6 \cite{Naess2025,Louis2025}, result in a tighter $S_8$ constraint than \planck\ alone, our cross-correlation measurement does not have enough SNR to take advantage of this additional constraining power. For all of the results reported in this section, using the SPA likelihood instead of \planck\ results in negligible changes in the reported constraints.

\begin{figure*}
    \centering
    \includegraphics[width=5in]{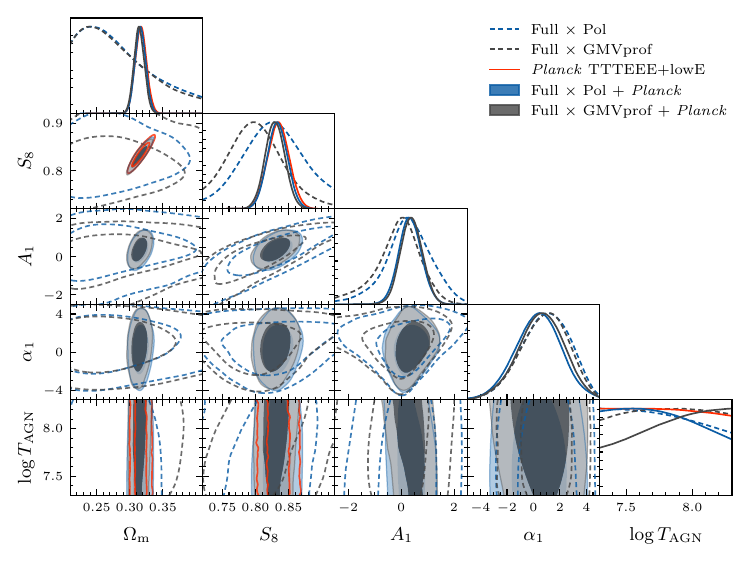}
    \caption{Cosmological and astrophysical constraints from the combination of \planck\ with the \desfull$\times$\pol\ and \desfull$\times$\prof\ cross-correlations.}
    \label{fig:pol_prof_planck}
\end{figure*}

\cref{fig:pol_prof_planck} shows the marginalized posteriors resulting from the combination of the \desfull$\times$\pol\ and \desfull$\times$\prof\ cross-correlations with \planck. We focus specifically on these two data combinations since \desfull$\times$\pol\ is our fiducial result that is expected to be robust against foreground biases, while \desfull$\times$\prof\ is our highest SNR measurement that does not show significant signs of bias at the current level of statistical noise. By combining with \planck, we are able to significantly tighten the constraints on the IA amplitude of the DES Y3 sample. From \desfull$\times$\pol\ + \planck\ we obtain a constraint of $A_1 = 0.34\pm 0.43$. We get a similar constraint using \desfull$\times$\prof\ + \planck: $A_1 = 0.39\pm0.41$. In both cases, we see no evidence for any redshift evolution of the IA amplitude. It is noteworthy that these constraints on the IA amplitude are fully consistent with and in some cases tighter than the constraints on the NLA model found in shear auto-correlation analyses \cite{Doux2022,Secco2022,DESandKIDS2023}.

By combining with \planck, we can also test our assumption of no IA in the DES \desblue\ sample. When analyzing \desblue$\times$\pol\ + \planck\ assuming NLA and using the uninformative amplitude prior, we find an IA amplitude of $A_1 = -0.25^{+0.55}_{-0.46}$, consistent with zero and with the constraint found by \cite{McCullough2024}.

We find that our data have some mild sensitivity to the strength of baryonic feedback as parameterized by $\logTagn$. The \pol\ lensing maps are too noisy on small scales to provide constraining power on the strength of baryonic feedback, even when mitigating IA using the blue sample selection. On the other hand, by combining the \desfull$\times$\prof\ cross-correlation with \planck, we begin to see evidence for small-scale suppression of the matter power spectrum from baryonic feedback. We get a 68\% credible lower limit of $\logTagn > 7.70$ which is consistent with the amount of suppression generally found in shear auto-correlation analyses \cite{Amon2022}.

Finally, we note that measurements of the cross-correlation between CMB lensing and cosmic shear have the potential to help calibrate the cosmic shear nuisance parameters $m_i$ and $\Delta z_i$ \citep[e.g.,][]{Das2013,Vallinotto2012,Schaan2017}. When assuming wide, uniform priors for these parameters, we find that the shear calibration parameters $m_i$ are essentially unconstrained using the combination \desfull$\times$\pol\ + \planck. On the other hand, the photometric redshift calibration parameters $\Delta z_i$ are mildly constrained. We focus specifically on the highest-redshift DES bin since it is likely the hardest to calibrate directly \citep[e.g.,][]{Gomes2025}, but is the most correlated with CMB lensing. Keeping the fiducial Gaussian priors on the other calibration parameters while sampling $\Delta z_4$ from a wide, uniform prior, we find $\Delta z_4 = -0.018 \pm 0.052$. This constraint is consistent with, but roughly 3.3 times wider than the fiducial prior. As future galaxy lensing surveys will probe higher redshifts, the potential for CMB lensing to provide constraints on calibration parameters is expexcted to increase significantly.

\section{Conclusions}\label{sec:conclusion}
We have measured the cross-correlation between CMB lensing ($\kcmb$) from SPT-3G and DES Y3 cosmic shear ($\gamma$) over an overlapping sky area of approximately 1,300 deg$^2$. This cross-correlation provides a useful consistency test that is less sensitive to the measurement systematics of each survey. Based on this measurement, we found cosmological constraints that are consistent with expectations based on previous measurements of the primary CMB and cosmic shear.

The main findings of this analysis are as follows.
\begin{itemize}
    \item We measure the $\kappa\gamma$ cross-correlation at a significance of $14\sigma$ when only using CMB polarization data in the lensing reconstruction. The significance increases to $\sim 15 - 16\sigma$ when including temperature data and using various foreground mitigation techniques.
    \item We fit a model to the data, marginalizing over uncertainties in galaxy intrinsic alignments, baryonic feedback, and cosmic shear nuisance parameters, and find a 6.5\% constraint on the amplitude of matter clustering, $S_8 = 0.833^{+0.047}_{-0.061}$, that is consistent with both results from \planck\ and from cosmic shear surveys.
    \item Using a high-purity sub-selection of blue star-forming galaxies from the full DES Y3 catalog that is expected to be less impacted by IA than the full sample \cite{McCullough2024,Siegel2025}, we show that reducing the uncertainty on the IA amplitude significantly improves the constraining power on $S_8$. When modeling the $\kappa\gamma$ cross-correlation with the blue sample with an informative IA prior, we obtain a $\sim$5\% constraint on $S_8$ that is still consistent with both the primary CMB and cosmic shear surveys.
    \item We use the \planck\ 2018 results to put a tight prior on cosmological parameters and show that the measured $\kappa\gamma$ cross-correlation has some sensitivity to small-scale physics. Using the \desfull$\times$\pol\ data vector, we find a constraint on the IA amplitude that is comparable to what is found in the full DES Y3 cosmic shear analysis. We find no evidence for significant redshift evolution of this amplitude. Using the \desfull$\times$\prof\ data vector, we additionally find hints of a small-scale suppression of power due to baryonic feedback.
\end{itemize}
These results represent the first high-significance measurement of the $\kappa\gamma$ cross-correlation using a polarization-only lensing reconstruction that is expected to be robust against biases induced by extragalactic foregrounds. This is a significant improvement over the first such measurement with POLARBEAR and HSC \cite{Namikawa2019} and highlights the power of low-noise CMB polarization observations. In future work, we hope to implement foreground marginalization for cross-correlations, based on the results of \cite[][O26]{Baleato-Lizancos2025}, on top of the \gmv\ lensing estimator which will allow full use of the available lensing reconstruction SNR. This work also emphasizes the use of multicomponent simulations and data consistency checks to understand the impact of foregrounds in cross-correlation measurements. This will be especially important for future measurements with lower statistical noise.

Future measurements of this cross-correlation will benefit from significantly deeper cosmic shear surveys such as LSST \cite{Ivezic2019} and \euclid\ \cite{Mellier2025} as well as future CMB data releases from SPT-3G \cite{Prabhu2024} that aim to cover 25\% of the sky, and the Simons Observatory \cite{Abitbol2025} which aims to cover $\sim$60\% of the sky. Looking forward to future analyses using upcoming data, more careful work will need to be done to ensure robust modeling on small scales. In particular, more advanced IA modeling (potentially some combination of perturbative \cite[e.g.,][]{Chen2024} and non-perturbative \cite[e.g.,][]{Fortuna2021} techniques) and more flexible baryonic feedback models \cite[e.g.,][]{Arico2021} will be necessary to robustly fit small scales. More constraining data will allow these more flexible models to provide new insights into cosmology and the astrophysical processes that affect matter clustering on small scales.
This work also sets the stage for an updated 6$\times$2-point analysis combining data from the full six years of DES \cite{Bechtol2026} and SPT-3G, which will provide powerful consistency tests and probe the growth of structure across a wide range of cosmic time \cite{Abbott2023}.

\section*{Acknowledgments}
The South Pole Telescope program is supported by the National Science Foundation (NSF) through awards OPP-1852617 and OPP-2332483. Partial support is also provided by the Kavli Institute of Cosmological Physics at the University of Chicago.
YO and CC were supported by NSF grant AST-2306166 and AST-2508321.
Argonne National Laboratory’s work was supported by the U.S. Department of Energy, Office of High Energy Physics, under contract DE-AC02-06CH11357.
The UC Davis group acknowledges support from Michael and Ester Vaida.
Work at the Fermi National Accelerator Laboratory (Fermilab), a U.S. Department of Energy, Office of Science, Office of High Energy Physics HEP User Facility, is managed by Fermi Forward Discovery Group, LLC, acting under Contract No. 89243024CSC000002.
The Melbourne authors acknowledge support from the Australian Research Council’s Discovery Project scheme (No. DP260100705).
The Paris group has received funding from the European Research Council (ERC) under the European Union’s Horizon 2020 research and innovation program (grant agreement No 101001897), and funding from the Centre National d’Etudes Spatiales.
The SLAC group is supported in part by the Department of Energy at SLAC National Accelerator Laboratory, under contract DE-AC02-76SF00515.
This work was partially supported by the Center for AstroPhysical Surveys (CAPS) at the National Center for Supercomputing Applications (NCSA), University of Illinois Urbana-Champaign. This work made use of the following computing resources: the Illinois Campus Cluster, a computing resource that is operated by the Illinois Campus Cluster Program (ICCP) in conjunction with the National Center for Supercomputing Applications (NCSA) and which is supported by funds from the University of Illinois Urbana-Champaign; and Crossover, a high-performance computing cluster operated by the Laboratory Computing Resource Center at Argonne National Laboratory.
This research has made use of the Astrophysics Data System, funded by NASA under Cooperative Agreement 80NSSC21M0056.
This work relied on \textsc{NumPy} \cite{Harris2020} and \textsc{SciPy} \cite{Virtanen2020} for numerical calculations, \textsc{Matplotlib} \cite{Hunter2007} for plotting, and \textsc{GetDist} \cite{Lewis2019} for plotting and analyzing posterior distributions.

\appendix

\section{Shear $B$-modes}\label{app:bmodes}
\begin{figure*}
    \centering
    \includegraphics[width=\linewidth]{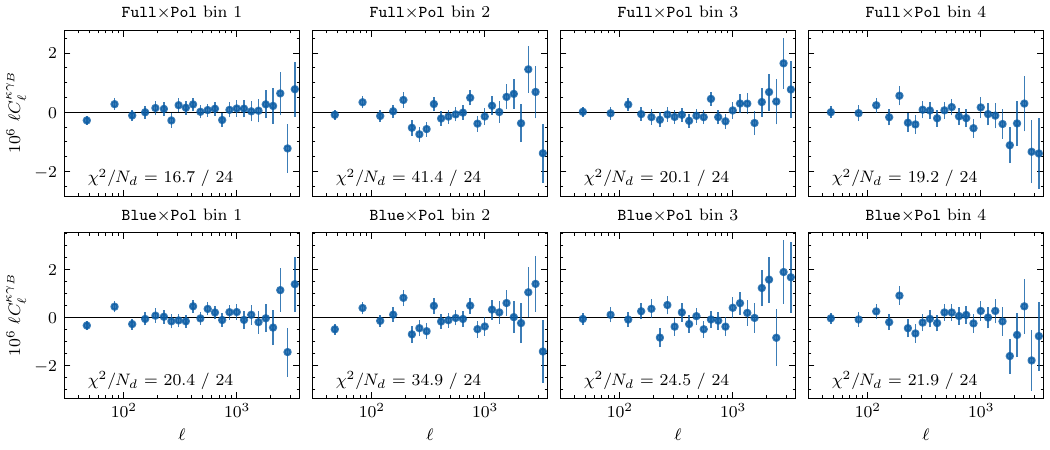}
    \caption{Top row: $B$-mode $\kappa\gamma$ cross-correlations for the \desfull\ shear sample and the \pol\ lensing reconstruction. Bottom row: same as top row, but for the \desblue\ shear subsample. We list the $\chi^2$ values for each data vector relative to zero.}
    \label{fig:bmodes}
\end{figure*}

In \cref{fig:bmodes} we show the \desfull$\times$\pol\ and \desblue$\times$\pol\ $B$-mode cross-correlations and list the $\chi^2$ values for each data vector relative to zero. We find no significant evidence for non-zero $B$-modes. Out of all of the data vectors (including the \gmv, \prof, \mh, and \xilc\ lensing reconstructions), the worst $\chi^2$ value is 41.4 with 24 degrees of freedom for bin 2 of the \desfull$\times$\pol\ data vector. While this is somewhat high, it is within the expected distribution when considering four independent shear bins. A Kolmogorov-Smirnov test on the $\chi^2$ values of the four bins of the \desfull$\times$\pol\ data vector with respect to a $\chi^2$ distribution yields a $p$-value of 0.32. Additionally, the $\chi^2$ of the full data vector, taking into account the covariance between bins, is 97.3 ($N_d = 96$) resulting in a PTE of 0.44. We find similar results for all of the other data vectors, indicating that the cross-correlation $B$-modes are consistent with zero.

\section{Full posteriors}\label{app:full_posts}
\begin{figure*}
    \centering
    \includegraphics[width=\linewidth]{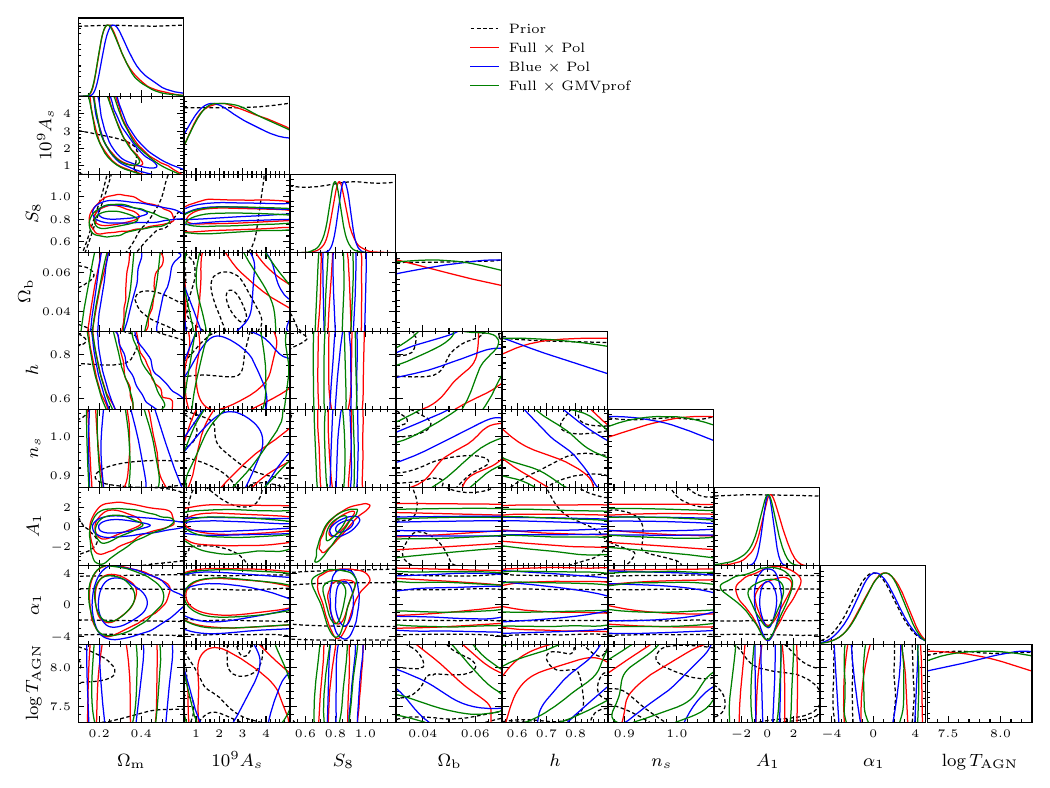}
    \caption{Full posteriors for the \desfull$\times$\pol, \desblue$\times$\pol, and \desfull$\times$\prof\ $\kappa\gamma$ cross-correlations. The black dashed contours show the result of sampling the prior as specified in \cref{tab:priors}. The photo-$z$ nuisance parameters $\Delta z_i$ (not shown) are fully prior-dominated.}
    \label{fig:full_post}
\end{figure*}

For completeness, we show the full posteriors for the \desfull$\times$\pol, \desfull$\times$\prof, and \desblue$\times$\pol\ data combinations in \cref{fig:full_post}.

\bibliographystyle{prd}
\bibliography{references}

@string{june = {June}}

@article{Abbott2018,
 adsurl = {https://ui.adsabs.harvard.edu/abs/2018ApJS..239...18A},
 archiveprefix = {arXiv},
 author = {{Abbott}, T.~M.~C. and {Abdalla}, F.~B. and {Allam}, S. and {Amara}, A. and {Annis}, J. and {Asorey}, J. and {Avila}, S. and {Ballester}, O. and {Banerji}, M. and {Barkhouse}, W. and {Baruah}, L. and {Baumer}, M. and {Bechtol}, K. and {Becker}, M.~R. and {Benoit-L{\'e}vy}, A. and {Bernstein}, G.~M. and {Bertin}, E. and {Blazek}, J. and {Bocquet}, S. and {Brooks}, D. and {Brout}, D. and {Buckley-Geer}, E. and {Burke}, D.~L. and {Busti}, V. and {Campisano}, R. and {Cardiel-Sas}, L. and {Carnero Rosell}, A. and {Carrasco Kind}, M. and {Carretero}, J. and {Castander}, F.~J. and {Cawthon}, R. and {Chang}, C. and {Chen}, X. and {Conselice}, C. and {Costa}, G. and {Crocce}, M. and {Cunha}, C.~E. and {D'Andrea}, C.~B. and {da Costa}, L.~N. and {Das}, R. and {Daues}, G. and {Davis}, T.~M. and {Davis}, C. and {De Vicente}, J. and {DePoy}, D.~L. and {DeRose}, J. and {Desai}, S. and {Diehl}, H.~T. and {Dietrich}, J.~P. and {Dodelson}, S. and {Doel}, P. and {Drlica-Wagner}, A. and {Eifler}, T.~F. and {Elliott}, A.~E. and {Evrard}, A.~E. and {Farahi}, A. and {Fausti Neto}, A. and {Fernandez}, E. and {Finley}, D.~A. and {Flaugher}, B. and {Foley}, R.~J. and {Fosalba}, P. and {Friedel}, D.~N. and {Frieman}, J. and {Garc{\'\i}a-Bellido}, J. and {Gaztanaga}, E. and {Gerdes}, D.~W. and {Giannantonio}, T. and {Gill}, M.~S.~S. and {Glazebrook}, K. and {Goldstein}, D.~A. and {Gower}, M. and {Gruen}, D. and {Gruendl}, R.~A. and {Gschwend}, J. and {Gupta}, R.~R. and {Gutierrez}, G. and {Hamilton}, S. and {Hartley}, W.~G. and {Hinton}, S.~R. and {Hislop}, J.~M. and {Hollowood}, D. and {Honscheid}, K. and {Hoyle}, B. and {Huterer}, D. and {Jain}, B. and {James}, D.~J. and {Jeltema}, T. and {Johnson}, M.~W.~G. and {Johnson}, M.~D. and {Kacprzak}, T. and {Kent}, S. and {Khullar}, G. and {Klein}, M. and {Kovacs}, A. and {Koziol}, A.~M.~G. and {Krause}, E. and {Kremin}, A. and {Kron}, R. and {Kuehn}, K. and {Kuhlmann}, S. and {Kuropatkin}, N. and {Lahav}, O. and {Lasker}, J. and {Li}, T.~S. and {Li}, R.~T. and {Liddle}, A.~R. and {Lima}, M. and {Lin}, H. and {L{\'o}pez-Reyes}, P. and {MacCrann}, N. and {Maia}, M.~A.~G. and {Maloney}, J.~D. and {Manera}, M. and {March}, M. and {Marriner}, J. and {Marshall}, J.~L. and {Martini}, P. and {McClintock}, T. and {McKay}, T. and {McMahon}, R.~G. and {Melchior}, P. and {Menanteau}, F. and {Miller}, C.~J. and {Miquel}, R. and {Mohr}, J.~J. and {Morganson}, E. and {Mould}, J. and {Neilsen}, E. and {Nichol}, R.~C. and {Nogueira}, F. and {Nord}, B. and {Nugent}, P. and {Nunes}, L. and {Ogando}, R.~L.~C. and {Old}, L. and {Pace}, A.~B. and {Palmese}, A. and {Paz-Chinch{\'o}n}, F. and {Peiris}, H.~V. and {Percival}, W.~J. and {Petravick}, D. and {Plazas}, A.~A. and {Poh}, J. and {Pond}, C. and {Porredon}, A. and {Pujol}, A. and {Refregier}, A. and {Reil}, K. and {Ricker}, P.~M. and {Rollins}, R.~P. and {Romer}, A.~K. and {Roodman}, A. and {Rooney}, P. and {Ross}, A.~J. and {Rykoff}, E.~S. and {Sako}, M. and {Sanchez}, M.~L. and {Sanchez}, E. and {Santiago}, B. and {Saro}, A. and {Scarpine}, V. and {Scolnic}, D. and {Serrano}, S. and {Sevilla-Noarbe}, I. and {Sheldon}, E. and {Shipp}, N. and {Silveira}, M.~L. and {Smith}, M. and {Smith}, R.~C. and {Smith}, J.~A. and {Soares-Santos}, M. and {Sobreira}, F. and {Song}, J. and {Stebbins}, A. and {Suchyta}, E. and {Sullivan}, M. and {Swanson}, M.~E.~C. and {Tarle}, G. and {Thaler}, J. and {Thomas}, D. and {Thomas}, R.~C. and {Troxel}, M.~A. and {Tucker}, D.~L. and {Vikram}, V. and {Vivas}, A.~K. and {Walker}, A.~R. and {Wechsler}, R.~H. and {Weller}, J. and {Wester}, W. and {Wolf}, R.~C. and {Wu}, H. and {Yanny}, B. and {Zenteno}, A. and {Zhang}, Y. and {Zuntz}, J. and {DES Collaboration} and {Juneau}, S. and {Fitzpatrick}, M. and {Nikutta}, R.},
 doi = {10.3847/1538-4365/aae9f0},
 eid = {18},
 eprint = {1801.03181},
 journal = {\apjs},
 month = {December},
 number = {2},
 pages = {18},
 primaryclass = {astro-ph.IM},
 title = {{The Dark Energy Survey: Data Release 1}},
 volume = {239},
 year = {2018}
}

@article{Abbott2018b,
 adsurl = {https://ui.adsabs.harvard.edu/abs/2018PhRvD..98d3526A},
 archiveprefix = {arXiv},
 author = {{Abbott}, T.~M.~C. and {Abdalla}, F.~B. and {Alarcon}, A. and {Aleksi{\'c}}, J. and {Allam}, S. and {Allen}, S. and {Amara}, A. and {Annis}, J. and {Asorey}, J. and {Avila}, S. and {Bacon}, D. and {Balbinot}, E. and {Banerji}, M. and {Banik}, N. and {Barkhouse}, W. and {Baumer}, M. and {Baxter}, E. and {Bechtol}, K. and {Becker}, M.~R. and {Benoit-L{\'e}vy}, A. and {Benson}, B.~A. and {Bernstein}, G.~M. and {Bertin}, E. and {Blazek}, J. and {Bridle}, S.~L. and {Brooks}, D. and {Brout}, D. and {Buckley-Geer}, E. and {Burke}, D.~L. and {Busha}, M.~T. and {Campos}, A. and {Capozzi}, D. and {Carnero Rosell}, A. and {Carrasco Kind}, M. and {Carretero}, J. and {Castander}, F.~J. and {Cawthon}, R. and {Chang}, C. and {Chen}, N. and {Childress}, M. and {Choi}, A. and {Conselice}, C. and {Crittenden}, R. and {Crocce}, M. and {Cunha}, C.~E. and {D'Andrea}, C.~B. and {da Costa}, L.~N. and {Das}, R. and {Davis}, T.~M. and {Davis}, C. and {De Vicente}, J. and {DePoy}, D.~L. and {DeRose}, J. and {Desai}, S. and {Diehl}, H.~T. and {Dietrich}, J.~P. and {Dodelson}, S. and {Doel}, P. and {Drlica-Wagner}, A. and {Eifler}, T.~F. and {Elliott}, A.~E. and {Elsner}, F. and {Elvin-Poole}, J. and {Estrada}, J. and {Evrard}, A.~E. and {Fang}, Y. and {Fernandez}, E. and {Fert{\'e}}, A. and {Finley}, D.~A. and {Flaugher}, B. and {Fosalba}, P. and {Friedrich}, O. and {Frieman}, J. and {Garc{\'\i}a-Bellido}, J. and {Garcia-Fernandez}, M. and {Gatti}, M. and {Gaztanaga}, E. and {Gerdes}, D.~W. and {Giannantonio}, T. and {Gill}, M.~S.~S. and {Glazebrook}, K. and {Goldstein}, D.~A. and {Gruen}, D. and {Gruendl}, R.~A. and {Gschwend}, J. and {Gutierrez}, G. and {Hamilton}, S. and {Hartley}, W.~G. and {Hinton}, S.~R. and {Honscheid}, K. and {Hoyle}, B. and {Huterer}, D. and {Jain}, B. and {James}, D.~J. and {Jarvis}, M. and {Jeltema}, T. and {Johnson}, M.~D. and {Johnson}, M.~W.~G. and {Kacprzak}, T. and {Kent}, S. and {Kim}, A.~G. and {King}, A. and {Kirk}, D. and {Kokron}, N. and {Kovacs}, A. and {Krause}, E. and {Krawiec}, C. and {Kremin}, A. and {Kuehn}, K. and {Kuhlmann}, S. and {Kuropatkin}, N. and {Lacasa}, F. and {Lahav}, O. and {Li}, T.~S. and {Liddle}, A.~R. and {Lidman}, C. and {Lima}, M. and {Lin}, H. and {MacCrann}, N. and {Maia}, M.~A.~G. and {Makler}, M. and {Manera}, M. and {March}, M. and {Marshall}, J.~L. and {Martini}, P. and {McMahon}, R.~G. and {Melchior}, P. and {Menanteau}, F. and {Miquel}, R. and {Miranda}, V. and {Mudd}, D. and {Muir}, J. and {M{\"o}ller}, A. and {Neilsen}, E. and {Nichol}, R.~C. and {Nord}, B. and {Nugent}, P. and {Ogando}, R.~L.~C. and {Palmese}, A. and {Peacock}, J. and {Peiris}, H.~V. and {Peoples}, J. and {Percival}, W.~J. and {Petravick}, D. and {Plazas}, A.~A. and {Porredon}, A. and {Prat}, J. and {Pujol}, A. and {Rau}, M.~M. and {Refregier}, A. and {Ricker}, P.~M. and {Roe}, N. and {Rollins}, R.~P. and {Romer}, A.~K. and {Roodman}, A. and {Rosenfeld}, R. and {Ross}, A.~J. and {Rozo}, E. and {Rykoff}, E.~S. and {Sako}, M. and {Salvador}, A.~I. and {Samuroff}, S. and {S{\'a}nchez}, C. and {Sanchez}, E. and {Santiago}, B. and {Scarpine}, V. and {Schindler}, R. and {Scolnic}, D. and {Secco}, L.~F. and {Serrano}, S. and {Sevilla-Noarbe}, I. and {Sheldon}, E. and {Smith}, R.~C. and {Smith}, M. and {Smith}, J. and {Soares-Santos}, M. and {Sobreira}, F. and {Suchyta}, E. and {Tarle}, G. and {Thomas}, D. and {Troxel}, M.~A. and {Tucker}, D.~L. and {Tucker}, B.~E. and {Uddin}, S.~A. and {Varga}, T.~N. and {Vielzeuf}, P. and {Vikram}, V. and {Vivas}, A.~K. and {Walker}, A.~R. and {Wang}, M. and {Wechsler}, R.~H. and {Weller}, J. and {Wester}, W. and {Wolf}, R.~C. and {Yanny}, B. and {Yuan}, F. and {Zenteno}, A. and {Zhang}, B. and {Zhang}, Y. and {Zuntz}, J.},
 doi = {10.1103/PhysRevD.98.043526},
 eid = {043526},
 eprint = {1708.01530},
 journal = {\prd},
 month = {August},
 number = {4},
 pages = {043526},
 primaryclass = {astro-ph.CO},
 title = {{Dark Energy Survey year 1 results: Cosmological constraints from galaxy clustering and weak lensing}},
 volume = {98},
 year = {2018}
}

@article{Abbott2021,
 adsurl = {https://ui.adsabs.harvard.edu/abs/2021ApJS..255...20A},
 archiveprefix = {arXiv},
 author = {{Abbott}, T.~M.~C. and {Adam{\'o}w}, M. and {Aguena}, M. and {Allam}, S. and {Amon}, A. and {Annis}, J. and {Avila}, S. and {Bacon}, D. and {Banerji}, M. and {Bechtol}, K. and {Becker}, M.~R. and {Bernstein}, G.~M. and {Bertin}, E. and {Bhargava}, S. and {Bridle}, S.~L. and {Brooks}, D. and {Burke}, D.~L. and {Carnero Rosell}, A. and {Carrasco Kind}, M. and {Carretero}, J. and {Castander}, F.~J. and {Cawthon}, R. and {Chang}, C. and {Choi}, A. and {Conselice}, C. and {Costanzi}, M. and {Crocce}, M. and {da Costa}, L.~N. and {Davis}, T.~M. and {De Vicente}, J. and {DeRose}, J. and {Desai}, S. and {Diehl}, H.~T. and {Dietrich}, J.~P. and {Drlica-Wagner}, A. and {Eckert}, K. and {Elvin-Poole}, J. and {Everett}, S. and {Evrard}, A.~E. and {Ferrero}, I. and {Fert{\'e}}, A. and {Flaugher}, B. and {Fosalba}, P. and {Friedel}, D. and {Frieman}, J. and {Garc{\'\i}a-Bellido}, J. and {Gaztanaga}, E. and {Gelman}, L. and {Gerdes}, D.~W. and {Giannantonio}, T. and {Gill}, M.~S.~S. and {Gruen}, D. and {Gruendl}, R.~A. and {Gschwend}, J. and {Gutierrez}, G. and {Hartley}, W.~G. and {Hinton}, S.~R. and {Hollowood}, D.~L. and {Honscheid}, K. and {Huterer}, D. and {James}, D.~J. and {Jeltema}, T. and {Johnson}, M.~D. and {Kent}, S. and {Kron}, R. and {Kuehn}, K. and {Kuropatkin}, N. and {Lahav}, O. and {Li}, T.~S. and {Lidman}, C. and {Lin}, H. and {MacCrann}, N. and {Maia}, M.~A.~G. and {Manning}, T.~A. and {Maloney}, J.~D. and {March}, M. and {Marshall}, J.~L. and {Martini}, P. and {Melchior}, P. and {Menanteau}, F. and {Miquel}, R. and {Morgan}, R. and {Myles}, J. and {Neilsen}, E. and {Ogando}, R.~L.~C. and {Palmese}, A. and {Paz-Chinch{\'o}n}, F. and {Petravick}, D. and {Pieres}, A. and {Plazas}, A.~A. and {Pond}, C. and {Rodriguez-Monroy}, M. and {Romer}, A.~K. and {Roodman}, A. and {Rykoff}, E.~S. and {Sako}, M. and {Sanchez}, E. and {Santiago}, B. and {Scarpine}, V. and {Serrano}, S. and {Sevilla-Noarbe}, I. and {Smith}, J. Allyn and {Smith}, M. and {Soares-Santos}, M. and {Suchyta}, E. and {Swanson}, M.~E.~C. and {Tarle}, G. and {Thomas}, D. and {To}, C. and {Tremblay}, P.~E. and {Troxel}, M.~A. and {Tucker}, D.~L. and {Turner}, D.~J. and {Varga}, T.~N. and {Walker}, A.~R. and {Wechsler}, R.~H. and {Weller}, J. and {Wester}, W. and {Wilkinson}, R.~D. and {Yanny}, B. and {Zhang}, Y. and {Nikutta}, R. and {Fitzpatrick}, M. and {Jacques}, A. and {Scott}, A. and {Olsen}, K. and {Huang}, L. and {Herrera}, D. and {Juneau}, S. and {Nidever}, D. and {Weaver}, B.~A. and {Adean}, C. and {Correia}, V. and {de Freitas}, M. and {Freitas}, F.~N. and {Singulani}, C. and {Vila-Verde}, G. and {Linea Science Server}},
 doi = {10.3847/1538-4365/ac00b3},
 eid = {20},
 eprint = {2101.05765},
 journal = {\apjs},
 month = {August},
 number = {2},
 pages = {20},
 primaryclass = {astro-ph.IM},
 title = {{The Dark Energy Survey Data Release 2}},
 volume = {255},
 year = {2021}
}

@article{Abbott2022,
 adsurl = {https://ui.adsabs.harvard.edu/abs/2022PhRvD.105b3520A},
 archiveprefix = {arXiv},
 author = {{Abbott}, T.~M.~C. and {Aguena}, M. and {Alarcon}, A. and {Allam}, S. and {Alves}, O. and {Amon}, A. and {Andrade-Oliveira}, F. and {Annis}, J. and {Avila}, S. and {Bacon}, D. and {Baxter}, E. and {Bechtol}, K. and {Becker}, M.~R. and {Bernstein}, G.~M. and {Bhargava}, S. and {Birrer}, S. and {Blazek}, J. and {Brandao-Souza}, A. and {Bridle}, S.~L. and {Brooks}, D. and {Buckley-Geer}, E. and {Burke}, D.~L. and {Camacho}, H. and {Campos}, A. and {Carnero Rosell}, A. and {Carrasco Kind}, M. and {Carretero}, J. and {Castander}, F.~J. and {Cawthon}, R. and {Chang}, C. and {Chen}, A. and {Chen}, R. and {Choi}, A. and {Conselice}, C. and {Cordero}, J. and {Costanzi}, M. and {Crocce}, M. and {da Costa}, L.~N. and {da Silva Pereira}, M.~E. and {Davis}, C. and {Davis}, T.~M. and {De Vicente}, J. and {DeRose}, J. and {Desai}, S. and {Di Valentino}, E. and {Diehl}, H.~T. and {Dietrich}, J.~P. and {Dodelson}, S. and {Doel}, P. and {Doux}, C. and {Drlica-Wagner}, A. and {Eckert}, K. and {Eifler}, T.~F. and {Elsner}, F. and {Elvin-Poole}, J. and {Everett}, S. and {Evrard}, A.~E. and {Fang}, X. and {Farahi}, A. and {Fernandez}, E. and {Ferrero}, I. and {Fert{\'e}}, A. and {Fosalba}, P. and {Friedrich}, O. and {Frieman}, J. and {Garc{\'\i}a-Bellido}, J. and {Gatti}, M. and {Gaztanaga}, E. and {Gerdes}, D.~W. and {Giannantonio}, T. and {Giannini}, G. and {Gruen}, D. and {Gruendl}, R.~A. and {Gschwend}, J. and {Gutierrez}, G. and {Harrison}, I. and {Hartley}, W.~G. and {Herner}, K. and {Hinton}, S.~R. and {Hollowood}, D.~L. and {Honscheid}, K. and {Hoyle}, B. and {Huff}, E.~M. and {Huterer}, D. and {Jain}, B. and {James}, D.~J. and {Jarvis}, M. and {Jeffrey}, N. and {Jeltema}, T. and {Kovacs}, A. and {Krause}, E. and {Kron}, R. and {Kuehn}, K. and {Kuropatkin}, N. and {Lahav}, O. and {Leget}, P. -F. and {Lemos}, P. and {Liddle}, A.~R. and {Lidman}, C. and {Lima}, M. and {Lin}, H. and {MacCrann}, N. and {Maia}, M.~A.~G. and {Marshall}, J.~L. and {Martini}, P. and {McCullough}, J. and {Melchior}, P. and {Mena-Fern{\'a}ndez}, J. and {Menanteau}, F. and {Miquel}, R. and {Mohr}, J.~J. and {Morgan}, R. and {Muir}, J. and {Myles}, J. and {Nadathur}, S. and {Navarro-Alsina}, A. and {Nichol}, R.~C. and {Ogando}, R.~L.~C. and {Omori}, Y. and {Palmese}, A. and {Pandey}, S. and {Park}, Y. and {Paz-Chinch{\'o}n}, F. and {Petravick}, D. and {Pieres}, A. and {Plazas Malag{\'o}n}, A.~A. and {Porredon}, A. and {Prat}, J. and {Raveri}, M. and {Rodriguez-Monroy}, M. and {Rollins}, R.~P. and {Romer}, A.~K. and {Roodman}, A. and {Rosenfeld}, R. and {Ross}, A.~J. and {Rykoff}, E.~S. and {Samuroff}, S. and {S{\'a}nchez}, C. and {Sanchez}, E. and {Sanchez}, J. and {Sanchez Cid}, D. and {Scarpine}, V. and {Schubnell}, M. and {Scolnic}, D. and {Secco}, L.~F. and {Serrano}, S. and {Sevilla-Noarbe}, I. and {Sheldon}, E. and {Shin}, T. and {Smith}, M. and {Soares-Santos}, M. and {Suchyta}, E. and {Swanson}, M.~E.~C. and {Tabbutt}, M. and {Tarle}, G. and {Thomas}, D. and {To}, C. and {Troja}, A. and {Troxel}, M.~A. and {Tucker}, D.~L. and {Tutusaus}, I. and {Varga}, T.~N. and {Walker}, A.~R. and {Weaverdyck}, N. and {Wechsler}, R. and {Weller}, J. and {Yanny}, B. and {Yin}, B. and {Zhang}, Y. and {Zuntz}, J. and {DES Collaboration}},
 doi = {10.1103/PhysRevD.105.023520},
 eid = {023520},
 eprint = {2105.13549},
 journal = {\prd},
 month = {January},
 number = {2},
 pages = {023520},
 primaryclass = {astro-ph.CO},
 title = {{Dark Energy Survey Year 3 results: Cosmological constraints from galaxy clustering and weak lensing}},
 volume = {105},
 year = {2022}
}

@article{Abbott2023,
 adsurl = {https://ui.adsabs.harvard.edu/abs/2023PhRvD.107b3531A},
 archiveprefix = {arXiv},
 author = {{Abbott}, T.~M.~C. and {Aguena}, M. and {Alarcon}, A. and {Alves}, O. and {Amon}, A. and {Andrade-Oliveira}, F. and {Annis}, J. and {Ansarinejad}, B. and {Avila}, S. and {Bacon}, D. and {Baxter}, E.~J. and {Bechtol}, K. and {Becker}, M.~R. and {Benson}, B.~A. and {Bernstein}, G.~M. and {Bertin}, E. and {Blazek}, J. and {Bleem}, L.~E. and {Bocquet}, S. and {Brooks}, D. and {Buckley-Geer}, E. and {Burke}, D.~L. and {Camacho}, H. and {Campos}, A. and {Carlstrom}, J.~E. and {Carnero Rosell}, A. and {Carrasco Kind}, M. and {Carretero}, J. and {Cawthon}, R. and {Chang}, C. and {Chang}, C.~L. and {Chen}, R. and {Choi}, A. and {Chown}, R. and {Conselice}, C. and {Cordero}, J. and {Costanzi}, M. and {Crawford}, T. and {Crites}, A.~T. and {Crocce}, M. and {da Costa}, L.~N. and {Davis}, C. and {Davis}, T.~M. and {de Haan}, T. and {De Vicente}, J. and {DeRose}, J. and {Desai}, S. and {Diehl}, H.~T. and {Dobbs}, M.~A. and {Dodelson}, S. and {Doel}, P. and {Doux}, C. and {Drlica-Wagner}, A. and {Eckert}, K. and {Eifler}, T.~F. and {Elsner}, F. and {Elvin-Poole}, J. and {Everett}, S. and {Everett}, W. and {Fang}, X. and {Ferrero}, I. and {Fert{\'e}}, A. and {Flaugher}, B. and {Fosalba}, P. and {Friedrich}, O. and {Frieman}, J. and {Garc{\'\i}a-Bellido}, J. and {Gatti}, M. and {George}, E.~M. and {Giannantonio}, T. and {Giannini}, G. and {Gruen}, D. and {Gruendl}, R.~A. and {Gschwend}, J. and {Gutierrez}, G. and {Halverson}, N.~W. and {Harrison}, I. and {Herner}, K. and {Hinton}, S.~R. and {Holder}, G.~P. and {Hollowood}, D.~L. and {Holzapfel}, W.~L. and {Honscheid}, K. and {Hrubes}, J.~D. and {Huang}, H. and {Huff}, E.~M. and {Huterer}, D. and {Jain}, B. and {James}, D.~J. and {Jarvis}, M. and {Jeltema}, T. and {Kent}, S. and {Knox}, L. and {Kovacs}, A. and {Krause}, E. and {Kuehn}, K. and {Kuropatkin}, N. and {Lahav}, O. and {Lee}, A.~T. and {Leget}, P. -F. and {Lemos}, P. and {Liddle}, A.~R. and {Lidman}, C. and {Luong-Van}, D. and {McMahon}, J.~J. and {MacCrann}, N. and {March}, M. and {Marshall}, J.~L. and {Martini}, P. and {McCullough}, J. and {Melchior}, P. and {Menanteau}, F. and {Meyer}, S.~S. and {Miquel}, R. and {Mocanu}, L. and {Mohr}, J.~J. and {Morgan}, R. and {Muir}, J. and {Myles}, J. and {Natoli}, T. and {Navarro-Alsina}, A. and {Nichol}, R.~C. and {Omori}, Y. and {Padin}, S. and {Pandey}, S. and {Park}, Y. and {Paz-Chinch{\'o}n}, F. and {Pereira}, M.~E.~S. and {Pieres}, A. and {Plazas Malag{\'o}n}, A.~A. and {Porredon}, A. and {Prat}, J. and {Pryke}, C. and {Raveri}, M. and {Reichardt}, C.~L. and {Rollins}, R.~P. and {Romer}, A.~K. and {Roodman}, A. and {Rosenfeld}, R. and {Ross}, A.~J. and {Ruhl}, J.~E. and {Rykoff}, E.~S. and {S{\'a}nchez}, C. and {Sanchez}, E. and {Sanchez}, J. and {Schaffer}, K.~K. and {Secco}, L.~F. and {Sevilla-Noarbe}, I. and {Sheldon}, E. and {Shin}, T. and {Shirokoff}, E. and {Smith}, M. and {Staniszewski}, Z. and {Stark}, A.~A. and {Suchyta}, E. and {Swanson}, M.~E.~C. and {Tarle}, G. and {To}, C. and {Troxel}, M.~A. and {Tutusaus}, I. and {Varga}, T.~N. and {Vieira}, J.~D. and {Weaverdyck}, N. and {Wechsler}, R.~H. and {Weller}, J. and {Williamson}, R. and {Wu}, W.~L.~K. and {Yanny}, B. and {Yin}, B. and {Zhang}, Y. and {Zuntz}, J. and {DES} and {SPT Collaborations}},
 doi = {10.1103/PhysRevD.107.023531},
 eid = {023531},
 eprint = {2206.10824},
 journal = {\prd},
 month = {January},
 number = {2},
 pages = {023531},
 primaryclass = {astro-ph.CO},
 title = {{Joint analysis of Dark Energy Survey Year 3 data and CMB lensing from SPT and Planck. III. Combined cosmological constraints}},
 volume = {107},
 year = {2023}
}

@article{Abitbol2025,
 adsurl = {https://ui.adsabs.harvard.edu/abs/2025JCAP...08..034A},
 archiveprefix = {arXiv},
 author = {{Abitbol}, M. and {Abril-Cabezas}, I. and {Adachi}, S. and {Ade}, P. and {Adler}, A.~E. and {Agrawal}, P. and {Aguirre}, J. and {Ahmed}, Z. and {Aiola}, S. and {Alford}, T. and {Ali}, A. and {Alonso}, D. and {Alvarez}, M.~A. and {An}, R. and {Arnold}, K. and {Ashton}, P. and {Atkins}, Z. and {Austermann}, J. and {Azzoni}, S. and {Baccigalupi}, C. and {Baleato Lizancos}, A. and {Barron}, D. and {Barry}, P. and {Bartlett}, J. and {Battaglia}, N. and {Battye}, R. and {Baxter}, E. and {Bazarko}, A. and {Beall}, J.~A. and {Bean}, R. and {Beck}, D. and {Beckman}, S. and {Begin}, J. and {Beheshti}, A. and {Beringue}, B. and {Bhandarkar}, T. and {Bhimani}, S. and {Bianchini}, F. and {Biermann}, E. and {Biquard}, S. and {Bixler}, B. and {Boada}, S. and {Boettger}, D. and {Bolliet}, B. and {Bond}, J.~R. and {Borrill}, J. and {Borrow}, J. and {Braithwaite}, C. and {Brien}, T.~L.~R. and {Brown}, M.~L. and {Bruno}, S.~M. and {Bryan}, S. and {Bustos}, R. and {Cai}, H. and {Calabrese}, E. and {Calafut}, V. and {Carl}, F.~M. and {Carones}, A. and {Carron}, J. and {Challinor}, A. and {Chanial}, P. and {Chen}, N. and {Cheung}, K. and {Chiang}, B. and {Chinone}, Y. and {Chluba}, J. and {Cho}, H.~S. and {Choi}, S.~K. and {Chu}, M. and {Clancy}, J. and {Clark}, S.~E. and {Clarke}, P. and {Cleary}, J. and {Clements}, D.~L. and {Connors}, J. and {Contaldi}, C. and {Coppi}, G. and {Corbett}, L. and {Cothard}, N.~F. and {Coulton}, W. and {Crowley}, K.~D. and {Crowley}, K.~T. and {Cukierman}, A. and {D'Ewart}, J.~M. and {Dachlythra}, K. and {Datta}, R. and {Day-Weiss}, S. and {de Haan}, T. and {Devlin}, M. and {Di Mascolo}, L. and {Dicker}, S. and {Dober}, B. and {Doux}, C. and {Dow}, P. and {Doyle}, S. and {Duell}, C.~J. and {Duff}, S.~M. and {Duivenvoorden}, A.~J. and {Dunkley}, J. and {Dutcher}, D. and {D{\"u}nner}, R. and {Edenton}, M. and {El Bouhargani}, H. and {Errard}, J. and {Fabbian}, G. and {Fanfani}, V. and {Farren}, G.~S. and {Fergusson}, J. and {Ferraro}, S. and {Flauger}, R. and {Foster}, A. and {Freese}, K. and {Frisch}, J.~C. and {Frolov}, A. and {Fuller}, G. and {Galitzki}, N. and {Gallardo}, P.~A. and {Galvez Ghersi}, J.~T. and {Ganga}, K. and {Gao}, J. and {Garrido}, X. and {Gawiser}, E. and {Gerbino}, M. and {Gerras}, R. and {Giardiello}, S. and {Gill}, A. and {Gilles}, V. and {Giri}, U. and {Gleave}, E. and {Gluscevic}, V. and {Goeckner-Wald}, N. and {Golec}, J.~E. and {Gordon}, S. and {Gralla}, M. and {Gratton}, S. and {Green}, D. and {Groh}, J.~C. and {Groppi}, C. and {Guan}, Y. and {Gupta}, N. and {Gudmundsson}, J.~E. and {Hagstotz}, S. and {Hargrave}, P. and {Haridas}, S. and {Harrington}, K. and {Harrison}, I. and {Hasegawa}, M. and {Hasselfield}, M. and {Haynes}, V. and {Hazumi}, M. and {He}, A. and {Healy}, E. and {Henderson}, S.~W. and {Hensley}, B.~S. and {Hertig}, E. and {Herv{\'\i}as-Caimapo}, C. and {Higuchi}, M. and {Hill}, C.~A. and {Hill}, J.~C. and {Hilton}, G. and {Hilton}, M. and {Hincks}, A.~D. and {Hinshaw}, G. and {Hlo{\v{z}}ek}, R. and {Ho}, A.~Y.~Q. and {Ho}, S. and {Ho}, S.~P. and {Hoang}, T.~D. and {Hoh}, J. and {Hornecker}, E. and {Hornsby}, A.~L. and {Hotinli}, S.~C. and {Huang}, Z. and {Huber}, Z.~B. and {Hubmayr}, J. and {Huffenberger}, K. and {Hughes}, J.~P. and {Idicherian Lonappan}, A. and {Ikape}, M. and {Irwin}, K. and {Iuliano}, J. and {Jaffe}, A.~H. and {Jain}, B. and {Jense}, H.~T. and {Jeong}, O. and {Johnson}, A. and {Johnson}, B.~R. and {Johnson}, M. and {Jones}, M. and {Jost}, B. and {Kaneko}, D. and {Karpel}, E.~D. and {Kasai}, Y. and {Katayama}, N. and {Keating}, B. and {Keller}, B. and {Keskitalo}, R. and {Kim}, J. and {Kisner}, T. and {Kiuchi}, K.},
 doi = {10.1088/1475-7516/2025/08/034},
 eid = {034},
 eprint = {2503.00636},
 journal = {\jcap},
 month = {August},
 number = {8},
 pages = {034},
 primaryclass = {astro-ph.IM},
 title = {{The Simons Observatory: science goals and forecasts for the enhanced Large Aperture Telescope}},
 volume = {2025},
 year = {2025}
}

@article{Abril-Cabezas2025,
 adsurl = {https://ui.adsabs.harvard.edu/abs/2025PhRvD.112b3522A},
 archiveprefix = {arXiv},
 author = {{Abril-Cabezas}, Irene and {Qu}, Frank J. and {Sherwin}, Blake D. and {van Engelen}, Alexander and {MacCrann}, Niall and {Herv{\'\i}as-Caimapo}, Carlos and {Darwish}, Omar and {Hill}, J. Colin and {Madhavacheril}, Mathew S. and {Sehgal}, Neelima},
 doi = {10.1103/jhmr-mg6w},
 eid = {023522},
 eprint = {2505.03737},
 journal = {\prd},
 month = {July},
 number = {2},
 pages = {023522},
 primaryclass = {astro-ph.CO},
 title = {{Impact of Galactic non-Gaussian foregrounds on CMB lensing measurements}},
 volume = {112},
 year = {2025}
}

@article{Alonso2019,
 adsurl = {https://ui.adsabs.harvard.edu/abs/2019MNRAS.484.4127A},
 archiveprefix = {arXiv},
 author = {{Alonso}, David and {Sanchez}, Javier and {Slosar}, An{\v{z}}e and {LSST Dark Energy Science Collaboration}},
 doi = {10.1093/mnras/stz093},
 eprint = {1809.09603},
 journal = {\mnras},
 month = {April},
 number = {3},
 pages = {4127-4151},
 primaryclass = {astro-ph.CO},
 title = {{A unified pseudo-C$_{{\ensuremath{\ell}}}$ framework}},
 volume = {484},
 year = {2019}
}

@article{Amon2022,
 adsurl = {https://ui.adsabs.harvard.edu/abs/2022MNRAS.516.5355A},
 archiveprefix = {arXiv},
 author = {{Amon}, Alexandra and {Efstathiou}, George},
 doi = {10.1093/mnras/stac2429},
 eprint = {2206.11794},
 journal = {\mnras},
 month = {November},
 number = {4},
 pages = {5355-5366},
 primaryclass = {astro-ph.CO},
 title = {{A non-linear solution to the S$_{8}$ tension?}},
 volume = {516},
 year = {2022}
}

@article{Amon2022b,
 adsurl = {https://ui.adsabs.harvard.edu/abs/2022PhRvD.105b3514A},
 archiveprefix = {arXiv},
 author = {{Amon}, A. and {Gruen}, D. and {Troxel}, M.~A. and {MacCrann}, N. and {Dodelson}, S. and {Choi}, A. and {Doux}, C. and {Secco}, L.~F. and {Samuroff}, S. and {Krause}, E. and {Cordero}, J. and {Myles}, J. and {DeRose}, J. and {Wechsler}, R.~H. and {Gatti}, M. and {Navarro-Alsina}, A. and {Bernstein}, G.~M. and {Jain}, B. and {Blazek}, J. and {Alarcon}, A. and {Fert{\'e}}, A. and {Lemos}, P. and {Raveri}, M. and {Campos}, A. and {Prat}, J. and {S{\'a}nchez}, C. and {Jarvis}, M. and {Alves}, O. and {Andrade-Oliveira}, F. and {Baxter}, E. and {Bechtol}, K. and {Becker}, M.~R. and {Bridle}, S.~L. and {Camacho}, H. and {Carnero Rosell}, A. and {Carrasco Kind}, M. and {Cawthon}, R. and {Chang}, C. and {Chen}, R. and {Chintalapati}, P. and {Crocce}, M. and {Davis}, C. and {Diehl}, H.~T. and {Drlica-Wagner}, A. and {Eckert}, K. and {Eifler}, T.~F. and {Elvin-Poole}, J. and {Everett}, S. and {Fang}, X. and {Fosalba}, P. and {Friedrich}, O. and {Gaztanaga}, E. and {Giannini}, G. and {Gruendl}, R.~A. and {Harrison}, I. and {Hartley}, W.~G. and {Herner}, K. and {Huang}, H. and {Huff}, E.~M. and {Huterer}, D. and {Kuropatkin}, N. and {Leget}, P. and {Liddle}, A.~R. and {McCullough}, J. and {Muir}, J. and {Pandey}, S. and {Park}, Y. and {Porredon}, A. and {Refregier}, A. and {Rollins}, R.~P. and {Roodman}, A. and {Rosenfeld}, R. and {Ross}, A.~J. and {Rykoff}, E.~S. and {Sanchez}, J. and {Sevilla-Noarbe}, I. and {Sheldon}, E. and {Shin}, T. and {Troja}, A. and {Tutusaus}, I. and {Tutusaus}, I. and {Varga}, T.~N. and {Weaverdyck}, N. and {Yanny}, B. and {Yin}, B. and {Zhang}, Y. and {Zuntz}, J. and {Aguena}, M. and {Allam}, S. and {Annis}, J. and {Bacon}, D. and {Bertin}, E. and {Bhargava}, S. and {Brooks}, D. and {Buckley-Geer}, E. and {Burke}, D.~L. and {Carretero}, J. and {Costanzi}, M. and {da Costa}, L.~N. and {Pereira}, M.~E.~S. and {De Vicente}, J. and {Desai}, S. and {Dietrich}, J.~P. and {Doel}, P. and {Ferrero}, I. and {Flaugher}, B. and {Frieman}, J. and {Garc{\'\i}a-Bellido}, J. and {Gaztanaga}, E. and {Gerdes}, D.~W. and {Giannantonio}, T. and {Gschwend}, J. and {Gutierrez}, G. and {Hinton}, S.~R. and {Hollowood}, D.~L. and {Honscheid}, K. and {Hoyle}, B. and {James}, D.~J. and {Kron}, R. and {Kuehn}, K. and {Lahav}, O. and {Lima}, M. and {Lin}, H. and {Maia}, M.~A.~G. and {Marshall}, J.~L. and {Martini}, P. and {Melchior}, P. and {Menanteau}, F. and {Miquel}, R. and {Mohr}, J.~J. and {Morgan}, R. and {Ogando}, R.~L.~C. and {Palmese}, A. and {Paz-Chinch{\'o}n}, F. and {Petravick}, D. and {Pieres}, A. and {Romer}, A.~K. and {Sanchez}, E. and {Scarpine}, V. and {Schubnell}, M. and {Serrano}, S. and {Smith}, M. and {Soares-Santos}, M. and {Tarle}, G. and {Thomas}, D. and {To}, C. and {Weller}, J. and {DES Collaboration}},
 doi = {10.1103/PhysRevD.105.023514},
 eid = {023514},
 eprint = {2105.13543},
 journal = {\prd},
 month = {January},
 number = {2},
 pages = {023514},
 primaryclass = {astro-ph.CO},
 title = {{Dark Energy Survey Year 3 results: Cosmology from cosmic shear and robustness to data calibration}},
 volume = {105},
 year = {2022}
}

@article{Anbajagane2025,
 adsurl = {https://ui.adsabs.harvard.edu/abs/2025OJAp....846161A},
 archiveprefix = {arXiv},
 author = {{Anbajagane}, D. and {Chang}, C. and {Drlica-Wagner}, A. and {Tan}, C.~Y. and {Adamow}, M. and {Gruendl}, R.~A. and {Secco}, L.~F. and {Zhang}, Z. and {Becker}, M.~R. and {Ferguson}, P.~S. and {Chicoine}, N. and {Herron}, K. and {Alarcon}, A. and {Teixeira}, R. and {Suson}, D. and {Alsina}, A.~N. and {Amon}, A. and {Andrade-Oliveira}, F. and {Blazek}, J. and {Bom}, C.~R. and {Camacho}, H. and {Carballo-Bello}, J.~A. and {Rosell}, A. Carnero and {Cawthon}, R. and {Cerny}, W. and {Choi}, A. and {Choi}, Y. and {Dodelson}, S. and {Doux}, C. and {Eckert}, K. and {Elvin-Poole}, J. and {Esteves}, J. and {Gatti}, M. and {Giannini}, G. and {Hartley}, W.~G. and {Gruen}, D. and {Herner}, K. and {Huff}, E.~M. and {James}, D.~J. and {Jarvis}, M. and {Krause}, E. and {Kuropatkin}, N. and {Mart{\'\i}nez-V{\'a}zquez}, C.~E. and {Massana}, P. and {Mau}, S. and {McCullough}, J. and {Medina}, G.~E. and {Mutlu-Pakdil}, B. and {Myles}, J. and {Navabi}, M. and {No{\"e}l}, N.~E.~D. and {Pace}, A.~B. and {Porredon}, A. and {Prat}, J. and {Raveri}, M. and {Riley}, A.~H. and {Rykoff}, E.~S. and {Sakowska}, J.~D. and {Samuroff}, S. and {Sanchez-Cid}, D. and {Sand}, D.~J. and {Santana-Silva}, L. and {Sevilla-Noarbe}, I. and {Shin}, T. and {Soares-Santos}, M. and {Stringfellow}, G.~S. and {To}, C. and {Tong}, A. and {Troxel}, M.~A. and {Vivas}, A.~K. and {Yamamoto}, M. and {Yanny}, B. and {Yin}, B. and {Zhang}, Y. and {Zuntz}, J.},
 doi = {10.33232/001c.146161},
 eprint = {2502.17677},
 journal = {The Open Journal of Astrophysics},
 month = {October},
 pages = {46161},
 primaryclass = {astro-ph.CO},
 title = {{The DECADE cosmic shear project IV: cosmological constraints from 107 million galaxies across 5,400 deg 2 of the sky}},
 volume = {8},
 year = {2025}
}

@article{Arico2021,
 adsurl = {https://ui.adsabs.harvard.edu/abs/2021MNRAS.506.4070A},
 archiveprefix = {arXiv},
 author = {{Aric{\`o}}, Giovanni and {Angulo}, Raul E. and {Contreras}, Sergio and {Ondaro-Mallea}, Lurdes and {Pellejero-Iba{\~n}ez}, Marcos and {Zennaro}, Matteo},
 doi = {10.1093/mnras/stab1911},
 eprint = {2011.15018},
 journal = {\mnras},
 month = {September},
 number = {3},
 pages = {4070-4082},
 primaryclass = {astro-ph.CO},
 title = {{The BACCO simulation project: a baryonification emulator with neural networks}},
 volume = {506},
 year = {2021}
}

@article{Arico2023,
 adsurl = {https://ui.adsabs.harvard.edu/abs/2023A&A...678A.109A},
 archiveprefix = {arXiv},
 author = {{Aric{\`o}}, Giovanni and {Angulo}, Raul E. and {Zennaro}, Matteo and {Contreras}, Sergio and {Chen}, Angela and {Hern{\'a}ndez-Monteagudo}, Carlos},
 doi = {10.1051/0004-6361/202346539},
 eid = {A109},
 eprint = {2303.05537},
 journal = {\aap},
 month = {October},
 pages = {A109},
 primaryclass = {astro-ph.CO},
 title = {{DES Y3 cosmic shear down to small scales: Constraints on cosmology and baryons}},
 volume = {678},
 year = {2023}
}

@article{Baleato-Lizancos2025,
 adsurl = {https://ui.adsabs.harvard.edu/abs/2025JCAP...11..031B},
 archiveprefix = {arXiv},
 author = {{Baleato Lizancos}, A. and {Coulton}, W. and {Challinor}, A. and {Sherwin}, B.~D. and {Mehta}, Y.},
 doi = {10.1088/1475-7516/2025/11/031},
 eid = {031},
 eprint = {2507.03859},
 journal = {\jcap},
 month = {November},
 number = {11},
 pages = {031},
 primaryclass = {astro-ph.CO},
 title = {{A halo model of extragalactic contamination to CMB lensing, delensing, and cross-correlations}},
 volume = {2025},
 year = {2025}
}

@article{Bartelmann2001,
 adsurl = {https://ui.adsabs.harvard.edu/abs/2001PhR...340..291B},
 archiveprefix = {arXiv},
 author = {{Bartelmann}, M. and {Schneider}, P.},
 doi = {10.1016/S0370-1573(00)00082-X},
 eprint = {astro-ph/9912508},
 journal = {\physrep},
 month = {January},
 number = {4-5},
 pages = {291-472},
 primaryclass = {astro-ph},
 title = {{Weak gravitational lensing}},
 volume = {340},
 year = {2001}
}

@article{Bechtol2026,
 adsurl = {https://ui.adsabs.harvard.edu/abs/2026ApJS..282...62B},
 archiveprefix = {arXiv},
 author = {{Bechtol}, K. and {Sevilla-Noarbe}, I. and {Drlica-Wagner}, A. and {Yanny}, B. and {Gruendl}, R.~A. and {Sheldon}, E. and {Rykoff}, E.~S. and {De Vicente}, J. and {Adamow}, M. and {Anbajagane}, D. and {Becker}, M.~R. and {Bernstein}, G.~M. and {Carnero Rosell}, A. and {Gschwend}, J. and {Gorsuch}, M. and {Hartley}, W.~G. and {Jarvis}, M. and {Jeltema}, T. and {Kron}, R. and {Manning}, T.~A. and {O'Donnell}, J. and {Pieres}, A. and {Rodr{\'\i}guez-Monroy}, M. and {Sanchez Cid}, D. and {Tabbutt}, M. and {Tan}, C.~Y. and {Toribio San Cipriano}, L. and {Tucker}, D.~L. and {Weaverdyck}, N. and {Yamamoto}, M. and {Abbott}, T.~M.~C. and {Aguena}, M. and {Alarc{\'o}n}, A. and {Allam}, S. and {Amon}, A. and {Andrade-Oliveira}, F. and {Avila}, S. and {Bernardinelli}, P.~H. and {Bertin}, E. and {Blazek}, J. and {Brooks}, D. and {Burke}, D.~L. and {Carretero}, J. and {Castander}, F.~J. and {Cawthon}, R. and {Chang}, C. and {Choi}, A. and {Conselice}, C. and {Costanzi}, M. and {Crocce}, M. and {da Costa}, L.~N. and {Davis}, T.~M. and {Desai}, S. and {Diehl}, H.~T. and {Dodelson}, S. and {Doel}, P. and {Doux}, C. and {Fert{\'e}}, A. and {Flaugher}, B. and {Fosalba}, P. and {Frieman}, J. and {Garc{\'\i}a-Bellido}, J. and {Gatti}, M. and {Gaztanaga}, E. and {Giannini}, G. and {Gruen}, D. and {Gutierrez}, G. and {Herner}, K. and {Hinton}, S.~R. and {Hollowood}, D.~L. and {Honscheid}, K. and {Huterer}, D. and {Jeffrey}, N. and {Krause}, E. and {Kuehn}, K. and {Lahav}, O. and {Lee}, S. and {Lidman}, C. and {Lima}, M. and {Lin}, H. and {Marshall}, J.~L. and {Mena-Fern{\'a}ndez}, J. and {Miquel}, R. and {Mohr}, J.~J. and {Muir}, J. and {Myles}, J. and {Ogando}, R.~L.~C. and {Palmese}, A. and {Plazas Malag{\'o}n}, A.~A. and {Porredon}, A. and {Prat}, J. and {Raveri}, M. and {Romer}, A.~K. and {Roodman}, A. and {Samuroff}, S. and {Sanchez}, E. and {Scarpine}, V. and {Smith}, M. and {Soares-Santos}, M. and {Suchyta}, E. and {Tarle}, G. and {Troxel}, M.~A. and {Vikram}, V. and {Walker}, A.~R. and {Weller}, J. and {Wiseman}, P. and {Zhang}, Y. and {DES Collaboration}},
 doi = {10.3847/1538-4365/ae18d3},
 eid = {62},
 eprint = {2501.05739},
 journal = {\apjs},
 month = {February},
 number = {2},
 pages = {62},
 primaryclass = {astro-ph.CO},
 title = {{Dark Energy Survey Year 6 Results: Photometric Dataset for Cosmology}},
 volume = {282},
 year = {2026}
}

@article{Bigwood2025,
 adsurl = {https://ui.adsabs.harvard.edu/abs/2025arXiv251204209B},
 archiveprefix = {arXiv},
 author = {{Bigwood}, Leah and {McCullough}, Jamie and {Siegel}, Jared and {Amon}, Alexandra and {Efstathiou}, George and {Sanchez-Cid}, David and {Legnani}, Elisa and {Gruen}, Daniel and {Blazek}, Jonathan and {Doux}, Cyrille and {Carnero Rosell}, Aurelio and {Gatti}, Marco and {Huff}, Eric and {MacCrann}, Niall and {Porredon}, Anna and {Prat Marti}, Judit and {Soares dos Santos}, Marcelle and {Myles}, Justin and {Samuroff}, Simon and {Yamamoto}, Masaya and {Yin}, Boyan and {Zuntz}, Joe},
 doi = {10.48550/arXiv.2512.04209},
 eid = {arXiv:2512.04209},
 eprint = {2512.04209},
 journal = {arXiv e-prints},
 month = {December},
 pages = {arXiv:2512.04209},
 primaryclass = {astro-ph.CO},
 title = {{Confronting cosmic shear astrophysical uncertainties: DES Year 3 revisited}},
 year = {2025}
}

@article{Blazek2019,
 adsurl = {https://ui.adsabs.harvard.edu/abs/2019PhRvD.100j3506B},
 archiveprefix = {arXiv},
 author = {{Blazek}, Jonathan A. and {MacCrann}, Niall and {Troxel}, M.~A. and {Fang}, Xiao},
 doi = {10.1103/PhysRevD.100.103506},
 eid = {103506},
 eprint = {1708.09247},
 journal = {\prd},
 month = {November},
 number = {10},
 pages = {103506},
 primaryclass = {astro-ph.CO},
 title = {{Beyond linear galaxy alignments}},
 volume = {100},
 year = {2019}
}

@article{Bridle2007,
 adsurl = {https://ui.adsabs.harvard.edu/abs/2007NJPh....9..444B},
 archiveprefix = {arXiv},
 author = {{Bridle}, Sarah and {King}, Lindsay},
 doi = {10.1088/1367-2630/9/12/444},
 eprint = {0705.0166},
 journal = {New Journal of Physics},
 month = {December},
 number = {12},
 pages = {444},
 primaryclass = {astro-ph},
 title = {{Dark energy constraints from cosmic shear power spectra: impact of intrinsic alignments on photometric redshift requirements}},
 volume = {9},
 year = {2007}
}

@article{Camphuis2025,
 adsurl = {https://ui.adsabs.harvard.edu/abs/2026PhRvD.113h3504C},
 archiveprefix = {arXiv},
 author = {{Camphuis}, E. and {Quan}, W. and {Balkenhol}, L. and {Khalife}, A.~R. and {Ge}, F. and {Guidi}, F. and {Huang}, N. and {Lynch}, G.~P. and {Omori}, Y. and {Trendafilova}, C. and {Anderson}, A.~J. and {Ansarinejad}, B. and {Archipley}, M. and {Barry}, P.~S. and {Benabed}, K. and {Bender}, A.~N. and {Benson}, B.~A. and {Bianchini}, F. and {Bleem}, L.~E. and {Bouchet}, F.~R. and {Bryant}, L. and {Campitiello}, M.~G. and {Carlstrom}, J.~E. and {Chang}, C.~L. and {Chaubal}, P. and {Chichura}, P.~M. and {Chokshi}, A. and {Chou}, T.-L. and {Coerver}, A. and {Crawford}, T.~M. and {Daley}, C. and {de Haan}, T. and {Dibert}, K.~R. and {Dobbs}, M.~A. and {Doohan}, M. and {Doussot}, A. and {Dutcher}, D. and {Everett}, W. and {Feng}, C. and {Ferguson}, K.~R. and {Fichman}, K. and {Foster}, A. and {Galli}, S. and {Gambrel}, A.~E. and {Gardner}, R.~W. and {Goeckner-Wald}, N. and {Gualtieri}, R. and {Guns}, S. and {Halverson}, N.~W. and {Hivon}, E. and {Holder}, G.~P. and {Holzapfel}, W.~L. and {Hood}, J.~C. and {Hryciuk}, A. and {K{\'e}ruzor{\'e}}, F. and {Knox}, L. and {Korman}, M. and {Kornoelje}, K. and {Kuo}, C.-L. and {Levy}, K. and {Lowitz}, A.~E. and {Lu}, C. and {Maniyar}, A. and {Martsen}, E.~S. and {Menanteau}, F. and {Millea}, M. and {Montgomery}, J. and {Nakato}, Y. and {Natoli}, T. and {Noble}, G.~I. and {Ouellette}, A. and {Pan}, Z. and {Paschos}, P. and {Phadke}, K.~A. and {Pollak}, A.~W. and {Prabhu}, K. and {Raghunathan}, S. and {Rahimi}, M. and {Rahlin}, A. and {Reichardt}, C.~L. and {Rouble}, M. and {Ruhl}, J.~E. and {Schiappucci}, E. and {Simpson}, A. and {Sobrin}, J.~A. and {Stark}, A.~A. and {Stephen}, J. and {Tandoi}, C. and {Thorne}, B. and {Umilta}, C. and {Vieira}, J.~D. and {Vitrier}, A. and {Wan}, Y. and {Whitehorn}, N. and {Wu}, W.~L.~K. and {Young}, M.~R. and {Zebrowski}, J.~A. and {SPT-3G Collaboration}},
 doi = {10.1103/7wt3-9v2y},
 eid = {083504},
 eprint = {2506.20707},
 journal = {\prd},
 month = {April},
 number = {8},
 pages = {083504},
 primaryclass = {astro-ph.CO},
 title = {{SPT-3G D1: CMB temperature and polarization power spectra and cosmology from 2019 and 2020 observations of the SPT-3G main field}},
 volume = {113},
 year = {2026}
}

@article{Carlstrom2011,
 adsurl = {https://ui.adsabs.harvard.edu/abs/2011PASP..123..568C},
 archiveprefix = {arXiv},
 author = {{Carlstrom}, J.~E. and {Ade}, P.~A.~R. and {Aird}, K.~A. and {Benson}, B.~A. and {Bleem}, L.~E. and {Busetti}, S. and {Chang}, C.~L. and {Chauvin}, E. and {Cho}, H. -M. and {Crawford}, T.~M. and {Crites}, A.~T. and {Dobbs}, M.~A. and {Halverson}, N.~W. and {Heimsath}, S. and {Holzapfel}, W.~L. and {Hrubes}, J.~D. and {Joy}, M. and {Keisler}, R. and {Lanting}, T.~M. and {Lee}, A.~T. and {Leitch}, E.~M. and {Leong}, J. and {Lu}, W. and {Lueker}, M. and {Luong-Van}, D. and {McMahon}, J.~J. and {Mehl}, J. and {Meyer}, S.~S. and {Mohr}, J.~J. and {Montroy}, T.~E. and {Padin}, S. and {Plagge}, T. and {Pryke}, C. and {Ruhl}, J.~E. and {Schaffer}, K.~K. and {Schwan}, D. and {Shirokoff}, E. and {Spieler}, H.~G. and {Staniszewski}, Z. and {Stark}, A.~A. and {Tucker}, C. and {Vanderlinde}, K. and {Vieira}, J.~D. and {Williamson}, R.},
 doi = {10.1086/659879},
 eprint = {0907.4445},
 journal = {\pasp},
 month = {May},
 number = {903},
 pages = {568},
 primaryclass = {astro-ph.IM},
 title = {{The 10 Meter South Pole Telescope}},
 volume = {123},
 year = {2011}
}

@article{Carron2017,
 adsurl = {https://ui.adsabs.harvard.edu/abs/2017PhRvD..96f3510C},
 archiveprefix = {arXiv},
 author = {{Carron}, Julien and {Lewis}, Antony},
 doi = {10.1103/PhysRevD.96.063510},
 eid = {063510},
 eprint = {1704.08230},
 journal = {\prd},
 month = {September},
 number = {6},
 pages = {063510},
 primaryclass = {astro-ph.CO},
 title = {{Maximum a posteriori CMB lensing reconstruction}},
 volume = {96},
 year = {2017}
}

@article{Carron2022,
 adsurl = {https://ui.adsabs.harvard.edu/abs/2022JCAP...09..039C},
 archiveprefix = {arXiv},
 author = {{Carron}, Julien and {Mirmelstein}, Mark and {Lewis}, Antony},
 doi = {10.1088/1475-7516/2022/09/039},
 eid = {039},
 eprint = {2206.07773},
 journal = {\jcap},
 month = {September},
 number = {9},
 pages = {039},
 primaryclass = {astro-ph.CO},
 title = {{CMB lensing from Planck PR4 maps}},
 volume = {2022},
 year = {2022}
}

@article{Chang2023,
 adsurl = {https://ui.adsabs.harvard.edu/abs/2023PhRvD.107b3530C},
 archiveprefix = {arXiv},
 author = {{Chang}, C. and {Omori}, Y. and {Baxter}, E.~J. and {Doux}, C. and {Choi}, A. and {Pandey}, S. and {Alarcon}, A. and {Alves}, O. and {Amon}, A. and {Andrade-Oliveira}, F. and {Bechtol}, K. and {Becker}, M.~R. and {Bernstein}, G.~M. and {Bianchini}, F. and {Blazek}, J. and {Bleem}, L.~E. and {Camacho}, H. and {Campos}, A. and {Carnero Rosell}, A. and {Carrasco Kind}, M. and {Cawthon}, R. and {Chen}, R. and {Cordero}, J. and {Crawford}, T.~M. and {Crocce}, M. and {Davis}, C. and {DeRose}, J. and {Dodelson}, S. and {Drlica-Wagner}, A. and {Eckert}, K. and {Eifler}, T.~F. and {Elsner}, F. and {Elvin-Poole}, J. and {Everett}, S. and {Fang}, X. and {Fert{\'e}}, A. and {Fosalba}, P. and {Friedrich}, O. and {Gatti}, M. and {Giannini}, G. and {Gruen}, D. and {Gruendl}, R.~A. and {Harrison}, I. and {Herner}, K. and {Huang}, H. and {Huff}, E.~M. and {Huterer}, D. and {Jarvis}, M. and {Kovacs}, A. and {Krause}, E. and {Kuropatkin}, N. and {Leget}, P. -F. and {Lemos}, P. and {Liddle}, A.~R. and {MacCrann}, N. and {McCullough}, J. and {Muir}, J. and {Myles}, J. and {Navarro-Alsina}, A. and {Park}, Y. and {Porredon}, A. and {Prat}, J. and {Raveri}, M. and {Rollins}, R.~P. and {Roodman}, A. and {Rosenfeld}, R. and {Ross}, A.~J. and {Rykoff}, E.~S. and {S{\'a}nchez}, C. and {Sanchez}, J. and {Secco}, L.~F. and {Sevilla-Noarbe}, I. and {Sheldon}, E. and {Shin}, T. and {Troxel}, M.~A. and {Tutusaus}, I. and {Varga}, T.~N. and {Weaverdyck}, N. and {Wechsler}, R.~H. and {Wu}, W.~L.~K. and {Yanny}, B. and {Yin}, B. and {Zhang}, Y. and {Zuntz}, J. and {Abbott}, T.~M.~C. and {Aguena}, M. and {Allam}, S. and {Annis}, J. and {Bacon}, D. and {Benson}, B.~A. and {Bertin}, E. and {Bocquet}, S. and {Brooks}, D. and {Burke}, D.~L. and {Carlstrom}, J.~E. and {Carretero}, J. and {Chang}, C.~L. and {Chown}, R. and {Costanzi}, M. and {da Costa}, L.~N. and {Crites}, A.~T. and {Pereira}, M.~E.~S. and {de Haan}, T. and {De Vicente}, J. and {Desai}, S. and {Diehl}, H.~T. and {Dobbs}, M.~A. and {Doel}, P. and {Everett}, W. and {Ferrero}, I. and {Flaugher}, B. and {Friedel}, D. and {Frieman}, J. and {Garc{\'\i}a-Bellido}, J. and {Gaztanaga}, E. and {George}, E.~M. and {Giannantonio}, T. and {Halverson}, N.~W. and {Hinton}, S.~R. and {Holder}, G.~P. and {Hollowood}, D.~L. and {Holzapfel}, W.~L. and {Honscheid}, K. and {Hrubes}, J.~D. and {James}, D.~J. and {Knox}, L. and {Kuehn}, K. and {Lahav}, O. and {Lee}, A.~T. and {Lima}, M. and {Luong-Van}, D. and {March}, M. and {McMahon}, J.~J. and {Melchior}, P. and {Menanteau}, F. and {Meyer}, S.~S. and {Miquel}, R. and {Mocanu}, L. and {Mohr}, J.~J. and {Morgan}, R. and {Natoli}, T. and {Padin}, S. and {Palmese}, A. and {Paz-Chinch{\'o}n}, F. and {Pieres}, A. and {Plazas Malag{\'o}n}, A.~A. and {Pryke}, C. and {Reichardt}, C.~L. and {Rodr{\'\i}guez-Monroy}, M. and {Romer}, A.~K. and {Ruhl}, J.~E. and {Sanchez}, E. and {Schaffer}, K.~K. and {Schubnell}, M. and {Serrano}, S. and {Shirokoff}, E. and {Smith}, M. and {Staniszewski}, Z. and {Stark}, A.~A. and {Suchyta}, E. and {Tarle}, G. and {Thomas}, D. and {To}, C. and {Vieira}, J.~D. and {Weller}, J. and {Williamson}, R. and {DES} and {SPT Collaborations}},
 doi = {10.1103/PhysRevD.107.023530},
 eid = {023530},
 eprint = {2203.12440},
 journal = {\prd},
 month = {January},
 number = {2},
 pages = {023530},
 primaryclass = {astro-ph.CO},
 title = {{Joint analysis of Dark Energy Survey Year 3 data and CMB lensing from SPT and Planck. II. Cross-correlation measurements and cosmological constraints}},
 volume = {107},
 year = {2023}
}

@article{Chen2024,
 adsurl = {https://ui.adsabs.harvard.edu/abs/2024PhRvD.110j3518C},
 archiveprefix = {arXiv},
 author = {{Chen}, S. and {DeRose}, J. and {Zhou}, R. and {White}, M. and {Ferraro}, S. and {Blake}, C. and {Lange}, J.~U. and {Wechsler}, R.~H. and {Aguilar}, J. and {Ahlen}, S. and {Brooks}, D. and {Claybaugh}, T. and {Dawson}, K. and {de la Macorra}, A. and {Doel}, P. and {Font-Ribera}, A. and {Gazta{\~n}aga}, E. and {Gontcho A Gontcho}, S. and {Gutierrez}, G. and {Honscheid}, K. and {Howlett}, C. and {Kehoe}, R. and {Kirkby}, D. and {Kisner}, T. and {Kremin}, A. and {Landriau}, M. and {Le Guillou}, L. and {Manera}, M. and {Meisner}, A. and {Miquel}, R. and {Newman}, J.~A. and {Niz}, G. and {Palanque-Delabrouille}, N. and {Percival}, W.~J. and {Prada}, F. and {Rossi}, G. and {Sanchez}, E. and {Schlegel}, D. and {Schubnell}, M. and {Sprayberry}, D. and {Tarl{\'e}}, G. and {Weaver}, B.~A.},
 doi = {10.1103/PhysRevD.110.103518},
 eid = {103518},
 eprint = {2407.04795},
 journal = {\prd},
 month = {November},
 number = {10},
 pages = {103518},
 primaryclass = {astro-ph.CO},
 title = {{Analysis of DESI{\texttimes}DES using the Lagrangian effective theory of LSS}},
 volume = {110},
 year = {2024}
}

@article{Chisari2019,
 adsurl = {https://ui.adsabs.harvard.edu/abs/2019ApJS..242....2C},
 archiveprefix = {arXiv},
 author = {{Chisari}, Nora Elisa and {Alonso}, David and {Krause}, Elisabeth and {Leonard}, C. Danielle and {Bull}, Philip and {Neveu}, J{\'e}r{\'e}my and {Villarreal}, Antonia Sierra and {Singh}, Sukhdeep and {McClintock}, Thomas and {Ellison}, John and {Du}, Zilong and {Zuntz}, Joe and {Mead}, Alexander and {Joudaki}, Shahab and {Lorenz}, Christiane S. and {Tr{\"o}ster}, Tilman and {Sanchez}, Javier and {Lanusse}, Francois and {Ishak}, Mustapha and {Hlozek}, Ren{\'e}e and {Blazek}, Jonathan and {Campagne}, Jean-Eric and {Almoubayyed}, Husni and {Eifler}, Tim and {Kirby}, Matthew and {Kirkby}, David and {Plaszczynski}, St{\'e}phane and {Slosar}, An{\v{z}}e and {Vrastil}, Michal and {Wagoner}, Erika L. and {LSST Dark Energy Science Collaboration}},
 doi = {10.3847/1538-4365/ab1658},
 eid = {2},
 eprint = {1812.05995},
 journal = {\apjs},
 month = {May},
 number = {1},
 pages = {2},
 primaryclass = {astro-ph.CO},
 title = {{Core Cosmology Library: Precision Cosmological Predictions for LSST}},
 volume = {242},
 year = {2019}
}

@article{Chisari2019b,
 adsurl = {https://ui.adsabs.harvard.edu/abs/2019OJAp....2E...4C},
 archiveprefix = {arXiv},
 author = {{Chisari}, Nora Elisa and {Mead}, Alexander J. and {Joudaki}, Shahab and {Ferreira}, Pedro G. and {Schneider}, Aurel and {Mohr}, Joseph and {Tr{\"o}ster}, Tilman and {Alonso}, David and {McCarthy}, Ian G. and {Martin-Alvarez}, Sergio and {Devriendt}, Julien and {Slyz}, Adrianne and {van Daalen}, Marcel P.},
 doi = {10.21105/astro.1905.06082},
 eid = {4},
 eprint = {1905.06082},
 journal = {The Open Journal of Astrophysics},
 month = {June},
 number = {1},
 pages = {4},
 primaryclass = {astro-ph.CO},
 title = {{Modelling baryonic feedback for survey cosmology}},
 volume = {2},
 year = {2019}
}

@article{Chisari2025,
 adsurl = {https://ui.adsabs.harvard.edu/abs/2025A&ARv..33....5C},
 archiveprefix = {arXiv},
 author = {{Chisari}, Nora Elisa},
 doi = {10.1007/s00159-025-00161-8},
 eid = {5},
 eprint = {2510.15738},
 journal = {\aapr},
 month = {October},
 number = {1},
 pages = {5},
 primaryclass = {astro-ph.CO},
 title = {{A rising tide: intrinsic alignments since the turn of the millennium}},
 volume = {33},
 year = {2025}
}

@article{Dalal2023,
 adsurl = {https://ui.adsabs.harvard.edu/abs/2023PhRvD.108l3519D},
 archiveprefix = {arXiv},
 author = {{Dalal}, Roohi and {Li}, Xiangchong and {Nicola}, Andrina and {Zuntz}, Joe and {Strauss}, Michael A. and {Sugiyama}, Sunao and {Zhang}, Tianqing and {Rau}, Markus M. and {Mandelbaum}, Rachel and {Takada}, Masahiro and {More}, Surhud and {Miyatake}, Hironao and {Kannawadi}, Arun and {Shirasaki}, Masato and {Taniguchi}, Takanori and {Takahashi}, Ryuichi and {Osato}, Ken and {Hamana}, Takashi and {Oguri}, Masamune and {Nishizawa}, Atsushi J. and {Malag{\'o}n}, Andr{\'e}s A. Plazas and {Sunayama}, Tomomi and {Alonso}, David and {Slosar}, An{\v{z}}e and {Luo}, Wentao and {Armstrong}, Robert and {Bosch}, James and {Hsieh}, Bau-Ching and {Komiyama}, Yutaka and {Lupton}, Robert H. and {Lust}, Nate B. and {MacArthur}, Lauren A. and {Miyazaki}, Satoshi and {Murayama}, Hitoshi and {Nishimichi}, Takahiro and {Okura}, Yuki and {Price}, Paul A. and {Tait}, Philip J. and {Tanaka}, Masayuki and {Wang}, Shiang-Yu},
 doi = {10.1103/PhysRevD.108.123519},
 eid = {123519},
 eprint = {2304.00701},
 journal = {\prd},
 month = {December},
 number = {12},
 pages = {123519},
 primaryclass = {astro-ph.CO},
 title = {{Hyper Suprime-Cam Year 3 results: Cosmology from cosmic shear power spectra}},
 volume = {108},
 year = {2023}
}

@article{Das2013,
 adsurl = {https://ui.adsabs.harvard.edu/abs/2013arXiv1311.2338D},
 archiveprefix = {arXiv},
 author = {{Das}, Sudeep and {Errard}, Josquin and {Spergel}, David},
 doi = {10.48550/arXiv.1311.2338},
 eid = {arXiv:1311.2338},
 eprint = {1311.2338},
 journal = {arXiv e-prints},
 month = {November},
 pages = {arXiv:1311.2338},
 primaryclass = {astro-ph.CO},
 title = {{Can CMB Lensing Help Cosmic Shear Surveys?}},
 year = {2013}
}

@article{DeRose2025,
 adsurl = {https://ui.adsabs.harvard.edu/abs/2025arXiv251018981D},
 archiveprefix = {arXiv},
 author = {{DeRose}, Joseph and {Chen}, Shi-Fan},
 doi = {10.48550/arXiv.2510.18981},
 eid = {arXiv:2510.18981},
 eprint = {2510.18981},
 journal = {arXiv e-prints},
 month = {October},
 pages = {arXiv:2510.18981},
 primaryclass = {astro-ph.CO},
 title = {{The Lensing Counter Narrative: An Effective Description of Small-Scale Clustering in Weak Lensing Power Spectra}},
 year = {2025}
}

@article{DES2005,
 adsurl = {https://ui.adsabs.harvard.edu/abs/2005astro.ph.10346T},
 archiveprefix = {arXiv},
 author = {{The Dark Energy Survey Collaboration}},
 doi = {10.48550/arXiv.astro-ph/0510346},
 eid = {astro-ph/0510346},
 eprint = {astro-ph/0510346},
 journal = {arXiv e-prints},
 month = {October},
 pages = {astro-ph/0510346},
 primaryclass = {astro-ph},
 title = {{The Dark Energy Survey}},
 year = {2005}
}

@article{DESandKIDS2023,
 adsurl = {https://ui.adsabs.harvard.edu/abs/2023OJAp....6E..36D},
 archiveprefix = {arXiv},
 author = {{Dark Energy Survey and Kilo-Degree Survey Collaboration} and {Abbott}, T.~M.~C. and {Aguena}, M. and {Alarcon}, A. and {Alves}, O. and {Amon}, A. and {Andrade-Oliveira}, F. and {Asgari}, M. and {Avila}, S. and {Bacon}, D. and {Bechtol}, K. and {Becker}, M.~R. and {Bernstein}, G.~M. and {Bertin}, E. and {Bilicki}, M. and {Blazek}, J. and {Bocquet}, S. and {Brooks}, D. and {Burger}, P. and {Burke}, D.~L. and {Camacho}, H. and {Campos}, A. and {Carnero Rosell}, A. and {Carrasco Kind}, M. and {Carretero}, J. and {Castander}, F.~J. and {Cawthon}, R. and {Chang}, C. and {Chen}, R. and {Choi}, A. and {Conselice}, C. and {Cordero}, J. and {Crocce}, M. and {da Costa}, L.~N. and {da Silva Pereira}, M.~E. and {Dalal}, R. and {Davis}, C. and {de Jong}, J.~T.~A. and {DeRose}, J. and {Desai}, S. and {Diehl}, H.~T. and {Dodelson}, S. and {Doel}, P. and {Doux}, C. and {Drlica-Wagner}, A. and {Dvornik}, A. and {Eckert}, K. and {Eifler}, T.~F. and {Elvin-Poole}, J. and {Everett}, S. and {Fang}, X. and {Ferrero}, I. and {Fert{\'e}}, A. and {Flaugher}, B. and {Friedrich}, O. and {Frieman}, J. and {Garc{\'\i}a-Bellido}, J. and {Gatti}, M. and {Giannini}, G. and {Giblin}, B. and {Gruen}, D. and {Gruendl}, R.~A. and {Gutierrez}, G. and {Harrison}, I. and {Hartley}, W.~G. and {Herner}, K. and {Heymans}, C. and {Hildebrandt}, H. and {Hinton}, S.~R. and {Hoekstra}, H. and {Hollowood}, D.~L. and {Honscheid}, K. and {Huang}, H. and {Huff}, E.~M. and {Huterer}, D. and {James}, D.~J. and {Jarvis}, M. and {Jeffrey}, N. and {Jeltema}, T. and {Joachimi}, B. and {Joudaki}, S. and {Kannawadi}, A. and {Krause}, E. and {Kuehn}, K. and {Kuijken}, K. and {Kuropatkin}, N. and {Lahav}, O. and {Leget}, P. -F. and {Lemos}, P. and {Li}, S. -S. and {Li}, X. and {Liddle}, A.~R. and {Lima}, M. and {Lin}, C. -A. and {Lin}, H. and {MacCrann}, N. and {Mahony}, C. and {Marshall}, J.~L. and {McCullough}, J. and {Mena-Fern{\'a}ndez}, J. and {Menanteau}, F. and {Miquel}, R. and {Mohr}, J.~J. and {Muir}, J. and {Myles}, J. and {Napolitano}, N. and {Navarro-Alsina}, A. and {Ogando}, R.~L.~C. and {Palmese}, A. and {Pandey}, S. and {Park}, Y. and {Paterno}, M. and {Peacock}, J.~A. and {Petravick}, D. and {Pieres}, A. and {Plazas Malag{\'o}n}, A.~A. and {Porredon}, A. and {Prat}, J. and {Radovich}, M. and {Raveri}, M. and {Reischke}, R. and {Robertson}, N.~C. and {Rollins}, R.~P. and {Romer}, A.~K. and {Roodman}, A. and {Rykoff}, E.~S. and {Samuroff}, S. and {S{\'a}nchez}, C. and {Sanchez}, E. and {Sanchez}, J. and {Schneider}, P. and {Secco}, L.~F. and {Sevilla-Noarbe}, I. and {Shan}, H. -Y. and {Sheldon}, E. and {Shin}, T. and {Sif{\'o}n}, C. and {Smith}, M. and {Soares-Santos}, M. and {St{\"o}lzner}, B. and {Suchyta}, E. and {Swanson}, M.~E.~C. and {Tarle}, G. and {Thomas}, D. and {To}, C. and {Troxel}, M.~A. and {Tr{\"o}ster}, T. and {Tutusaus}, I. and {van den Busch}, J.~L. and {Varga}, T.~N. and {Walker}, A.~R. and {Weaverdyck}, N. and {Wechsler}, R.~H. and {Weller}, J. and {Wiseman}, P. and {Wright}, A.~H. and {Yanny}, B. and {Yin}, B. and {Yoon}, M. and {Zhang}, Y. and {Zuntz}, J.},
 doi = {10.21105/astro.2305.17173},
 eid = {36},
 eprint = {2305.17173},
 journal = {The Open Journal of Astrophysics},
 month = {October},
 pages = {36},
 primaryclass = {astro-ph.CO},
 title = {{DES Y3 + KiDS-1000: Consistent cosmology combining cosmic shear surveys}},
 volume = {6},
 year = {2023}
}

@article{Deutsch2018,
 adsurl = {https://ui.adsabs.harvard.edu/abs/2018JCAP...04..034D},
 archiveprefix = {arXiv},
 author = {{Deutsch}, Anne-Sylvie and {Johnson}, Matthew C. and {M{\"u}nchmeyer}, Moritz and {Terrana}, Alexandra},
 doi = {10.1088/1475-7516/2018/04/034},
 eid = {034},
 eprint = {1705.08907},
 journal = {\jcap},
 month = {April},
 number = {4},
 pages = {034},
 primaryclass = {astro-ph.CO},
 title = {{Polarized Sunyaev Zel'dovich tomography}},
 volume = {2018},
 year = {2018}
}

@article{DiValentino2025,
 adsurl = {https://ui.adsabs.harvard.edu/abs/2025PDU....4901965D},
 archiveprefix = {arXiv},
 author = {{Di Valentino}, Eleonora and {Said}, Jackson Levi and {Riess}, Adam and {Pollo}, Agnieszka and {Poulin}, Vivian and {G{\'o}mez-Valent}, Adri{\`a} and {Weltman}, Amanda and {Palmese}, Antonella and {Huang}, Caroline D. and {van de Bruck}, Carsten and {Saraf}, Chandra Shekhar and {Kuo}, Cheng-Yu and {Uhlemann}, Cora and {Grand{\'o}n}, Daniela and {Paz}, Dante and {Eckert}, Dominique and {Teixeira}, Elsa M. and {Saridakis}, Emmanuel N. and {Colg{\'a}in}, Eoin {\'O}. and {Beutler}, Florian and {Niedermann}, Florian and {Bajardi}, Francesco and {Barenboim}, Gabriela and {Gubitosi}, Giulia and {Musella}, Ilaria and {Banik}, Indranil and {Szapudi}, Istvan and {Singal}, Jack and {Cases}, Jaume Haro and {Chluba}, Jens and {Torrado}, Jes{\'u}s and {Mifsud}, Jurgen and {Jedamzik}, Karsten and {Said}, Khaled and {Dialektopoulos}, Konstantinos and {Herold}, Laura and {Perivolaropoulos}, Leandros and {Zu}, Lei and {Galbany}, Llu{\'\i}s and {Breuval}, Louise and {Visinelli}, Luca and {Escamilla}, Luis A. and {Anchordoqui}, Luis A. and {Sheikh-Jabbari}, M.~M. and {Lembo}, Margherita and {Dainotti}, Maria Giovanna and {Vincenzi}, Maria and {Asgari}, Marika and {Gerbino}, Martina and {Forconi}, Matteo and {Cantiello}, Michele and {Moresco}, Michele and {Benetti}, Micol and {Sch{\"o}neberg}, Nils and {Akarsu}, {\"O}zg{\"u}r and {Nunes}, Rafael C. and {Bernardo}, Reginald Christian and {Ch{\'a}vez}, Ricardo and {Anderson}, Richard I. and {Watkins}, Richard and {Capozziello}, Salvatore and {Li}, Siyang and {Vagnozzi}, Sunny and {Pan}, Supriya and {Treu}, Tommaso and {Irsic}, Vid and {Handley}, Will and {Giar{\`e}}, William and {Murakami}, Yukei and {Banihashemi}, Abdolali and {Poudou}, Ad{\`e}le and {Heavens}, Alan and {Kogut}, Alan and {Domi}, Alba and {Lenart}, Aleksander {\L}ukasz and {Melchiorri}, Alessandro and {Vadal{\`a}}, Alessandro and {Amon}, Alexandra and {Rivera}, Alexander Bonilla and {Reeves}, Alexander and {Zhuk}, Alexander and {Bonanno}, Alfio and {{\"O}vg{\"u}n}, Ali and {Pisani}, Alice and {Talebian}, Alireza and {Abebe}, Amare and {Aboubrahim}, Amin and {Gonz{\'a}lez Mor{\'a}n}, Ana Luisa and {Kov{\'a}cs}, Andr{\'a}s and {Lymperis}, Andreas and {Papatriantafyllou}, Andreas and {Liddle}, Andrew R. and {Paliathanasis}, Andronikos and {Borowiec}, Andrzej and {Yadav}, Anil Kumar and {Yadav}, Anita and {Sen}, Anjan Ananda and {William}, Anjitha John and {Davis}, Anne Christine and {Shajib}, Anowar J. and {Walters}, Anthony and {Lonappan}, Anto Idicherian and {Chudaykin}, Anton and {Capodagli}, Antonio and {da Silva}, Antonio and {De Felice}, Antonio and {Racioppi}, Antonio and {Oficial}, Araceli Soler and {Montiel}, Ariadna and {Favale}, Arianna and {Bernui}, Armando and {Velasco}, Arrianne Crystal and {Heinesen}, Asta and {Bakopoulos}, Athanasios and {Chatzistavrakidis}, Athanasios and {Khanpour}, Bahman and {Sathyaprakash}, Bangalore S. and {Zgirski}, Bartek and {L'Huillier}, Benjamin and {Famaey}, Benoit and {Jain}, Bhuvnesh and {Zhang}, Bing and {Karmakar}, Biswajit and {Dragovich}, Branko and {Thomas}, Brooks and {Correa}, Carlos and {Boiza}, Carlos G. and {Marques}, Catarina and {Escamilla-Rivera}, Celia and {Tzerefos}, Charalampos and {Zhang}, Chi and {De Leo}, Chiara and {Pfeifer}, Christian and {Lee}, Christine and {Venter}, Christo and {Gomes}, Cl{\'a}udio and {Roque De bom}, Clecio and {Moreno-Pulido}, Cristian and {Iosifidis}, Damianos and {Grin}, Dan and {Blixt}, Daniel and {Scolnic}, Dan and {Oriti}, Daniele and {Dobrycheva}, Daria and {Bettoni}, Dario and {Benisty}, David and {Fern{\'a}ndez-Arenas}, David and {Wiltshire}, David L. and {Sanchez Cid}, David and {Tamayo}, David and {Valls-Gabaud}, David and {Pedrotti}, Davide and {Wang}, Deng and {Staicova}, Denitsa and {Totolou}, Despoina and {Rubiera-Garcia}, Diego and {Milakovi{\'c}}, Dinko and {Pesce}, Dominic W. and {Sluse}, Dominique and {Borka}, Du{\v{s}}ko and {Yusofi}, Ebrahim and {Giusarma}, Elena and {Terlevich}, Elena and {Tomasetti}, Elena and {Vagenas}, Elias C. and {Fazzari}, Elisa and {Ferreira}, Elisa G.~M. and {Barakovic}, Elvis and {Dimastrogiovanni}, Emanuela and {Holm}, Emil Brinch and {Mottola}, Emil and {{\"O}z{\"u}lker}, Emre and {Specogna}, Enrico and {Brocato}, Enzo and {Jensko}, Erik and {Enriquez}, Erika Antonette and {Bhatia}, Esha and {Bresolin}, Fabio and {Avila}, Felipe and {Bouch{\`e}}, Filippo and {Bombacigno}, Flavio and {Anagnostopoulos}, Fotios K. and {Pace}, Francesco and {Sorrenti}, Francesco and {Lobo}, Francisco S.~N. and {Courbin}, Fr{\'e}d{\'e}ric and {Hansen}, Frode K. and {Sloan}, Greg and {Farrugia}, Gabriel and {Lynch}, Gabriel and {Garcia-Arroyo}, Gabriela and {Raimondo}, Gabriella and {Lambiase}, Gaetano and {Anand}, Gagandeep S. and {Poulot}, Gaspard and {Leon}, Genly and {Kouniatalis}, Gerasimos and {Nardini}, Germano and {Cs{\"o}rnyei}, G{\'e}za and {Galloni}, Giacomo},
 doi = {10.1016/j.dark.2025.101965},
 eid = {101965},
 eprint = {2504.01669},
 journal = {Physics of the Dark Universe},
 month = {September},
 pages = {101965},
 primaryclass = {astro-ph.CO},
 title = {{The CosmoVerse White Paper: Addressing observational tensions in cosmology with systematics and fundamental physics}},
 volume = {49},
 year = {2025}
}

@article{Doux2021,
 adsurl = {https://ui.adsabs.harvard.edu/abs/2021MNRAS.503.3796D},
 archiveprefix = {arXiv},
 author = {{Doux}, C. and {Chang}, C. and {Jain}, B. and {Blazek}, J. and {Camacho}, H. and {Fang}, X. and {Gatti}, M. and {Krause}, E. and {MacCrann}, N. and {Samuroff}, S. and {Secco}, L.~F. and {Troxel}, M.~A. and {Zuntz}, J. and {Aguena}, M. and {Allam}, S. and {Amon}, A. and {Avila}, S. and {Bacon}, D. and {Bertin}, E. and {Brooks}, D. and {Burke}, D.~L. and {Carnero Rosell}, A. and {Carrasco Kind}, M. and {Carretero}, J. and {Choi}, A. and {Costanzi}, M. and {Crocce}, M. and {da Costa}, L.~N. and {Pereira}, M.~E.~S. and {Davis}, T.~M. and {Dietrich}, J.~P. and {Doel}, P. and {Ferrero}, I. and {Fert{\'e}}, A. and {Fosalba}, P. and {Garc{\'\i}a-Bellido}, J. and {Gaztanaga}, E. and {Gerdes}, D.~W. and {Gruen}, D. and {Gruendl}, R.~A. and {Gschwend}, J. and {Gutierrez}, G. and {Hartley}, W.~G. and {Hinton}, S.~R. and {Hollowood}, D.~L. and {Huterer}, D. and {James}, D.~J. and {Kuehn}, K. and {Kuropatkin}, N. and {Maia}, M.~A.~G. and {Marshall}, J.~L. and {Menanteau}, F. and {Miquel}, R. and {Morgan}, R. and {Palmese}, A. and {Paz-Chinch{\'o}n}, F. and {Plazas}, A.~A. and {Roodman}, A. and {Sanchez}, E. and {Schubnell}, M. and {Serrano}, S. and {Sevilla-Noarbe}, I. and {Smith}, M. and {Soares-Santos}, M. and {Suchyta}, E. and {Tarle}, G. and {To}, C. and {Varga}, T.~N. and {Weller}, J. and {Wilkinson}, R.~D. and {DES Collaboration}},
 doi = {10.1093/mnras/stab661},
 eprint = {2011.06469},
 journal = {\mnras},
 month = {May},
 number = {3},
 pages = {3796-3817},
 primaryclass = {astro-ph.CO},
 title = {{Consistency of cosmic shear analyses in harmonic and real space}},
 volume = {503},
 year = {2021}
}

@article{Doux2022,
 adsurl = {https://ui.adsabs.harvard.edu/abs/2022MNRAS.515.1942D},
 archiveprefix = {arXiv},
 author = {{Doux}, C. and {Jain}, B. and {Zeurcher}, D. and {Lee}, J. and {Fang}, X. and {Rosenfeld}, R. and {Amon}, A. and {Camacho}, H. and {Choi}, A. and {Secco}, L.~F. and {Blazek}, J. and {Chang}, C. and {Gatti}, M. and {Gaztanaga}, E. and {Jeffrey}, N. and {Raveri}, M. and {Samuroff}, S. and {Alarcon}, A. and {Alves}, O. and {Andrade-Oliveira}, F. and {Baxter}, E. and {Bechtol}, K. and {Becker}, M.~R. and {Bernstein}, G.~M. and {Campos}, A. and {Carnero Rosell}, A. and {Carrasco Kind}, M. and {Cawthon}, R. and {Chen}, R. and {Cordero}, J. and {Crocce}, M. and {Davis}, C. and {DeRose}, J. and {Dodelson}, S. and {Drlica-Wagner}, A. and {Eckert}, K. and {Eifler}, T.~F. and {Elsner}, F. and {Elvin-Poole}, J. and {Everett}, S. and {Fert{\'e}}, A. and {Fosalba}, P. and {Friedrich}, O. and {Giannini}, G. and {Gruen}, D. and {Gruendl}, R.~A. and {Harrison}, I. and {Hartley}, W.~G. and {Herner}, K. and {Huang}, H. and {Huff}, E.~M. and {Huterer}, D. and {Jarvis}, M. and {Krause}, E. and {Kuropatkin}, N. and {Leget}, P. -F. and {Lemos}, P. and {Liddle}, A.~R. and {MacCrann}, N. and {McCullough}, J. and {Muir}, J. and {Myles}, J. and {Navarro-Alsina}, A. and {Pandey}, S. and {Park}, Y. and {Porredon}, A. and {Prat}, J. and {Rodriguez-Monroy}, M. and {Rollins}, R.~P. and {Roodman}, A. and {Ross}, A.~J. and {Rykoff}, E.~S. and {S{\'a}nchez}, C. and {Sanchez}, J. and {Sevilla-Noarbe}, I. and {Sheldon}, E. and {Shin}, T. and {Troja}, A. and {Troxel}, M.~A. and {Tutusaus}, I. and {Varga}, T.~N. and {Weaverdyck}, N. and {Wechsler}, R.~H. and {Yanny}, B. and {Yin}, B. and {Zhang}, Y. and {Zuntz}, J. and {Abbott}, T.~M.~C. and {Aguena}, M. and {Allam}, S. and {Annis}, J. and {Bacon}, D. and {Bertin}, E. and {Bocquet}, S. and {Brooks}, D. and {Burke}, D.~L. and {Carretero}, J. and {Costanzi}, M. and {da Costa}, L.~N. and {Pereira}, M.~E.~S. and {De Vicente}, J. and {Desai}, S. and {Diehl}, H.~T. and {Doel}, P. and {Ferrero}, I. and {Flaugher}, B. and {Frieman}, J. and {Garc{\'\i}a-Bellido}, J. and {Gerdes}, D.~W. and {Giannantonio}, T. and {Gschwend}, J. and {Gutierrez}, G. and {Hinton}, S.~R. and {Hollowood}, D.~L. and {Honscheid}, K. and {James}, D.~J. and {Kim}, A.~G. and {Kuehn}, K. and {Lahav}, O. and {Marshall}, J.~L. and {Menanteau}, F. and {Miquel}, R. and {Morgan}, R. and {Ogando}, R.~L.~C. and {Palmese}, A. and {Paz-Chinch{\'o}n}, F. and {Pieres}, A. and {Plazas Malag{\'o}n}, A.~A. and {Reil}, K. and {Sanchez}, E. and {Scarpine}, V. and {Serrano}, S. and {Smith}, M. and {Suchyta}, E. and {Swanson}, M.~E.~C. and {Tarle}, G. and {Thomas}, D. and {To}, C. and {Weller}, J. and {DES Collaboration}},
 doi = {10.1093/mnras/stac1826},
 eprint = {2203.07128},
 journal = {\mnras},
 month = {September},
 number = {2},
 pages = {1942-1972},
 primaryclass = {astro-ph.CO},
 title = {{Dark energy survey year 3 results: cosmological constraints from the analysis of cosmic shear in harmonic space}},
 volume = {515},
 year = {2022}
}

@article{Doux2026,
 adsurl = {https://ui.adsabs.harvard.edu/abs/2026OJAp....955045D},
 archiveprefix = {arXiv},
 author = {{Doux}, Cyrille and {Karwal}, Tanvi},
 doi = {10.33232/001c.155045},
 eprint = {2506.16434},
 journal = {The Open Journal of Astrophysics},
 month = {January},
 pages = {55045},
 primaryclass = {astro-ph.CO},
 title = {{Going beyond $S_8$: fast inference of the matter power spectrum from weak-lensing surveys}},
 volume = {9},
 year = {2026}
}

@article{Duffy2008,
 adsurl = {https://ui.adsabs.harvard.edu/abs/2008MNRAS.390L..64D},
 archiveprefix = {arXiv},
 author = {{Duffy}, Alan R. and {Schaye}, Joop and {Kay}, Scott T. and {Dalla Vecchia}, Claudio},
 doi = {10.1111/j.1745-3933.2008.00537.x},
 eprint = {0804.2486},
 journal = {\mnras},
 month = {October},
 number = {1},
 pages = {L64-L68},
 primaryclass = {astro-ph},
 title = {{Dark matter halo concentrations in the Wilkinson Microwave Anisotropy Probe year 5 cosmology}},
 volume = {390},
 year = {2008}
}

@article{Feng2020,
 adsurl = {https://ui.adsabs.harvard.edu/abs/2020ApJ...897..140F},
 archiveprefix = {arXiv},
 author = {{Feng}, Chang and {Holder}, Gilbert},
 doi = {10.3847/1538-4357/ab9013},
 eid = {140},
 eprint = {1907.12085},
 journal = {\apj},
 month = {July},
 number = {2},
 pages = {140},
 primaryclass = {astro-ph.CO},
 title = {{Polarization of the Cosmic Infrared Background Fluctuations}},
 volume = {897},
 year = {2020}
}

@article{Flaugher2015,
 adsurl = {https://ui.adsabs.harvard.edu/abs/2015AJ....150..150F},
 archiveprefix = {arXiv},
 author = {{Flaugher}, B. and {Diehl}, H.~T. and {Honscheid}, K. and {Abbott}, T.~M.~C. and {Alvarez}, O. and {Angstadt}, R. and {Annis}, J.~T. and {Antonik}, M. and {Ballester}, O. and {Beaufore}, L. and {Bernstein}, G.~M. and {Bernstein}, R.~A. and {Bigelow}, B. and {Bonati}, M. and {Boprie}, D. and {Brooks}, D. and {Buckley-Geer}, E.~J. and {Campa}, J. and {Cardiel-Sas}, L. and {Castander}, F.~J. and {Castilla}, J. and {Cease}, H. and {Cela-Ruiz}, J.~M. and {Chappa}, S. and {Chi}, E. and {Cooper}, C. and {da Costa}, L.~N. and {Dede}, E. and {Derylo}, G. and {DePoy}, D.~L. and {de Vicente}, J. and {Doel}, P. and {Drlica-Wagner}, A. and {Eiting}, J. and {Elliott}, A.~E. and {Emes}, J. and {Estrada}, J. and {Fausti Neto}, A. and {Finley}, D.~A. and {Flores}, R. and {Frieman}, J. and {Gerdes}, D. and {Gladders}, M.~D. and {Gregory}, B. and {Gutierrez}, G.~R. and {Hao}, J. and {Holland}, S.~E. and {Holm}, S. and {Huffman}, D. and {Jackson}, C. and {James}, D.~J. and {Jonas}, M. and {Karcher}, A. and {Karliner}, I. and {Kent}, S. and {Kessler}, R. and {Kozlovsky}, M. and {Kron}, R.~G. and {Kubik}, D. and {Kuehn}, K. and {Kuhlmann}, S. and {Kuk}, K. and {Lahav}, O. and {Lathrop}, A. and {Lee}, J. and {Levi}, M.~E. and {Lewis}, P. and {Li}, T.~S. and {Mandrichenko}, I. and {Marshall}, J.~L. and {Martinez}, G. and {Merritt}, K.~W. and {Miquel}, R. and {Mu{\~n}oz}, F. and {Neilsen}, E.~H. and {Nichol}, R.~C. and {Nord}, B. and {Ogando}, R. and {Olsen}, J. and {Palaio}, N. and {Patton}, K. and {Peoples}, J. and {Plazas}, A.~A. and {Rauch}, J. and {Reil}, K. and {Rheault}, J. -P. and {Roe}, N.~A. and {Rogers}, H. and {Roodman}, A. and {Sanchez}, E. and {Scarpine}, V. and {Schindler}, R.~H. and {Schmidt}, R. and {Schmitt}, R. and {Schubnell}, M. and {Schultz}, K. and {Schurter}, P. and {Scott}, L. and {Serrano}, S. and {Shaw}, T.~M. and {Smith}, R.~C. and {Soares-Santos}, M. and {Stefanik}, A. and {Stuermer}, W. and {Suchyta}, E. and {Sypniewski}, A. and {Tarle}, G. and {Thaler}, J. and {Tighe}, R. and {Tran}, C. and {Tucker}, D. and {Walker}, A.~R. and {Wang}, G. and {Watson}, M. and {Weaverdyck}, C. and {Wester}, W. and {Woods}, R. and {Yanny}, B. and {DES Collaboration}},
 doi = {10.1088/0004-6256/150/5/150},
 eid = {150},
 eprint = {1504.02900},
 journal = {\aj},
 month = {November},
 number = {5},
 pages = {150},
 primaryclass = {astro-ph.IM},
 title = {{The Dark Energy Camera}},
 volume = {150},
 year = {2015}
}

@article{Fortuna2021,
 adsurl = {https://ui.adsabs.harvard.edu/abs/2021MNRAS.501.2983F},
 archiveprefix = {arXiv},
 author = {{Fortuna}, Maria Cristina and {Hoekstra}, Henk and {Joachimi}, Benjamin and {Johnston}, Harry and {Chisari}, Nora Elisa and {Georgiou}, Christos and {Mahony}, Constance},
 doi = {10.1093/mnras/staa3802},
 eprint = {2003.02700},
 journal = {\mnras},
 month = {February},
 number = {2},
 pages = {2983-3002},
 primaryclass = {astro-ph.CO},
 title = {{The halo model as a versatile tool to predict intrinsic alignments}},
 volume = {501},
 year = {2021}
}

@article{Garcia-Garcia2019,
 adsurl = {https://ui.adsabs.harvard.edu/abs/2019JCAP...11..043G},
 archiveprefix = {arXiv},
 author = {{Garc{\'\i}a-Garc{\'\i}a}, Carlos and {Alonso}, David and {Bellini}, Emilio},
 doi = {10.1088/1475-7516/2019/11/043},
 eid = {043},
 eprint = {1906.11765},
 journal = {\jcap},
 month = {November},
 number = {11},
 pages = {043},
 primaryclass = {astro-ph.CO},
 title = {{Disconnected pseudo-C$_{l}$ covariances for projected large-scale structure data}},
 volume = {2019},
 year = {2019}
}

@article{Gatti2021,
 adsurl = {https://ui.adsabs.harvard.edu/abs/2021MNRAS.504.4312G},
 archiveprefix = {arXiv},
 author = {{Gatti}, M. and {Sheldon}, E. and {Amon}, A. and {Becker}, M. and {Troxel}, M. and {Choi}, A. and {Doux}, C. and {MacCrann}, N. and {Navarro-Alsina}, A. and {Harrison}, I. and {Gruen}, D. and {Bernstein}, G. and {Jarvis}, M. and {Secco}, L.~F. and {Fert{\'e}}, A. and {Shin}, T. and {McCullough}, J. and {Rollins}, R.~P. and {Chen}, R. and {Chang}, C. and {Pandey}, S. and {Tutusaus}, I. and {Prat}, J. and {Elvin-Poole}, J. and {Sanchez}, C. and {Plazas}, A.~A. and {Roodman}, A. and {Zuntz}, J. and {Abbott}, T.~M.~C. and {Aguena}, M. and {Allam}, S. and {Annis}, J. and {Avila}, S. and {Bacon}, D. and {Bertin}, E. and {Bhargava}, S. and {Brooks}, D. and {Burke}, D.~L. and {Carnero Rosell}, A. and {Carrasco Kind}, M. and {Carretero}, J. and {Castander}, F.~J. and {Conselice}, C. and {Costanzi}, M. and {Crocce}, M. and {da Costa}, L.~N. and {Davis}, T.~M. and {De Vicente}, J. and {Desai}, S. and {Diehl}, H.~T. and {Dietrich}, J.~P. and {Doel}, P. and {Drlica-Wagner}, A. and {Eckert}, K. and {Everett}, S. and {Ferrero}, I. and {Frieman}, J. and {Garc{\'\i}a-Bellido}, J. and {Gerdes}, D.~W. and {Giannantonio}, T. and {Gruendl}, R.~A. and {Gschwend}, J. and {Gutierrez}, G. and {Hartley}, W.~G. and {Hinton}, S.~R. and {Hollowood}, D.~L. and {Honscheid}, K. and {Hoyle}, B. and {Huff}, E.~M. and {Huterer}, D. and {Jain}, B. and {James}, D.~J. and {Jeltema}, T. and {Krause}, E. and {Kron}, R. and {Kuropatkin}, N. and {Lima}, M. and {Maia}, M.~A.~G. and {Marshall}, J.~L. and {Miquel}, R. and {Morgan}, R. and {Myles}, J. and {Palmese}, A. and {Paz-Chinch{\'o}n}, F. and {Rykoff}, E.~S. and {Samuroff}, S. and {Sanchez}, E. and {Scarpine}, V. and {Schubnell}, M. and {Serrano}, S. and {Sevilla-Noarbe}, I. and {Smith}, M. and {Soares-Santos}, M. and {Suchyta}, E. and {Swanson}, M.~E.~C. and {Tarle}, G. and {Thomas}, D. and {To}, C. and {Tucker}, D.~L. and {Varga}, T.~N. and {Wechsler}, R.~H. and {Weller}, J. and {Wester}, W. and {Wilkinson}, R.~D.},
 doi = {10.1093/mnras/stab918},
 eprint = {2011.03408},
 journal = {\mnras},
 month = {July},
 number = {3},
 pages = {4312-4336},
 primaryclass = {astro-ph.CO},
 title = {{Dark energy survey year 3 results: weak lensing shape catalogue}},
 volume = {504},
 year = {2021}
}

@article{Ge2025,
 adsurl = {https://ui.adsabs.harvard.edu/abs/2025PhRvD.111h3534G},
 archiveprefix = {arXiv},
 author = {{Ge}, F. and {Millea}, M. and {Camphuis}, E. and {Daley}, C. and {Huang}, N. and {Omori}, Y. and {Quan}, W. and {Anderes}, E. and {Anderson}, A.~J. and {Ansarinejad}, B. and {Archipley}, M. and {Balkenhol}, L. and {Benabed}, K. and {Bender}, A.~N. and {Benson}, B.~A. and {Bianchini}, F. and {Bleem}, L.~E. and {Bouchet}, F.~R. and {Bryant}, L. and {Carlstrom}, J.~E. and {Chang}, C.~L. and {Chaubal}, P. and {Chen}, G. and {Chichura}, P.~M. and {Chokshi}, A. and {Chou}, T. -L. and {Coerver}, A. and {Crawford}, T.~M. and {de Haan}, T. and {Dibert}, K.~R. and {Dobbs}, M.~A. and {Doohan}, M. and {Doussot}, A. and {Dutcher}, D. and {Everett}, W. and {Feng}, C. and {Ferguson}, K.~R. and {Fichman}, K. and {Foster}, A. and {Galli}, S. and {Gambrel}, A.~E. and {Gardner}, R.~W. and {Goeckner-Wald}, N. and {Gualtieri}, R. and {Guidi}, F. and {Guns}, S. and {Halverson}, N.~W. and {Hivon}, E. and {Holder}, G.~P. and {Holzapfel}, W.~L. and {Hood}, J.~C. and {Howe}, D. and {Hryciuk}, A. and {K{\'e}ruzor{\'e}}, F. and {Khalife}, A.~R. and {Knox}, L. and {Korman}, M. and {Kornoelje}, K. and {Kuo}, C. -L. and {Lee}, A.~T. and {Levy}, K. and {Lowitz}, A.~E. and {Lu}, C. and {Maniyar}, A. and {Martsen}, E.~S. and {Menanteau}, F. and {Montgomery}, J. and {Nakato}, Y. and {Natoli}, T. and {Noble}, G.~I. and {Pan}, Z. and {Paschos}, P. and {Phadke}, K.~A. and {Pollak}, A.~W. and {Prabhu}, K. and {Rahimi}, M. and {Rahlin}, A. and {Reichardt}, C.~L. and {Riebel}, D. and {Rouble}, M. and {Ruhl}, J.~E. and {Schiappucci}, E. and {Sobrin}, J.~A. and {Stark}, A.~A. and {Stephen}, J. and {Tandoi}, C. and {Thorne}, B. and {Trendafilova}, C. and {Umilta}, C. and {Vieira}, J.~D. and {Vitrier}, A. and {Wan}, Y. and {Whitehorn}, N. and {Wu}, W.~L.~K. and {Young}, M.~R. and {Zebrowski}, J.~A. and {SPT-3G Collaboration}},
 doi = {10.1103/PhysRevD.111.083534},
 eid = {083534},
 eprint = {2411.06000},
 journal = {\prd},
 month = {April},
 number = {8},
 pages = {083534},
 primaryclass = {astro-ph.CO},
 title = {{Cosmology from CMB lensing and delensed EE power spectra using 2019{\textendash}2020 SPT-3G polarization data}},
 volume = {111},
 year = {2025}
}

@article{Gomes2025,
 adsurl = {https://ui.adsabs.harvard.edu/abs/2025PhRvD.112l3515G},
 archiveprefix = {arXiv},
 author = {{Gomes}, R.~C.~H. and {Sugiyama}, S. and {Jain}, B. and {Jarvis}, M. and {Anbajagane}, D. and {Halder}, A. and {Marques}, G.~A. and {Pandey}, S. and {Marshall}, J. and {Alarcon}, A. and {Amon}, A. and {Bechtol}, K. and {Becker}, M. and {Bernstein}, G. and {Campos}, A. and {Cawthon}, R. and {Chang}, C. and {Chen}, R. and {Choi}, A. and {Cordero}, J. and {Davis}, C. and {Derose}, J. and {Dodelson}, S. and {Doux}, C. and {Eckert}, K. and {Elsner}, F. and {Elvin-Poole}, J. and {Everett}, S. and {Fert{\'e}}, A. and {Gatti}, M. and {Giannini}, G. and {Gruen}, D. and {Harrison}, I. and {Herner}, K. and {Huff}, E.~M. and {Huterer}, D. and {Kuropatkin}, N. and {Leget}, P.~F. and {Maccrann}, N. and {Mccullough}, J. and {Muir}, J. and {Myles}, J. and {Navarro Alsina}, A. and {Prat}, J. and {Raveri}, M. and {Rollins}, R.~P. and {Roodman}, A. and {Ross}, A.~J. and {Rykoff}, E.~S. and {S{\'a}nchez}, C. and {Secco}, L.~F. and {Sheldon}, E. and {Shin}, T. and {Troxel}, M. and {Tutusaus}, I. and {Varga}, T.~N. and {Yanny}, B. and {Yin}, B. and {Zhang}, Y. and {Zuntz}, J. and {Aguena}, M. and {Andrade-Oliveira}, F. and {Bacon}, D. and {Blazek}, J. and {Bocquet}, S. and {Brooks}, D. and {Carnero Rosell}, A. and {Carretero}, J. and {Costanzi}, M. and {da Costa}, L. and {da Silva Pereira}, M.~E. and {Davis}, T.~M. and {De Vicente}, J. and {Diehl}, H.~T. and {Flaugher}, B. and {Frieman}, J. and {Gutierrez}, G. and {Hinton}, S.~R. and {Hollowood}, D.~L. and {Honscheid}, K. and {James}, D.~J. and {Jeffrey}, N. and {Lee}, S. and {Mena-Fern{\'a}ndez}, J. and {Miquel}, R. and {Ogando}, R.~L.~C. and {Plazas Malag{\'o}n}, A.~A. and {Porredon}, A. and {Sanchez}, E. and {Sanchez Cid}, D. and {Samuroff}, S. and {Smith}, M. and {Suchyta}, E. and {Swanson}, M.~E.~C. and {Thomas}, D. and {Vikram}, V. and {Weller}, J. and {Yamamoto}, M. and {DES Collaboration}},
 doi = {10.1103/sxlz-t9gb},
 eid = {123515},
 eprint = {2508.14018},
 journal = {\prd},
 month = {December},
 number = {12},
 pages = {123515},
 primaryclass = {astro-ph.CO},
 title = {{Dark Energy Survey Year 3 Results: Cosmological constraints from second- and third-order shear statistics}},
 volume = {112},
 year = {2025}
}

@article{Gorski2005,
 adsurl = {https://ui.adsabs.harvard.edu/abs/2005ApJ...622..759G},
 archiveprefix = {arXiv},
 author = {{G{\'o}rski}, K.~M. and {Hivon}, E. and {Banday}, A.~J. and {Wandelt}, B.~D. and {Hansen}, F.~K. and {Reinecke}, M. and {Bartelmann}, M.},
 doi = {10.1086/427976},
 eprint = {astro-ph/0409513},
 journal = {\apj},
 month = {April},
 number = {2},
 pages = {759-771},
 primaryclass = {astro-ph},
 title = {{HEALPix: A Framework for High-Resolution Discretization and Fast Analysis of Data Distributed on the Sphere}},
 volume = {622},
 year = {2005}
}

@article{Han2025,
 adsurl = {https://ui.adsabs.harvard.edu/abs/2025ApJ...990..197H},
 archiveprefix = {arXiv},
 author = {{Han}, Sanghyeon and {Tonegawa}, Motonari and {Hwang}, Ho Seong and {Dubois}, Yohan and {Kim}, Juhan and {Kim}, Yonghwi and {Kwon}, Oh-Kyoung and {Lee}, Jaehyun and {Snaith}, Owain N. and {Gibson}, Brad K. and {Park}, Changbom},
 doi = {10.3847/1538-4357/adf33b},
 eid = {197},
 eprint = {2505.10840},
 journal = {\apj},
 month = {September},
 number = {2},
 pages = {197},
 primaryclass = {astro-ph.CO},
 title = {{Redshift Evolution of the Intrinsic Alignments of Early-type Galaxies and Subhalos in the Horizon Run 5 Simulation}},
 volume = {990},
 year = {2025}
}

@article{Harris2020,
 adsurl = {https://ui.adsabs.harvard.edu/abs/2020Natur.585..357H},
 archiveprefix = {arXiv},
 author = {{Harris}, Charles R. and {Millman}, K. Jarrod and {van der Walt}, St{\'e}fan J. and {Gommers}, Ralf and {Virtanen}, Pauli and {Cournapeau}, David and {Wieser}, Eric and {Taylor}, Julian and {Berg}, Sebastian and {Smith}, Nathaniel J. and {Kern}, Robert and {Picus}, Matti and {Hoyer}, Stephan and {van Kerkwijk}, Marten H. and {Brett}, Matthew and {Haldane}, Allan and {del R{\'\i}o}, Jaime Fern{\'a}ndez and {Wiebe}, Mark and {Peterson}, Pearu and {G{\'e}rard-Marchant}, Pierre and {Sheppard}, Kevin and {Reddy}, Tyler and {Weckesser}, Warren and {Abbasi}, Hameer and {Gohlke}, Christoph and {Oliphant}, Travis E.},
 doi = {10.1038/s41586-020-2649-2},
 eprint = {2006.10256},
 journal = {\nat},
 month = {September},
 number = {7825},
 pages = {357-362},
 primaryclass = {cs.MS},
 title = {{Array programming with NumPy}},
 volume = {585},
 year = {2020}
}

@article{He2023,
 adsurl = {https://ui.adsabs.harvard.edu/abs/2023ApJ...954L...8H},
 archiveprefix = {arXiv},
 author = {{He}, Adam and {Ivanov}, Mikhail M. and {An}, Rui and {Gluscevic}, Vera},
 doi = {10.3847/2041-8213/acdb63},
 eid = {L8},
 eprint = {2301.08260},
 journal = {\apjl},
 month = {September},
 number = {1},
 pages = {L8},
 primaryclass = {astro-ph.CO},
 title = {{S$_{8}$ Tension in the Context of Dark Matter-Baryon Scattering}},
 volume = {954},
 year = {2023}
}

@article{Hirata2003,
 adsurl = {https://ui.adsabs.harvard.edu/abs/2003PhRvD..68h3002H},
 archiveprefix = {arXiv},
 author = {{Hirata}, Christopher M. and {Seljak}, Uro{\v{s}}},
 doi = {10.1103/PhysRevD.68.083002},
 eid = {083002},
 eprint = {astro-ph/0306354},
 journal = {\prd},
 month = {October},
 number = {8},
 pages = {083002},
 primaryclass = {astro-ph},
 title = {{Reconstruction of lensing from the cosmic microwave background polarization}},
 volume = {68},
 year = {2003}
}

@article{Hirata2003b,
 adsurl = {https://ui.adsabs.harvard.edu/abs/2003PhRvD..67d3001H},
 archiveprefix = {arXiv},
 author = {{Hirata}, Christopher M. and {Seljak}, Uro{\v{s}}},
 doi = {10.1103/PhysRevD.67.043001},
 eid = {043001},
 eprint = {astro-ph/0209489},
 journal = {\prd},
 month = {February},
 number = {4},
 pages = {043001},
 primaryclass = {astro-ph},
 title = {{Analyzing weak lensing of the cosmic microwave background using the likelihood function}},
 volume = {67},
 year = {2003}
}

@article{Hirata2004,
 adsurl = {https://ui.adsabs.harvard.edu/abs/2004PhRvD..70f3526H},
 archiveprefix = {arXiv},
 author = {{Hirata}, Christopher M. and {Seljak}, Uro{\v{s}}},
 doi = {10.1103/PhysRevD.70.063526},
 eid = {063526},
 eprint = {astro-ph/0406275},
 journal = {\prd},
 month = {September},
 number = {6},
 pages = {063526},
 primaryclass = {astro-ph},
 title = {{Intrinsic alignment-lensing interference as a contaminant of cosmic shear}},
 volume = {70},
 year = {2004}
}

@article{Hivon2002,
 adsurl = {https://ui.adsabs.harvard.edu/abs/2002ApJ...567....2H},
 archiveprefix = {arXiv},
 author = {{Hivon}, Eric and {G{\'o}rski}, Krzysztof M. and {Netterfield}, C. Barth and {Crill}, Brendan P. and {Prunet}, Simon and {Hansen}, Frode},
 doi = {10.1086/338126},
 eprint = {astro-ph/0105302},
 journal = {\apj},
 month = {March},
 number = {1},
 pages = {2-17},
 primaryclass = {astro-ph},
 title = {{MASTER of the Cosmic Microwave Background Anisotropy Power Spectrum: A Fast Method for Statistical Analysis of Large and Complex Cosmic Microwave Background Data Sets}},
 volume = {567},
 year = {2002}
}

@article{Hu2002,
 adsurl = {https://ui.adsabs.harvard.edu/abs/2002ApJ...574..566H},
 archiveprefix = {arXiv},
 author = {{Hu}, Wayne and {Okamoto}, Takemi},
 doi = {10.1086/341110},
 eprint = {astro-ph/0111606},
 journal = {\apj},
 month = {August},
 number = {2},
 pages = {566-574},
 primaryclass = {astro-ph},
 title = {{Mass Reconstruction with Cosmic Microwave Background Polarization}},
 volume = {574},
 year = {2002}
}

@article{Huff2017,
 adsurl = {https://ui.adsabs.harvard.edu/abs/2017arXiv170202600H},
 archiveprefix = {arXiv},
 author = {{Huff}, Eric and {Mandelbaum}, Rachel},
 doi = {10.48550/arXiv.1702.02600},
 eid = {arXiv:1702.02600},
 eprint = {1702.02600},
 journal = {arXiv e-prints},
 month = {February},
 pages = {arXiv:1702.02600},
 primaryclass = {astro-ph.CO},
 title = {{Metacalibration: Direct Self-Calibration of Biases in Shear Measurement}},
 year = {2017}
}

@article{Hunter2007,
 adsurl = {https://ui.adsabs.harvard.edu/abs/2007CSE.....9...90H},
 author = {{Hunter}, John D.},
 doi = {10.1109/MCSE.2007.55},
 journal = {Computing in Science and Engineering},
 month = {January},
 number = {3},
 pages = {90-95},
 title = {{Matplotlib: A 2D Graphics Environment}},
 volume = {9},
 year = {2007}
}

@article{Ivezic2019,
 adsurl = {https://ui.adsabs.harvard.edu/abs/2019ApJ...873..111I},
 archiveprefix = {arXiv},
 author = {{Ivezi{\'c}}, {\v{Z}}eljko and {Kahn}, Steven M. and {Tyson}, J. Anthony and {Abel}, Bob and {Acosta}, Emily and {Allsman}, Robyn and {Alonso}, David and {AlSayyad}, Yusra and {Anderson}, Scott F. and {Andrew}, John and {Angel}, James Roger P. and {Angeli}, George Z. and {Ansari}, Reza and {Antilogus}, Pierre and {Araujo}, Constanza and {Armstrong}, Robert and {Arndt}, Kirk T. and {Astier}, Pierre and {Aubourg}, {\'E}ric and {Auza}, Nicole and {Axelrod}, Tim S. and {Bard}, Deborah J. and {Barr}, Jeff D. and {Barrau}, Aurelian and {Bartlett}, James G. and {Bauer}, Amanda E. and {Bauman}, Brian J. and {Baumont}, Sylvain and {Bechtol}, Ellen and {Bechtol}, Keith and {Becker}, Andrew C. and {Becla}, Jacek and {Beldica}, Cristina and {Bellavia}, Steve and {Bianco}, Federica B. and {Biswas}, Rahul and {Blanc}, Guillaume and {Blazek}, Jonathan and {Blandford}, Roger D. and {Bloom}, Josh S. and {Bogart}, Joanne and {Bond}, Tim W. and {Booth}, Michael T. and {Borgland}, Anders W. and {Borne}, Kirk and {Bosch}, James F. and {Boutigny}, Dominique and {Brackett}, Craig A. and {Bradshaw}, Andrew and {Brandt}, William Nielsen and {Brown}, Michael E. and {Bullock}, James S. and {Burchat}, Patricia and {Burke}, David L. and {Cagnoli}, Gianpietro and {Calabrese}, Daniel and {Callahan}, Shawn and {Callen}, Alice L. and {Carlin}, Jeffrey L. and {Carlson}, Erin L. and {Chandrasekharan}, Srinivasan and {Charles-Emerson}, Glenaver and {Chesley}, Steve and {Cheu}, Elliott C. and {Chiang}, Hsin-Fang and {Chiang}, James and {Chirino}, Carol and {Chow}, Derek and {Ciardi}, David R. and {Claver}, Charles F. and {Cohen-Tanugi}, Johann and {Cockrum}, Joseph J. and {Coles}, Rebecca and {Connolly}, Andrew J. and {Cook}, Kem H. and {Cooray}, Asantha and {Covey}, Kevin R. and {Cribbs}, Chris and {Cui}, Wei and {Cutri}, Roc and {Daly}, Philip N. and {Daniel}, Scott F. and {Daruich}, Felipe and {Daubard}, Guillaume and {Daues}, Greg and {Dawson}, William and {Delgado}, Francisco and {Dellapenna}, Alfred and {de Peyster}, Robert and {de Val-Borro}, Miguel and {Digel}, Seth W. and {Doherty}, Peter and {Dubois}, Richard and {Dubois-Felsmann}, Gregory P. and {Durech}, Josef and {Economou}, Frossie and {Eifler}, Tim and {Eracleous}, Michael and {Emmons}, Benjamin L. and {Fausti Neto}, Angelo and {Ferguson}, Henry and {Figueroa}, Enrique and {Fisher-Levine}, Merlin and {Focke}, Warren and {Foss}, Michael D. and {Frank}, James and {Freemon}, Michael D. and {Gangler}, Emmanuel and {Gawiser}, Eric and {Geary}, John C. and {Gee}, Perry and {Geha}, Marla and {Gessner}, Charles J.~B. and {Gibson}, Robert R. and {Gilmore}, D. Kirk and {Glanzman}, Thomas and {Glick}, William and {Goldina}, Tatiana and {Goldstein}, Daniel A. and {Goodenow}, Iain and {Graham}, Melissa L. and {Gressler}, William J. and {Gris}, Philippe and {Guy}, Leanne P. and {Guyonnet}, Augustin and {Haller}, Gunther and {Harris}, Ron and {Hascall}, Patrick A. and {Haupt}, Justine and {Hernandez}, Fabio and {Herrmann}, Sven and {Hileman}, Edward and {Hoblitt}, Joshua and {Hodgson}, John A. and {Hogan}, Craig and {Howard}, James D. and {Huang}, Dajun and {Huffer}, Michael E. and {Ingraham}, Patrick and {Innes}, Walter R. and {Jacoby}, Suzanne H. and {Jain}, Bhuvnesh and {Jammes}, Fabrice and {Jee}, M. James and {Jenness}, Tim and {Jernigan}, Garrett and {Jevremovi{\'c}}, Darko and {Johns}, Kenneth and {Johnson}, Anthony S. and {Johnson}, Margaret W.~G. and {Jones}, R. Lynne and {Juramy-Gilles}, Claire and {Juri{\'c}}, Mario and {Kalirai}, Jason S. and {Kallivayalil}, Nitya J. and {Kalmbach}, Bryce and {Kantor}, Jeffrey P. and {Karst}, Pierre and {Kasliwal}, Mansi M. and {Kelly}, Heather and {Kessler}, Richard and {Kinnison}, Veronica and {Kirkby}, David and {Knox}, Lloyd and {Kotov}, Ivan V. and {Krabbendam}, Victor L. and {Krughoff}, K. Simon and {Kub{\'a}nek}, Petr and {Kuczewski}, John and {Kulkarni}, Shri and {Ku}, John and {Kurita}, Nadine R. and {Lage}, Craig S. and {Lambert}, Ron and {Lange}, Travis and {Langton}, J. Brian and {Le Guillou}, Laurent and {Levine}, Deborah and {Liang}, Ming and {Lim}, Kian-Tat and {Lintott}, Chris J. and {Long}, Kevin E. and {Lopez}, Margaux and {Lotz}, Paul J. and {Lupton}, Robert H. and {Lust}, Nate B. and {MacArthur}, Lauren A. and {Mahabal}, Ashish and {Mandelbaum}, Rachel and {Markiewicz}, Thomas W. and {Marsh}, Darren S. and {Marshall}, Philip J. and {Marshall}, Stuart and {May}, Morgan and {McKercher}, Robert and {McQueen}, Michelle and {Meyers}, Joshua and {Migliore}, Myriam and {Miller}, Michelle and {Mills}, David J.},
 doi = {10.3847/1538-4357/ab042c},
 eid = {111},
 eprint = {0805.2366},
 journal = {\apj},
 month = {March},
 number = {2},
 pages = {111},
 primaryclass = {astro-ph},
 title = {{LSST: From Science Drivers to Reference Design and Anticipated Data Products}},
 volume = {873},
 year = {2019}
}

@article{Joachimi2021,
 adsurl = {https://ui.adsabs.harvard.edu/abs/2021A&A...646A.129J},
 archiveprefix = {arXiv},
 author = {{Joachimi}, B. and {Lin}, C. -A. and {Asgari}, M. and {Tr{\"o}ster}, T. and {Heymans}, C. and {Hildebrandt}, H. and {K{\"o}hlinger}, F. and {S{\'a}nchez}, A.~G. and {Wright}, A.~H. and {Bilicki}, M. and {Blake}, C. and {van den Busch}, J.~L. and {Crocce}, M. and {Dvornik}, A. and {Erben}, T. and {Getman}, F. and {Giblin}, B. and {Hoekstra}, H. and {Kannawadi}, A. and {Kuijken}, K. and {Napolitano}, N.~R. and {Schneider}, P. and {Scoccimarro}, R. and {Sellentin}, E. and {Shan}, H.~Y. and {von Wietersheim-Kramsta}, M. and {Zuntz}, J.},
 doi = {10.1051/0004-6361/202038831},
 eid = {A129},
 eprint = {2007.01844},
 journal = {\aap},
 month = {February},
 pages = {A129},
 primaryclass = {astro-ph.CO},
 title = {{KiDS-1000 methodology: Modelling and inference for joint weak gravitational lensing and spectroscopic galaxy clustering analysis}},
 volume = {646},
 year = {2021}
}

@article{Johnston2019,
 adsurl = {https://ui.adsabs.harvard.edu/abs/2019A&A...624A..30J},
 archiveprefix = {arXiv},
 author = {{Johnston}, Harry and {Georgiou}, Christos and {Joachimi}, Benjamin and {Hoekstra}, Henk and {Chisari}, Nora Elisa and {Farrow}, Daniel and {Fortuna}, Maria Cristina and {Heymans}, Catherine and {Joudaki}, Shahab and {Kuijken}, Konrad and {Wright}, Angus},
 doi = {10.1051/0004-6361/201834714},
 eid = {A30},
 eprint = {1811.09598},
 journal = {\aap},
 month = {April},
 pages = {A30},
 primaryclass = {astro-ph.CO},
 title = {{KiDS+GAMA: Intrinsic alignment model constraints for current and future weak lensing cosmology}},
 volume = {624},
 year = {2019}
}

@inproceedings{Kermish2012,
 adsurl = {https://ui.adsabs.harvard.edu/abs/2012SPIE.8452E..1CK},
 archiveprefix = {arXiv},
 author = {{Kermish}, Zigmund D. and {Ade}, Peter and {Anthony}, Aubra and {Arnold}, Kam and {Barron}, Darcy and {Boettger}, David and {Borrill}, Julian and {Chapman}, Scott and {Chinone}, Yuji and {Dobbs}, Matt A. and {Errard}, Josquin and {Fabbian}, Giulio and {Flanigan}, Daniel and {Fuller}, George and {Ghribi}, Adnan and {Grainger}, Will and {Halverson}, Nils and {Hasegawa}, Masaya and {Hattori}, Kaori and {Hazumi}, Masashi and {Holzapfel}, William L. and {Howard}, Jacob and {Hyland}, Peter and {Jaffe}, Andrew and {Keating}, Brian and {Kisner}, Theodore and {Lee}, Adrian T. and {Le Jeune}, Maude and {Linder}, Eric and {Lungu}, Marius and {Matsuda}, Frederick and {Matsumura}, Tomotake and {Meng}, Xiaofan and {Miller}, Nathan J. and {Morii}, Hideki and {Moyerman}, Stephanie and {Myers}, Mike J. and {Nishino}, Haruki and {Paar}, Hans and {Quealy}, Erin and {Reichardt}, Christian L. and {Richards}, Paul L. and {Ross}, Colin and {Shimizu}, Akie and {Shimon}, Meir and {Shimmin}, Chase and {Sholl}, Mike and {Siritanasak}, Praween and {Spieler}, Helmuth and {Stebor}, Nathan and {Steinbach}, Bryan and {Stompor}, Radek and {Suzuki}, Aritoki and {Tomaru}, Takayuki and {Tucker}, Carole and {Zahn}, Oliver},
 booktitle = {Millimeter, Submillimeter, and Far-Infrared Detectors and Instrumentation for Astronomy VI},
 doi = {10.1117/12.926354},
 editor = {{Holland}, Wayne S. and {Zmuidzinas}, Jonas},
 eid = {84521C},
 eprint = {1210.7768},
 month = {September},
 pages = {84521C},
 primaryclass = {astro-ph.IM},
 series = {Society of Photo-Optical Instrumentation Engineers (SPIE) Conference Series},
 title = {{The POLARBEAR experiment}},
 volume = {8452},
 year = {2012}
}

@article{Kilbinger2015,
 adsurl = {https://ui.adsabs.harvard.edu/abs/2015RPPh...78h6901K},
 archiveprefix = {arXiv},
 author = {{Kilbinger}, Martin},
 doi = {10.1088/0034-4885/78/8/086901},
 eid = {086901},
 eprint = {1411.0115},
 journal = {Reports on Progress in Physics},
 month = {July},
 number = {8},
 pages = {086901},
 primaryclass = {astro-ph.CO},
 title = {{Cosmology with cosmic shear observations: a review}},
 volume = {78},
 year = {2015}
}

@article{Klypin2016,
 adsurl = {https://ui.adsabs.harvard.edu/abs/2016MNRAS.457.4340K},
 archiveprefix = {arXiv},
 author = {{Klypin}, Anatoly and {Yepes}, Gustavo and {Gottl{\"o}ber}, Stefan and {Prada}, Francisco and {He{\ss}}, Steffen},
 doi = {10.1093/mnras/stw248},
 eprint = {1411.4001},
 journal = {\mnras},
 month = {April},
 number = {4},
 pages = {4340-4359},
 primaryclass = {astro-ph.CO},
 title = {{MultiDark simulations: the story of dark matter halo concentrations and density profiles}},
 volume = {457},
 year = {2016}
}

@article{Lamman2024,
 adsurl = {https://ui.adsabs.harvard.edu/abs/2024OJAp....7E..14L},
 archiveprefix = {arXiv},
 author = {{Lamman}, Claire and {Tsaprazi}, Eleni and {Shi}, Jingjing and {{\v{S}}ar{\v{c}}evi{\'c}}, Nikolina Niko and {Pyne}, Susan and {Legnani}, Elisa and {Ferreira}, Tassia},
 doi = {10.21105/astro.2309.08605},
 eid = {14},
 eprint = {2309.08605},
 journal = {The Open Journal of Astrophysics},
 month = {February},
 pages = {14},
 primaryclass = {astro-ph.CO},
 title = {{The IA Guide: A Breakdown of Intrinsic Alignment Formalisms}},
 volume = {7},
 year = {2024}
}

@article{Lange2023,
 adsurl = {https://ui.adsabs.harvard.edu/abs/2023MNRAS.525.3181L},
 archiveprefix = {arXiv},
 author = {{Lange}, Johannes U.},
 doi = {10.1093/mnras/stad2441},
 eprint = {2306.16923},
 journal = {\mnras},
 month = {October},
 number = {2},
 pages = {3181-3194},
 primaryclass = {astro-ph.IM},
 title = {{NAUTILUS: boosting Bayesian importance nested sampling with deep learning}},
 volume = {525},
 year = {2023}
}

@article{Lewis2000,
 adsurl = {https://ui.adsabs.harvard.edu/abs/2000ApJ...538..473L},
 archiveprefix = {arXiv},
 author = {{Lewis}, Antony and {Challinor}, Anthony and {Lasenby}, Anthony},
 doi = {10.1086/309179},
 eprint = {astro-ph/9911177},
 journal = {\apj},
 month = {August},
 number = {2},
 pages = {473-476},
 primaryclass = {astro-ph},
 title = {{Efficient Computation of Cosmic Microwave Background Anisotropies in Closed Friedmann-Robertson-Walker Models}},
 volume = {538},
 year = {2000}
}

@software{Lewis2011,
 adsurl = {https://ui.adsabs.harvard.edu/abs/2011ascl.soft02026L},
 archiveprefix = {ascl},
 author = {{Lewis}, Antony and {Challinor}, Anthony},
 eid = {ascl:1102.026},
 eprint = {1102.026},
 howpublished = {Astrophysics Source Code Library, record ascl:1102.026},
 month = {February},
 title = {{CAMB: Code for Anisotropies in the Microwave Background}},
 year = {2011}
}

@article{Lewis2019,
 adsurl = {https://ui.adsabs.harvard.edu/abs/2025JCAP...08..025L},
 archiveprefix = {arXiv},
 author = {{Lewis}, Antony},
 doi = {10.1088/1475-7516/2025/08/025},
 eid = {025},
 eprint = {1910.13970},
 journal = {\jcap},
 month = {August},
 number = {8},
 pages = {025},
 primaryclass = {astro-ph.IM},
 title = {{GetDist: a Python package for analysing Monte Carlo samples}},
 volume = {2025},
 year = {2025}
}

@article{Limber1953,
 adsurl = {https://ui.adsabs.harvard.edu/abs/1953ApJ...117..134L},
 author = {{Limber}, D. Nelson},
 doi = {10.1086/145672},
 journal = {\apj},
 month = {January},
 pages = {134},
 title = {{The Analysis of Counts of the Extragalactic Nebulae in Terms of a Fluctuating Density Field.}},
 volume = {117},
 year = {1953}
}

@article{Louis2025,
 adsurl = {https://ui.adsabs.harvard.edu/abs/2025JCAP...11..062L},
 archiveprefix = {arXiv},
 author = {{Louis}, Thibaut and {La Posta}, Adrien and {Atkins}, Zachary and {Jense}, Hidde T. and {Abril-Cabezas}, Irene and {Addison}, Graeme E. and {Ade}, Peter A.~R. and {Aiola}, Simone and {Alford}, Tommy and {Alonso}, David and {Amiri}, Mandana and {An}, Rui and {Austermann}, Jason E. and {Barbavara}, Eleonora and {Battaglia}, Nicholas and {Battistelli}, Elia Stefano and {Beall}, James A. and {Bean}, Rachel and {Beheshti}, Ali and {Beringue}, Benjamin and {Bhandarkar}, Tanay and {Biermann}, Emily and {Bolliet}, Boris and {Bond}, J. Richard and {Calabrese}, Erminia and {Capalbo}, Valentina and {Carrero}, Felipe and {Chen}, Shi-Fan and {Chesmore}, Grace and {Cho}, Hsiao-mei and {Choi}, Steve K. and {Clark}, Susan E. and {Cothard}, Nicholas F. and {Coughlin}, Kevin and {Coulton}, William and {Crichton}, Devin and {Crowley}, Kevin T. and {Darwish}, Omar and {Devlin}, Mark J. and {Dicker}, Simon and {Duell}, Cody J. and {Duff}, Shannon M. and {Duivenvoorden}, Adriaan J. and {Dunkley}, Jo and {Dunner}, Rolando and {Embil Villagra}, Carmen and {Fankhanel}, Max and {Farren}, Gerrit S. and {Ferraro}, Simone and {Foster}, Allen and {Freundt}, Rodrigo and {Fuzia}, Brittany and {Gallardo}, Patricio A. and {Garrido}, Xavier and {Gerbino}, Martina and {Giardiello}, Serena and {Gill}, Ajay and {Givans}, Jahmour and {Gluscevic}, Vera and {Goldstein}, Samuel and {Golec}, Joseph E. and {Gong}, Yulin and {Guan}, Yilun and {Halpern}, Mark and {Harrison}, Ian and {Hasselfield}, Matthew and {Healy}, Erin and {Henderson}, Shawn and {Hensley}, Brandon and {Herv{\'\i}as-Caimapo}, Carlos and {Hill}, J. Colin and {Hilton}, Gene C. and {Hilton}, Matt and {Hincks}, Adam D. and {Hlo{\v{z}}ek}, Ren{\'e}e and {Ho}, Shuay-Pwu Patty and {Hood}, John and {Hornecker}, Erika and {Huber}, Zachary B. and {Hubmayr}, Johannes and {Huffenberger}, Kevin M. and {Hughes}, John P. and {Ikape}, Margaret and {Irwin}, Kent and {Isopi}, Giovanni and {Joshi}, Neha and {Keller}, Ben and {Kim}, Joshua and {Knowles}, Kenda and {Koopman}, Brian J. and {Kosowsky}, Arthur and {Kramer}, Darby and {Kusiak}, Aleksandra and {Lagu{\"e}}, Alex and {Lakey}, Victoria and {Lee}, Eunseong and {Li}, Yaqiong and {Li}, Zack and {Limon}, Michele and {Lokken}, Martine and {Lungu}, Marius and {MacCrann}, Niall and {MacInnis}, Amanda and {Madhavacheril}, Mathew S. and {Maldonado}, Diego and {Maldonado}, Felipe and {Mallaby-Kay}, Maya and {Marques}, Gabriela A. and {van Marrewijk}, Joshiwa and {McCarthy}, Fiona and {McMahon}, Jeff and {Mehta}, Yogesh and {Menanteau}, Felipe and {Moodley}, Kavilan and {Morris}, Thomas W. and {Mroczkowski}, Tony and {Naess}, Sigurd and {Namikawa}, Toshiya and {Nati}, Federico and {Nerval}, Simran K. and {Newburgh}, Laura and {Nicola}, Andrina and {Niemack}, Michael D. and {Nolta}, Michael R. and {Orlowski-Scherer}, John and {Pagano}, Luca and {Page}, Lyman A. and {Pandey}, Shivam and {Partridge}, Bruce and {Perez Sarmiento}, Karen and {Prince}, Heather and {Puddu}, Roberto and {Qu}, Frank J. and {Ragavan}, Damien C. and {Ried Guachalla}, Bernardita and {Rogers}, Keir K. and {Rojas}, Felipe and {Sakuma}, Tai and {Schaan}, Emmanuel and {Schmitt}, Benjamin L. and {Sehgal}, Neelima and {Shaikh}, Shabbir and {Sherwin}, Blake D. and {Sierra}, Carlos and {Sievers}, Jon and {Sif{\'o}n}, Crist{\'o}bal and {Simon}, Sara and {Sonka}, Rita and {Spergel}, David N. and {Staggs}, Suzanne T. and {Storer}, Emilie and {Surrao}, Kristen and {Switzer}, Eric R. and {Tampier}, Niklas and {Thornton}, Robert and {Trac}, Hy and {Tucker}, Carole and {Ullom}, Joel and {Vale}, Leila R. and {Van Engelen}, Alexander and {Van Lanen}, Jeff and {Vargas}, Cristian and {Vavagiakis}, Eve M. and {Wagoner}, Kasey and {Wang}, Yuhan and {Wenzl}, Lukas and {Wollack}, Edward J. and {Zheng}, Kaiwen and {The Atacama Cosmology Telescope collaboration}},
 doi = {10.1088/1475-7516/2025/11/062},
 eid = {062},
 eprint = {2503.14452},
 journal = {\jcap},
 month = {November},
 number = {11},
 pages = {062},
 primaryclass = {astro-ph.CO},
 title = {{The Atacama Cosmology Telescope: DR6 power spectra, likelihoods and {\ensuremath{\Lambda}}CDM parameters}},
 volume = {2025},
 year = {2025}
}

@article{Loverde2008,
 adsurl = {https://ui.adsabs.harvard.edu/abs/2008PhRvD..78l3506L},
 archiveprefix = {arXiv},
 author = {{LoVerde}, Marilena and {Afshordi}, Niayesh},
 doi = {10.1103/PhysRevD.78.123506},
 eid = {123506},
 eprint = {0809.5112},
 journal = {\prd},
 month = {December},
 number = {12},
 pages = {123506},
 primaryclass = {astro-ph},
 title = {{Extended Limber approximation}},
 volume = {78},
 year = {2008}
}

@article{Madhavacheril2018,
 adsurl = {https://ui.adsabs.harvard.edu/abs/2018PhRvD..98b3534M},
 archiveprefix = {arXiv},
 author = {{Madhavacheril}, Mathew S. and {Hill}, J. Colin},
 doi = {10.1103/PhysRevD.98.023534},
 eid = {023534},
 eprint = {1802.08230},
 journal = {\prd},
 month = {July},
 number = {2},
 pages = {023534},
 primaryclass = {astro-ph.CO},
 title = {{Mitigating foreground biases in CMB lensing reconstruction using cleaned gradi ents}},
 volume = {98},
 year = {2018}
}

@article{Madhavacheril2024,
 adsurl = {https://ui.adsabs.harvard.edu/abs/2024ApJ...962..113M},
 archiveprefix = {arXiv},
 author = {{Madhavacheril}, Mathew S. and {Qu}, Frank J. and {Sherwin}, Blake D. and {MacCrann}, Niall and {Li}, Yaqiong and {Abril-Cabezas}, Irene and {Ade}, Peter A.~R. and {Aiola}, Simone and {Alford}, Tommy and {Amiri}, Mandana and {Amodeo}, Stefania and {An}, Rui and {Atkins}, Zachary and {Austermann}, Jason E. and {Battaglia}, Nicholas and {Battistelli}, Elia Stefano and {Beall}, James A. and {Bean}, Rachel and {Beringue}, Benjamin and {Bhandarkar}, Tanay and {Biermann}, Emily and {Bolliet}, Boris and {Bond}, J. Richard and {Cai}, Hongbo and {Calabrese}, Erminia and {Calafut}, Victoria and {Capalbo}, Valentina and {Carrero}, Felipe and {Challinor}, Anthony and {Chesmore}, Grace E. and {Cho}, Hsiao-mei and {Choi}, Steve K. and {Clark}, Susan E. and {C{\'o}rdova Rosado}, Rodrigo and {Cothard}, Nicholas F. and {Coughlin}, Kevin and {Coulton}, William and {Crowley}, Kevin T. and {Dalal}, Roohi and {Darwish}, Omar and {Devlin}, Mark J. and {Dicker}, Simon and {Doze}, Peter and {Duell}, Cody J. and {Duff}, Shannon M. and {Duivenvoorden}, Adriaan J. and {Dunkley}, Jo and {D{\"u}nner}, Rolando and {Fanfani}, Valentina and {Fankhanel}, Max and {Farren}, Gerrit and {Ferraro}, Simone and {Freundt}, Rodrigo and {Fuzia}, Brittany and {Gallardo}, Patricio A. and {Garrido}, Xavier and {Givans}, Jahmour and {Gluscevic}, Vera and {Golec}, Joseph E. and {Guan}, Yilun and {Hall}, Kirsten R. and {Halpern}, Mark and {Han}, Dongwon and {Harrison}, Ian and {Hasselfield}, Matthew and {Healy}, Erin and {Henderson}, Shawn and {Hensley}, Brandon and {Herv{\'\i}as-Caimapo}, Carlos and {Hill}, J. Colin and {Hilton}, Gene C. and {Hilton}, Matt and {Hincks}, Adam D. and {Hlo{\v{z}}ek}, Ren{\'e}e and {Ho}, Shuay-Pwu Patty and {Huber}, Zachary B. and {Hubmayr}, Johannes and {Huffenberger}, Kevin M. and {Hughes}, John P. and {Irwin}, Kent and {Isopi}, Giovanni and {Jense}, Hidde T. and {Keller}, Ben and {Kim}, Joshua and {Knowles}, Kenda and {Koopman}, Brian J. and {Kosowsky}, Arthur and {Kramer}, Darby and {Kusiak}, Aleksandra and {La Posta}, Adrien and {Lague}, Alex and {Lakey}, Victoria and {Lee}, Eunseong and {Li}, Zack and {Limon}, Michele and {Lokken}, Martine and {Louis}, Thibaut and {Lungu}, Marius and {MacInnis}, Amanda and {Maldonado}, Diego and {Maldonado}, Felipe and {Mallaby-Kay}, Maya and {Marques}, Gabriela A. and {McMahon}, Jeff and {Mehta}, Yogesh and {Menanteau}, Felipe and {Moodley}, Kavilan and {Morris}, Thomas W. and {Mroczkowski}, Tony and {Naess}, Sigurd and {Namikawa}, Toshiya and {Nati}, Federico and {Newburgh}, Laura and {Nicola}, Andrina and {Niemack}, Michael D. and {Nolta}, Michael R. and {Orlowski-Scherer}, John and {Page}, Lyman A. and {Pandey}, Shivam and {Partridge}, Bruce and {Prince}, Heather and {Puddu}, Roberto and {Radiconi}, Federico and {Robertson}, Naomi and {Rojas}, Felipe and {Sakuma}, Tai and {Salatino}, Maria and {Schaan}, Emmanuel and {Schmitt}, Benjamin L. and {Sehgal}, Neelima and {Shaikh}, Shabbir and {Sierra}, Carlos and {Sievers}, Jon and {Sif{\'o}n}, Crist{\'o}bal and {Simon}, Sara and {Sonka}, Rita and {Spergel}, David N. and {Staggs}, Suzanne T. and {Storer}, Emilie and {Switzer}, Eric R. and {Tampier}, Niklas and {Thornton}, Robert and {Trac}, Hy and {Treu}, Jesse and {Tucker}, Carole and {Ullom}, Joel and {Vale}, Leila R. and {Van Engelen}, Alexander and {Van Lanen}, Jeff and {van Marrewijk}, Joshiwa and {Vargas}, Cristian and {Vavagiakis}, Eve M. and {Wagoner}, Kasey and {Wang}, Yuhan and {Wenzl}, Lukas and {Wollack}, Edward J. and {Xu}, Zhilei and {Zago}, Fernando and {Zheng}, Kaiwen},
 doi = {10.3847/1538-4357/acff5f},
 eid = {113},
 eprint = {2304.05203},
 journal = {\apj},
 month = {February},
 number = {2},
 pages = {113},
 primaryclass = {astro-ph.CO},
 title = {{The Atacama Cosmology Telescope: DR6 Gravitational Lensing Map and Cosmological Parameters}},
 volume = {962},
 year = {2024}
}

@article{Maniyar2021,
 adsurl = {https://ui.adsabs.harvard.edu/abs/2021PhRvD.103h3524M},
 archiveprefix = {arXiv},
 author = {{Maniyar}, Abhishek S. and {Ali-Ha{\"\i}moud}, Yacine and {Carron}, Julien and {Lewis}, Antony and {Madhavacheril}, Mathew S.},
 doi = {10.1103/PhysRevD.103.083524},
 eid = {083524},
 eprint = {2101.12193},
 journal = {\prd},
 month = {April},
 number = {8},
 pages = {083524},
 primaryclass = {astro-ph.CO},
 title = {{Quadratic estimators for CMB weak lensing}},
 volume = {103},
 year = {2021}
}

@article{Marques2020,
 adsurl = {https://ui.adsabs.harvard.edu/abs/2020ApJ...904..182M},
 archiveprefix = {arXiv},
 author = {{Marques}, Gabriela A. and {Liu}, Jia and {Huffenberger}, Kevin M. and {Colin Hill}, J.},
 doi = {10.3847/1538-4357/abc003},
 eid = {182},
 eprint = {2008.04369},
 journal = {\apj},
 month = {December},
 number = {2},
 pages = {182},
 primaryclass = {astro-ph.CO},
 title = {{Cross-correlation between Subaru Hyper Suprime-Cam Galaxy Weak Lensing and Planck Cosmic Microwave Background Lensing}},
 volume = {904},
 year = {2020}
}

@article{McCarthy2017,
 adsurl = {https://ui.adsabs.harvard.edu/abs/2017MNRAS.465.2936M},
 archiveprefix = {arXiv},
 author = {{McCarthy}, Ian G. and {Schaye}, Joop and {Bird}, Simeon and {Le Brun}, Amandine M.~C.},
 doi = {10.1093/mnras/stw2792},
 eprint = {1603.02702},
 journal = {\mnras},
 month = {March},
 number = {3},
 pages = {2936-2965},
 primaryclass = {astro-ph.CO},
 title = {{The BAHAMAS project: calibrated hydrodynamical simulations for large-scale structure cosmology}},
 volume = {465},
 year = {2017}
}

@article{McCullough2024,
 adsurl = {https://ui.adsabs.harvard.edu/abs/2026PhRvD.113j3509M},
 archiveprefix = {arXiv},
 author = {{McCullough}, J. and {Amon}, A. and {Legnani}, E. and {Gruen}, D. and {Roodman}, A. and {Friedrich}, O. and {MacCrann}, N. and {Becker}, M. and {Myles}, J. and {Dodelson}, S. and {Samuroff}, S. and {Blazek}, J. and {Prat}, J. and {Pieres}, A. and {Fert{\'e}}, A. and {Alarcon}, A. and {Drlica-Wagner}, A. and {Choi}, A. and {Navarro-Alsina}, A. and {Campos}, A. and {Plazas Malag{\'o}n}, A.~A. and {Porredon}, A. and {Ross}, A.~J. and {Carnero Rosell}, A. and {Yin}, B. and {Flaugher}, B. and {Yanny}, B. and {S{\'a}nchez}, C. and {Chang}, C. and {Davis}, C. and {To}, C. and {Doux}, C. and {Brooks}, D. and {James}, D.~J. and {Sanchez Cid}, D. and {Hollowood}, D.~L. and {Huterer}, D. and {Rykoff}, E.~S. and {Gaztanaga}, E. and {Huff}, E.~M. and {Suchyta}, E. and {Sheldon}, E. and {Sanchez}, E. and {Tarsitano}, F. and {Andrade-Oliveira}, F. and {Castander}, F.~J. and {Bernstein}, G.~M. and {Gutierrez}, G. and {Giannini}, G. and {Tarle}, G. and {Diehl}, H.~T. and {Huang}, H. and {Harrison}, I. and {Sevilla-Noarbe}, I. and {Tutusaus}, I. and {Ferrero}, I. and {Elvin-Poole}, J. and {Marshall}, J.~L. and {Muir}, J. and {Weller}, J. and {Zuntz}, J. and {Carretero}, J. and {DeRose}, J. and {Frieman}, J. and {Cordero}, J. and {De Vicente}, J. and {Garc{\'\i}a-Bellido}, J. and {Mena-Fern{\'a}ndez}, J. and {Eckert}, K. and {Romer}, A.~K. and {Bechtol}, K. and {Herner}, K. and {Honscheid}, K. and {Kuehn}, K. and {Secco}, L.~F. and {da Costa}, L.~N. and {Paterno}, M. and {Soares-Santos}, M. and {Gatti}, M. and {Raveri}, M. and {Yamamoto}, M. and {Smith}, M. and {Carrasco Kind}, M. and {Troxel}, M.~A. and {Jarvis}, M. and {Swanson}, M.~E.~C. and {Weaverdyck}, N. and {Lahav}, O. and {Doel}, P. and {Wiseman}, P. and {Miquel}, R. and {Gruendl}, R.~A. and {Cawthon}, R. and {Allam}, S. and {Hinton}, S.~R. and {Bridle}, S.~L. and {Bocquet}, S. and {Desai}, S. and {Pandey}, S. and {Everett}, S. and {Lee}, S. and {Shin}, T. and {Aguena}, M. and {Alves}, O. and {Buckley-Geer}, E. and {Burke}, D.~L. and {Conselice}, C. and {Crocce}, M. and {Pereira}, M.~E.~S. and {Farahi}, A. and {Jeffrey}, N. and {Lima}, M. and {Palmese}, A. and {Schubnell}, M. and {Vikram}, V. and {Vincenzi}, M. and {Zhang}, Y. and {DES Collaboration}},
 doi = {10.1103/wmp3-qc3p},
 eid = {103509},
 eprint = {2410.22272},
 journal = {\prd},
 month = {May},
 number = {10},
 pages = {103509},
 primaryclass = {astro-ph.CO},
 title = {{Dark Energy Survey Year 3: Blue shear}},
 volume = {113},
 year = {2026}
}

@article{Mead2021,
 adsurl = {https://ui.adsabs.harvard.edu/abs/2021MNRAS.502.1401M},
 archiveprefix = {arXiv},
 author = {{Mead}, A.~J. and {Brieden}, S. and {Tr{\"o}ster}, T. and {Heymans}, C.},
 doi = {10.1093/mnras/stab082},
 eprint = {2009.01858},
 journal = {\mnras},
 month = {March},
 number = {1},
 pages = {1401-1422},
 primaryclass = {astro-ph.CO},
 title = {{HMCODE-2020: improved modelling of non-linear cosmological power spectra with baryonic feedback}},
 volume = {502},
 year = {2021}
}

@article{Mellier2025,
 adsurl = {https://ui.adsabs.harvard.edu/abs/2025A&A...697A...1E},
 archiveprefix = {arXiv},
 author = {{Euclid Collaboration} and {Mellier}, Y. and {Abdurro'uf} and {Acevedo Barroso}, J.~A. and {Ach{\'u}carro}, A. and {Adamek}, J. and {Adam}, R. and {Addison}, G.~E. and {Aghanim}, N. and {Aguena}, M. and {Ajani}, V. and {Akrami}, Y. and {Al-Bahlawan}, A. and {Alavi}, A. and {Albuquerque}, I.~S. and {Alestas}, G. and {Alguero}, G. and {Allaoui}, A. and {Allen}, S.~W. and {Allevato}, V. and {Alonso-Tetilla}, A.~V. and {Altieri}, B. and {Alvarez-Candal}, A. and {Alvi}, S. and {Amara}, A. and {Amendola}, L. and {Amiaux}, J. and {Andika}, I.~T. and {Andreon}, S. and {Andrews}, A. and {Angora}, G. and {Angulo}, R.~E. and {Annibali}, F. and {Anselmi}, A. and {Anselmi}, S. and {Arcari}, S. and {Archidiacono}, M. and {Aric{\`o}}, G. and {Arnaud}, M. and {Arnouts}, S. and {Asgari}, M. and {Asorey}, J. and {Atayde}, L. and {Atek}, H. and {Atrio-Barandela}, F. and {Aubert}, M. and {Aubourg}, E. and {Auphan}, T. and {Auricchio}, N. and {Aussel}, B. and {Aussel}, H. and {Avelino}, P.~P. and {Avgoustidis}, A. and {Avila}, S. and {Awan}, S. and {Azzollini}, R. and {Baccigalupi}, C. and {Bachelet}, E. and {Bacon}, D. and {Baes}, M. and {Bagley}, M.~B. and {Bahr-Kalus}, B. and {Balaguera-Antolinez}, A. and {Balbinot}, E. and {Balcells}, M. and {Baldi}, M. and {Baldry}, I. and {Balestra}, A. and {Ballardini}, M. and {Ballester}, O. and {Balogh}, M. and {Ba{\~n}ados}, E. and {Barbier}, R. and {Bardelli}, S. and {Baron}, M. and {Barreiro}, T. and {Barrena}, R. and {Barriere}, J.-C. and {Barros}, B.~J. and {Barthelemy}, A. and {Bartolo}, N. and {Basset}, A. and {Battaglia}, P. and {Battisti}, A.~J. and {Baugh}, C.~M. and {Baumont}, L. and {Bazzanini}, L. and {Beaulieu}, J.-P. and {Beckmann}, V. and {Belikov}, A.~N. and {Bel}, J. and {Bellagamba}, F. and {Bella}, M. and {Bellini}, E. and {Benabed}, K. and {Bender}, R. and {Benevento}, G. and {Bennett}, C.~L. and {Benson}, K. and {Bergamini}, P. and {Bermejo-Climent}, J.~R. and {Bernardeau}, F. and {Bertacca}, D. and {Berthe}, M. and {Berthier}, J. and {Bethermin}, M. and {Beutler}, F. and {Bevillon}, C. and {Bhargava}, S. and {Bhatawdekar}, R. and {Bianchi}, D. and {Bisigello}, L. and {Biviano}, A. and {Blake}, R.~P. and {Blanchard}, A. and {Blazek}, J. and {Blot}, L. and {Bosco}, A. and {Bodendorf}, C. and {Boenke}, T. and {B{\"o}hringer}, H. and {Boldrini}, P. and {Bolzonella}, M. and {Bonchi}, A. and {Bonici}, M. and {Bonino}, D. and {Bonino}, L. and {Bonvin}, C. and {Bon}, W. and {Booth}, J.~T. and {Borgani}, S. and {Borlaff}, A.~S. and {Borsato}, E. and {Bose}, B. and {Botticella}, M.~T. and {Boucaud}, A. and {Bouche}, F. and {Boucher}, J.~S. and {Boutigny}, D. and {Bouvard}, T. and {Bouwens}, R. and {Bouy}, H. and {Bowler}, R.~A.~A. and {Bozza}, V. and {Bozzo}, E. and {Branchini}, E. and {Brando}, G. and {Brau-Nogue}, S. and {Brekke}, P. and {Bremer}, M.~N. and {Brescia}, M. and {Breton}, M.-A. and {Brinchmann}, J. and {Brinckmann}, T. and {Brockley-Blatt}, C. and {Brodwin}, M. and {Brouard}, L. and {Brown}, M.~L. and {Bruton}, S. and {Bucko}, J. and {Buddelmeijer}, H. and {Buenadicha}, G. and {Buitrago}, F. and {Burger}, P. and {Burigana}, C. and {Busillo}, V. and {Busonero}, D. and {Cabanac}, R. and {Cabayol-Garcia}, L. and {Cagliari}, M.~S. and {Caillat}, A. and {Caillat}, L. and {Calabrese}, M. and {Calabro}, A. and {Calderone}, G. and {Calura}, F. and {Camacho Quevedo}, B. and {Camera}, S. and {Campos}, L. and {Ca{\~n}as-Herrera}, G. and {Candini}, G.~P. and {Cantiello}, M. and {Capobianco}, V. and {Cappellaro}, E. and {Cappelluti}, N. and {Cappi}, A. and {Caputi}, K.~I. and {Cara}, C. and {Carbone}, C. and {Cardone}, V.~F. and {Carella}, E. and {Carlberg}, R.~G. and {Carle}, M. and {Carminati}, L. and {Caro}, F. and {Carrasco}, J.~M. and {Carretero}, J. and {Carrilho}, P. and {Carron Duque}, J. and {Carry}, B.},
 doi = {10.1051/0004-6361/202450810},
 eid = {A1},
 eprint = {2405.13491},
 journal = {\aap},
 month = {May},
 pages = {A1},
 primaryclass = {astro-ph.CO},
 title = {{Euclid: I. Overview of the Euclid mission}},
 volume = {697},
 year = {2025}
}

@article{Millea2022,
 adsurl = {https://ui.adsabs.harvard.edu/abs/2022PhRvD.105j3531M},
 archiveprefix = {arXiv},
 author = {{Millea}, Marius and {Seljak}, Uro{\v{s}}},
 doi = {10.1103/PhysRevD.105.103531},
 eid = {103531},
 eprint = {2112.09354},
 journal = {\prd},
 month = {May},
 number = {10},
 pages = {103531},
 primaryclass = {astro-ph.CO},
 title = {{Marginal unbiased score expansion and application to CMB lensing}},
 volume = {105},
 year = {2022}
}

@article{Myles2021,
 adsurl = {https://ui.adsabs.harvard.edu/abs/2021MNRAS.505.4249M},
 archiveprefix = {arXiv},
 author = {{Myles}, J. and {Alarcon}, A. and {Amon}, A. and {S{\'a}nchez}, C. and {Everett}, S. and {DeRose}, J. and {McCullough}, J. and {Gruen}, D. and {Bernstein}, G.~M. and {Troxel}, M.~A. and {Dodelson}, S. and {Campos}, A. and {MacCrann}, N. and {Yin}, B. and {Raveri}, M. and {Amara}, A. and {Becker}, M.~R. and {Choi}, A. and {Cordero}, J. and {Eckert}, K. and {Gatti}, M. and {Giannini}, G. and {Gschwend}, J. and {Gruendl}, R.~A. and {Harrison}, I. and {Hartley}, W.~G. and {Huff}, E.~M. and {Kuropatkin}, N. and {Lin}, H. and {Masters}, D. and {Miquel}, R. and {Prat}, J. and {Roodman}, A. and {Rykoff}, E.~S. and {Sevilla-Noarbe}, I. and {Sheldon}, E. and {Wechsler}, R.~H. and {Yanny}, B. and {Abbott}, T.~M.~C. and {Aguena}, M. and {Allam}, S. and {Annis}, J. and {Bacon}, D. and {Bertin}, E. and {Bhargava}, S. and {Bridle}, S.~L. and {Brooks}, D. and {Burke}, D.~L. and {Carnero Rosell}, A. and {Carrasco Kind}, M. and {Carretero}, J. and {Castander}, F.~J. and {Conselice}, C. and {Costanzi}, M. and {Crocce}, M. and {da Costa}, L.~N. and {Pereira}, M.~E.~S. and {Desai}, S. and {Diehl}, H.~T. and {Eifler}, T.~F. and {Elvin-Poole}, J. and {Evrard}, A.~E. and {Ferrero}, I. and {Fert{\'e}}, A. and {Flaugher}, B. and {Fosalba}, P. and {Frieman}, J. and {Garc{\'\i}a-Bellido}, J. and {Gaztanaga}, E. and {Giannantonio}, T. and {Hinton}, S.~R. and {Hollowood}, D.~L. and {Honscheid}, K. and {Hoyle}, B. and {Huterer}, D. and {James}, D.~J. and {Krause}, E. and {Kuehn}, K. and {Lahav}, O. and {Lima}, M. and {Maia}, M.~A.~G. and {Marshall}, J.~L. and {Martini}, P. and {Melchior}, P. and {Menanteau}, F. and {Mohr}, J.~J. and {Morgan}, R. and {Muir}, J. and {Ogando}, R.~L.~C. and {Palmese}, A. and {Paz-Chinch{\'o}n}, F. and {Plazas}, A.~A. and {Rodriguez-Monroy}, M. and {Samuroff}, S. and {Sanchez}, E. and {Scarpine}, V. and {Secco}, L.~F. and {Serrano}, S. and {Smith}, M. and {Soares-Santos}, M. and {Suchyta}, E. and {Swanson}, M.~E.~C. and {Tarle}, G. and {Thomas}, D. and {To}, C. and {Varga}, T.~N. and {Weller}, J. and {Wester}, W.},
 doi = {10.1093/mnras/stab1515},
 eprint = {2012.08566},
 journal = {\mnras},
 month = {August},
 number = {3},
 pages = {4249-4277},
 primaryclass = {astro-ph.CO},
 title = {{Dark Energy Survey Year 3 results: redshift calibration of the weak lensing source galaxies}},
 volume = {505},
 year = {2021}
}

@article{Nakato2025,
 adsurl = {https://ui.adsabs.harvard.edu/abs/2026JCAP...03..031N},
 archiveprefix = {arXiv},
 author = {{Nakato}, Yuka and {Wu}, W.~L. Kimmy and {Oliveira}, Ana Carolina Silva and {Omori}, Yuuki and {Maniyar}, Abhishek S.},
 doi = {10.1088/1475-7516/2026/03/031},
 eid = {031},
 eprint = {2512.08908},
 journal = {\jcap},
 month = {March},
 number = {3},
 pages = {031},
 primaryclass = {astro-ph.CO},
 title = {{Foreground mitigation for CMB lensing with the global minimum variance quadratic estimator}},
 volume = {2026},
 year = {2026}
}

\end{document}